\documentclass[12pt]{report}
\usepackage{graphicx}
\usepackage{epsfig}
\usepackage{amscd}
\usepackage{amsmath}
\usepackage{amsfonts}
\usepackage{amssymb}
\begin{document}
\vspace{3cm}
\begin{center}
{\bf \Huge {The Energy-Momentum  Problem in General Relativity}}\\
\end{center}
\vspace{2cm}
\begin{center}
{\bf \Large Sibusiso S. Xulu}
\end{center}
\vspace{6cm}
\begin{center}
{Thesis presented for the Degree of}\\
{\ }\\
{\Large DOCTOR OF PHILOSOPHY}\\
{in}\\
{\Large APPLIED MATHEMATICS}\\{\ }\\
{Department of Mathematical Sciences}\\
{University of Zululand}
\end{center}
\vspace{2cm}
\begin{center}
submitted on {18 November 2002}\\
\end{center}
\thispagestyle{empty}
\pagenumbering{roman}
\newpage
\begin{center}
\textbf{{\Large DEDICATION}}
\end{center}
This work is dedicated to my mother, Thokozile Maria
(ukaMdlalose), and to my sister, Nozipho, who were both called to
rest while I was writing this thesis. \thispagestyle{empty}
\newpage
\begin{center}
\textbf{{\Large ACKNOWLEDGEMENTS}}
\end{center}
\addcontentsline{toc}{section}{Acknowledgement} I wish to express
my thanks to the following individuals and organizations for their
valuable contribution towards the completion of my thesis.
\begin{itemize}
  \item My supervisor, Dr K. S. Virbhadra, for his excellent
        guidance, constructive criticism, and encouragement. He devoted
        his valuable time to introduce me to this
       subject. Dr Virbhadra thank you very much for many stimulating
       discussions
       and, for believing in me against the odds. Your thorough supervision
       and  care in  my work helped me complete seven
       articles  which were all accepted in international journals.  Your
       contribution in this regards is unequalled. You are the
       best teacher I have ever met in my life. It was indeed a privilege
       to be your student.  My thanks also goes to your family members
       for their understanding when I usurped your attention.
  \item My co-supervisor, Prof. T. A. Dube, for encouragement and
        assistance.
  \item Prof. G. F. R. Ellis for hospitality at the university of
        Cape Town.
  \item The NRF for financial support.
  \item The University of Zululand Research Committee for
        financial support.
  \item My wife, Ntombi, and children, Lungi, Khayo and Hla for
        patience and support.
  \item I am also thankful to the anonymous examiners for helpful
        comments and suggestions.
\end{itemize}
\newpage
\begin{center}
\textbf{{\Large DECLARATION}}
\end{center}
\addcontentsline{toc}{section}{Declaration} This thesis represents
research work carried out by the author and has not been submitted
in any form to another university for a degree. All the sources I
have used have been duly acknowledged in the text.
\newpage
\begin{center}
\textbf{{\Large ABSTRACT}}
\end{center}
\addcontentsline{toc}{section}{Abstract} Energy-momentum is an
important conserved quantity whose definition has been a focus of
many investigations in general relativity. Unfortunately, there is
still no generally accepted definition of energy and momentum in
general relativity. Attempts aimed at finding a quantity for
describing distribution of energy-momentum due to matter,
non-gravitational and gravitational fields only resulted in
various energy-momentum complexes (these are nontensorial under
general coordinate transformations) whose physical meaning  have
been questioned. The problems associated with energy-momentum
complexes resulted in some researchers even abandoning the concept
of energy-momentum localization in favor of the alternative
concept of quasi-localization. However, quasi-local masses have
their inadequacies, while the remarkable work of Virbhadra and
some others, and recent results of Cooperstock  and Chang {\it et al.} have
revived an interest in various energy-momentum complexes. Hence in
this work we use energy-momentum complexes to obtain the energy
distributions in various space-times.

We elaborate on the problem of energy localization in general
relativity and use energy-momentum prescriptions of Einstein,
Landau and Lifshitz, Papapetrou, Weinberg, and M\o ller to
investigate energy distributions in various space-times. It is
shown that several of these energy-momentum complexes give the
same and acceptable results for a given space-time. This shows the
importance of these energy-momentum complexes. Our results agree
with Virbhadra's conclusion that the Einstein's energy-momentum
complex is still the best tool for obtaining energy distribution
in a given space-time. The Cooperstock hypothesis (that energy and
momentum in a curved space-time are confined to the the regions of
non-vanishing energy-momentum of matter and the non-gravitational
field) is also supported.

\newpage
\tableofcontents
\newpage
\pagenumbering{arabic}
\chapter{Introduction}

\label{chapter-TWO}
\pagenumbering{arabic}

The notions of energy-momentum together with conservation laws
play a fundamental role in any physical theory. The importance of
conservation of energy and momentum concepts was first clearly
identified more than two centuries ago within the Newtonian
mechanics of a closed system of mass points. In this case both the
linear momentum and the sum of kinetic and potential energies (for
conservative forces) are conserved. The connection between these
conservation laws and their invariance under Galilei group was
only detected around the year 1900 (see in Havas\cite{Havas}). In
continuum mechanics of elastic bodies the formulation of energy
conservation laws necessitated the introduction of the concept of
strain energy. Maxwell and Poynting had to introduce the concepts
of electric and magnetic energy densities in classical
electrodynamics to retain energy conservation. Einstein's famous
result of the special theory of relativity (SR) that mass is
equivalent to energy is a consequence of the requirement that the
law of conservation of energy and momentum should hold in all
inertial frames. As a matter of fact, this process of introducing
new forms of energy in order to retain conservation laws
characterizes the whole development of physics. However, this
practice involving the introduction of new kinds of energy ran
into serious difficulties with the arrival of the general theory
of relativity (GR). The main difficulty is with the expression
defining the gravitational field energy part.

Now, after more than eighty years of the success story of the
theory of General Relativity \textit{(this includes, amongst other
things, verification of the deflection of light by the Sun, the
perihelion advance of Mercury, the gravitational red shift of
light, and discoveries of quasars, cosmic fireball radiation,
pulsars, X-ray sources that might contain black holes, and the
present interest in the imminent detection of gravitational
waves)}, there is still no general agreement on the definition of
energy, and more generally, of conserved quantities associated
with the gravitational field. This dilemma in GR is highlighted in
an important paper by Penrose\cite{Penrose} in the following way:
``It is perhaps ironic that \emph{energy conservation} \ldots
which now has found expression in the (covariant)  equation

\begin{equation}\label{CovTeq}
  \nabla_{a}T^{ab}=0,
\end{equation}
\ldots
should nevertheless have found no universally applicable
formulation, within Einstein's theory, incorporating the energy of
gravity itself''. Indeed, Einstein's search for his generally
covariant field equations was not only guided by the principle of
equivalence but also by conservation laws of energy-momentum.
Conservation laws of energy-momentum played a major role in the
development of Einstein and Grossmann's so-called \textit{Entwurf}
theory\cite{Norton}. Although the field equations of the
\textit{Entwurf} theory were only of limited covariance, the
theory had all the essential features of Einstein's final theory
of GR. Einstein and Grossmann had considered the use of the Ricci
tensor in deriving almost covariant field equations but had
rejected these equations because of misconceptions which were
based on Einstein's earlier work on static gravitational fields
(for details see in Norton \cite{Norton}).

In his\cite{Einst16} derivation of the generally covariant
gravitational field equations, Einstein formulated the
energy-momentum  conservation law in the form:
\begin{equation}\label{EMconEq}
  \frac{\partial}{\partial x^{i}}(\sqrt{-g}(T_{j}^{\ i}+t_{j}^{\ i}))=0.
\end{equation}
With $T_{j}^{\ i}$ representing the stress energy density of
matter\footnote{"Matter" includes the energy contribution of all
non-gravitational fields.}, Einstein identified $t_{j}^{\ i}$ as
representing the stress energy density of gravitation. Einstein
noted that $t_{j}^{\ i}$ was not a tensor, but concluded that the
above equations $(\ref{EMconEq})$ hold good in all coordinate
systems since they were directly obtained from the principle of
general relativity. The choice of a nontensorial quantity to
describe the gravitational field energy immediately attracted some
criticism. Levi-Civita not only attacked Einstein's use of a
pseudotensor quantity (which is only covariant under linear
transformations) in describing the gravitational field energy, but
also suggested an alternative gravitational energy tensor which
required that Einstein's field equations be also interpreted as
conservation laws (for details see in Cattani and De Maria
\cite{CattDeMaria}, and in Pauli \cite{Pauli} ). Both
Schr\"{o}dinger and Bauer came up with counter examples to
Einstein's choice of a nontensor. Schr\"{o}dinger showed that by a
suitable choice of a coordinate system the pseudotensor of a
Schwarzschild solution vanishes everywhere outside the
Schwarzschild radius. Bauer's example illustrated that a mere
introduction of polar coordinates instead of quasi-Cartesian
coordinates into a system of inertia without matter present would
create a nonvanishing energy density in space. By resorting to the
equivalence principle and physical arguments, Einstein vigorously
defended the use of his pseudotensor to represent gravitational
field\cite{CattDeMaria}. The problems associated with Einstein's
pseudotensor resulted in many alternative definitions of energy,
momentum and angular momentum being proposed for a general
relativistic system (see Aguirregabiria \textit{et al
}\cite{ACV96} and references therein).

The lack of a generally accepted definition of energy distribution
in curved space-times has  led to doubts regarding the idea of
energy localization. The ambiguity in the localization of energy
is not a new physics problem, peculiar to the theory of GR, but is
also present in classical electrodynamics (Feynmann \textit{et
al}\cite{Feynmann}. In GR there is a dispute with the importance
of non-tensorial energy-momentum complexes\footnote{We use the
term energy-momentum \emph{complex} for one which satisfies the
local conservation laws and gives the contribution from the matter
(including all nongravitational fields) as well as the
gravitational field.} whose physical interpretation has been
questioned by a number of scientists, including Weyl, Pauli and
Eddington (see in Chandrasekhar and Ferrari \cite{ChaFer}). There
are suspicions that, in a given space-time, different energy
distributions would be obtained from different energy-momentum
complexes. However, Virbhadra and co-workers investigated several
examples of particular space-times (the Kerr-Newman, the
Einstein-Rosen, and the Bonnor-Vaidya) and found that different
energy-momentum complexes give the same energy distribution for a
given space-time. Several energy-momentum complexes have been
shown to coincide for any Kerr-Schild class metric \cite{ACV96}.
In this thesis we are extending the work of Virbhadra and
co-workers by considering further space-times in showing that the
energy-momentum complexes are useful in obtaining meaningful
energy distribution in a given geometry. In the rest of the
chapter we give a brief review of both the theory of special
relativity and general relativity so as to establish both the
notation and terminology which will be used in the rest of our
discussion. We also give a list of formulas needed in our later
discussion.

 \section{Tensor equations}
Einstein's 1905 paper on the special theory of relativity which
demanded a complete change of attitude towards space and time
prompted Minkowski to declare with great elegance, in his famous
1908 speech, that ``Henceforth space by itself and time by itself,
are doomed to fade away into mere shadows, and only a kind of
union of the two will preserve an independent reality.'' The
importance of this  is supported by the simplicity
obtained by formulating the physical laws in four-dimensional
space-time. In special relativity this simplicity is obtained by
using the Minkowski space-time:
\begin{equation}
ds^{2}=\eta _{pq}dx^{p}dx^{q},  \label{teq1}
\end{equation}
where\ $\eta _{pq}=diag(1,-,1,-1,-1)$. ({\em Throughout we use the
convention that summation occurs over dummy indices,
Latin indices take values from 0 to 3 and Greek indices values
from 1 to 3, and take }$G=1${\em \ and }$c=1${\em \ units. The
comma and semi-colon indicate ordinary and covariant differentiation,
respectively}.)\ The presence of
gravitation necessitates a generalization of the Minkowski
space-time into the four dimensional Riemannian space-time:
\begin{equation}
ds^{2}=g_{pq}dx^{p}dx^{q},  \label{teq2}
\end{equation}
where $g_{ab}=g_{ab}(x^{i})$ is the metric tensor, symmetric in
its indices, which characterizes the space-time completely. In the
rest of this section we present important formulas/properties in
Riemannian spaces required for our later work.

A contravariant metric tensor $g^{ab}$ is defined as
\begin{equation}
g^{ab}=\frac{\Delta ^{ab}}{g},  \label{teq3}
\end{equation}
where $\Delta ^{ab}$ is a cofactor of $g_{ab}$, while $g=\det
(g_{pq})$. From this definition it is obvious that $g^{ab}$ is
also symmetric in its indices and that:
\begin{equation}
g_{bp}g^{pa}=\delta _{b}^{a}.  \label{teq4}
\end{equation}
The relationship:
\begin{equation}
dg=gg^{pq}\,dg_{pq}=-gg_{pq}\,dg^{pq}  \label{teq5}
\end{equation}
 follows from $(\ref {teq3})$ and $(\ref{teq4})$. The metric tensor
 $g_{ab}$ transforms under coordinates transformation
$x^{a}\rightarrow x^{a^{\prime }}$ as:
\begin{equation}
g_{a^{\prime }b^{\prime }}=\frac{\partial x^{p}}{\partial x^{a^{\prime }}}%
\frac{\partial x^{q}}{\partial x^{b^{\prime }}}g_{pq},
\label{teq6}
\end{equation}
whereas its determinant $g$ transforms under coordinates
transformation as a scalar density of weight $+2$:
\begin{equation}
g^{\prime }=\left| \frac{\partial x}{\partial x^{^{\prime
}}}\right| ^{2}g ,\label{teq7}
\end{equation}
where $\left| \frac{\partial x}{\partial x^{^{\prime }}}\right| $
is the Jacobian of the coordinates transformation. In general, a
quantity which transforms like:
\begin{equation}
\Re _{b^{\prime }...}^{a^{\prime }...}=\left| \frac{\partial
x}{\partial
x^{^{\prime }}}\right| ^{W}\frac{\partial x^{a^{\prime }}}{\partial x^{p}}...%
\frac{\partial x^{q}}{\partial x^{b^{\prime }}}...\Re
_{q...}^{p...}
\end{equation}
is called a {\em tensor density} of weight $W$. Using $%
dx^{a^{\prime }}=\frac{\partial x^{a^{\prime }}}{\partial
x^{p}}dx^{p}$\ and $(\ref{teq7})$\ we can deduce that $\left(
-g\right) ^{-w/2}\,\Re _{b^{\prime }...}^{a^{\prime }...}$\  is an
ordinary tensor if $\Re _{b^{\prime }...}^{a^{\prime }...}$ is a
tensor density of weight $W$. Thus for a four-dimensional volume
element\ $d^{4}x$\ the quantity $\sqrt{-g}\,d^{4}x$\ is an
invariant. An important tensor density of weight $+1$, called the
{\em Levi-Civita} contravariant tensor density,  is defined as:
\begin{equation}
\varepsilon ^{abcd}=\left\{
\begin{array}{clcr}
&+1,\text{ if }abcd\text{ is an even permutation of }0123 \\
&-1,\text{ if }abcd\text{ is an odd permutation of }0123 \\
 &\ \ 0,\text{ otherwise.}
\end{array}
\right.  \label{teq9}
\end{equation}
$\varepsilon ^{abcd}$ is totally anti-symmetric and has the useful
property that its components are the same in all coordinate
systems. Any totally skew-symmetric tensor of order $4$ is
proportional to this tensorial quantity. The covariant component
Levi-Civita tensor density $ \varepsilon _{abcd}$ may then be
defined in terms of the contravariant component by the usual
lowering of indices as:
\begin{equation}
\varepsilon _{abcd}=(-g)g_{ap}g_{bq}g_{cr}g_{ds}\varepsilon
^{pqrs} .\label{teq10}
\end{equation}
The tensor density $\varepsilon _{abcd}$ will be of weight $-1$.
The above tensor densities can also be used to define the
following totally antisymmetric unit tensors of rank four. The
contravariant tensor $\epsilon ^{abcd}$ is defined as:
\begin{equation}
\epsilon ^{abcd}=\frac{\varepsilon ^{abcd}}{\sqrt{-g}},
\label{teq11}
\end{equation}
while the corresponding covariant tensor is defined by:
\begin{equation}
\epsilon _{abcd}=\sqrt{-g}\,\varepsilon _{abcd}.  \label{teq12}
\end{equation}
If $A_{ij}$ is an antisymmetric tensor, then using $\varepsilon
^{abcd}$ and $\epsilon ^{abcd}$ we may define:
\begin{equation}
^{\ast }{\mathcal{A}}^{ij}=\frac{1}{2}\varepsilon ^{ijpq}A_{pq},
\label{teq13}
\end{equation}
and
\begin{equation}
^{\ast }{A}^{ij}=\frac{1}{2}\epsilon ^{ijpq}A_{pq}, \label{teq14}
\end{equation}
where $^{\ast }{\mathcal{A}}^{ij}$ and $^{\ast }{A}^{ij}$ are said
to be a {\em dual pseudotensor} and a {\em dual tensor}
respectively to $A_{ij}$. Using $A^{ij}$ both $^{\ast
}{\mathcal{A}}_{ij}$ and $^{\ast }{A}_{ij}$ may be defined in a
similar way in terms of $\varepsilon _{abcd}$ and $\epsilon
_{abcd}$ respectively.

The motion of an infinitesimally small particle moving in a
gravitational field is described by the {\em geodesic equation}:
\begin{equation}
\frac{d^{2}x^{a}}{ds^{2}}+\Gamma _{pq}^{a}\frac{dx^{p}}{ds}\frac{dx^{q}}{ds}%
=0 , \label{teq17}
\end{equation}
which can be deduced from $(\ref{teq2})$ by taking the deviation
of the action integral:
\begin{equation}
I=\int ds , \label{teq18}
\end{equation}
where $\Gamma _{bc}^{a}$ are the Christoffel symbols of the second
kind given by:
\begin{equation}
\Gamma _{bc}^{a}=\frac{1}{2}g^{ap}\left(
g_{bp,c}+g_{cp,b}-g_{bc,p}\right) . \label{teq19}
\end{equation}
By contracting the pair of indices $a$ and $c$ in the above
$(\ref{teq19})$ we obtain
\begin{equation}\label{teq19a}
   \Gamma _{bp}^{p}=\frac{1}{2}g^{pq}g_{pq,b},
\end{equation}
and combining this result with that of Eqn $(\ref{teq5})$ we
deduce that:
\begin{equation}\label{teq19b}
  \Gamma _{bp}^{p}=\frac{1}{2g}\frac{\partial g}{\partial x^{b}}=\frac{\partial\ln\sqrt{-g}}{\partial
  x^{b}}.
\end{equation}
The identity that the $g^{ab}_{~~;c}=0$ gives us the following
relationship:
\begin{equation}\label{teq19c}
  g^{ab}_{~~,c}=-g^{ap}\Gamma _{pc}^{b}-g^{pb}\Gamma _{pc}^{a}.
\end{equation}
>From the definition of covariant differentiation we may deduce the
following useful relation:
\begin{equation}\label{teq19d}
  A_{a;b}-A_{b;a}=A_{a,b}-A_{b,a}.
\end{equation}

The {\em Riemann curvature tensor} is defined using the above
Christoffel symbols as:
\begin{equation}
R_{~bcd}^{a}=\Gamma _{bd,c}^{a}-\Gamma _{bc,d}^{a}+\Gamma
_{pc}^{a}\Gamma _{bd}^{p}-\Gamma _{pd}^{a}\Gamma _{bc}^{p},
\label{teq20}
\end{equation}
which vanishes if and only if the space-time under consideration
is flat. In {\em geodesic coordinate system} (i.e. coordinate
system where components of Christoffel symbols vanish at the pole
of this special coordinate system) the Riemann curvature tensor
takes the simpler form:
\begin{equation}
R_{~bcd}^{a}=\Gamma _{bd,c}^{a}-\Gamma _{bc,d}^{a} . \label{RCTeq}
\end{equation}
By lowering the upper index of the Riemann tensor $(\ref{teq20})$ to $%
R_{abcd}=g_{ap}R_{~bcd}^{p}$ then its symmetry properties become
easy to
study. $R_{abcd}$ is anti-symmetric in each of the pair of indices $ab$ and $%
cd$, and is symmetric under the exchange of these pairs of indices
with each other. The cyclic sum of components of $R_{abcd}$\
formed by a permutation of any three indices is equal to zero.
Several important tensors may be constructed using this tensor. By
contraction of the Riemann curvature tensor we obtain the {\em
Ricci tensor} defined by
\begin{eqnarray}
R_{ab} &=&R_{~apb}^{p}=g^{pq}R_{paqb}  \label{teq22a} \\
&=&\Gamma _{ab,p}^{p}-\Gamma _{ap,b}^{p}+\Gamma _{ab}^{q}\Gamma
_{pq}^{p}-\Gamma _{ap}^{q}\Gamma _{bq}^{p},  \label{teq22b}
\end{eqnarray}
which is symmetric in its indices. The {\em Ricci scalar
curvature} is:
\begin{equation}
R=R_{~p}^{p}=g^{pq}R_{pq} . \label{Riccieq}
\end{equation}
The {\em Einstein tensor} is defined in terms of the Ricci tensor
as:
\begin{equation}
G_{ab}=R_{ab}-\frac{1}{2}g_{ab}\,R . \label{teq24}
\end{equation}
Similarly the above Einstein's tensor is symmetric in its indices.
The traceless {\em Weyl conformal tensor} $C_{abcd}$ is again
defined in terms of Riemann tensor as:
\begin{eqnarray}
C_{abcd} &=&R_{abcd}-\frac{1}{2}\left(
g_{ac}R_{bd}-g_{ad}R_{bc}-g_{bc}R_{ad}+g_{bd}R_{ac}\right)  \nonumber \\
&&+\frac{1}{6}\left( g_{ac}g_{bd}-g_{ad}g_{bc}\right) .
\label{teq25}
\end{eqnarray}
The symmetry properties of $C_{abcd}$ are similar to those of the
Riemann curvature tensor. The importance of the Weyl tensor is in
its invariance under conformal mapping of two Riemannian spaces.
The Bianchi identities are given by
\begin{equation}
\nabla _{a}R_{~qbc}^{p}+\nabla _{b}R_{~qca}^{p}+\nabla
_{c}R_{~qab}^{p}=0 .\label{Bianchi}
\end{equation}
These identities take a much simpler form if expressed in terms of the dual $%
^{\ast }R_{pqrs}$\ to the Riemann tensor $R_{abcd}$
\begin{equation}
\nabla ^{p}\,^{\ast }R_{abcp}=0.  \label{teq26b}
\end{equation}
By using the fact that the covariant derivatives of the metric
tensor vanishes and contracting the Bianchi identities
\ref{Bianchi} we get the following important relationship:
\begin{equation}\label{EinBianchi}
  \nabla_{p}G^{~p}_{a}=0,
\end{equation}
or the following equivalent identity:
\begin{equation}\label{RicciId}
    \nabla_{p}R^{~p}_{a}=\frac{1}{2}\frac{\partial R}{\partial
    x^{a}}.
\end{equation}

\section{Special relativity}


The idea of relativity can be traced back to Galileo who was the
first to state clearly the concept of relative motion. His example
of a boat which is in uniform motion illustrates that no
experiment performed in a sealed cabin can be able to detect
motion. According to the Galilean principle of relativity all laws
of mechanics are the same in all inertial reference frames. The
space $x^{\alpha}\rightarrow x^{\alpha\prime }$ and time
$t\rightarrow t^{\prime }$ coordinates transformations between any
two inertial frames, \emph{K} and \emph{K'}, may be written as:
\begin{eqnarray}
x^{\alpha\prime } &=&M^{\alpha}_{~\beta}\,x^\beta+v^{\alpha}\,t+d^\alpha,\nonumber \\
t^{\prime } &=&t+t_0,  \label{GalTrEq}
\end{eqnarray}
where constants $M^{\alpha}_{~\beta}$, $v^{\alpha}$, $d^\alpha$
and $t_0$ respectively represent rotations, uniform motion, space
and time translations of origins of reference systems.  The
so-called \textit{Galilean transformations} $(\ref{GalTrEq})$ form
a ten-parameters group of, namely three rotation Euler angles in
the orthogonal matrix $M$, three components in each of
$v^{\alpha}$ and $d^\alpha$, and lastly the constant $t_0$.
Newton's fundamental equations of classical mechanics, based on
the concepts of absolute space and absolute time, are invariant in
all inertial reference frames under Galilean coordinate
transformations. If we consider three inertial coordinates frames
$K_1$, $K_2$, and $K_3$, then using the group property of Galilean
transformations we obtain the  following velocity addition law:
\begin{equation}\label{GalVelEq}
  v_\mathbf{13}^\alpha=v_\mathbf{12}^\alpha+v_\mathbf{23}^\alpha ,
\end{equation}
where $v_\mathbf{ab}^\alpha$ is velocity of $K_b$ relative to
$K_a$. This velocity addition law (\ref{GalVelEq}) is valid in
Newtonian mechanics. However, Maxwell's theory of classical
electrodynamics implied that light travels in vacuum at a constant
speed $c$. Obviously, according to Galilean velocity addition law
light could not have the same speed $c$ with respect to arbitrary
inertial frames. One can also verify that the wave equation for
electrodynamics in free space is not invariant under Galilean
transformations. The fact that Newton's equations of mechanics are
invariant under Galilean transformations while the Maxwell
equations are not, leads one to enquire whether it is possible to
find some principle of relativity which holds for both mechanics
and electrodynamics, but where
\begin{itemize}
  \item Newton's laws of mechanics are not correct, or
  \item Maxwell's laws of electrodynamics are not correct?
\end{itemize}
Experimental evidence indicating deviations from either the
Newtonian or Maxwellian theory is required in order to decide
whether the laws of mechanics or electrodynamics need to be
reformulated.

The famous Michelson-Morley experiment showed that the velocity of
light is the same for light travelling along the direction of the
earth's orbital motion and transverse to it. Numerous attempts
were made  to explain the Michelson-Morley null result of the
earth's motion through ether. Fitzgerald and Lorentz independently
suggested that material bodies contract in the direction of the
their motion by a factor $\sqrt{1-v^{2}/c^{2}}$. Lorentz explained
the contraction hypothesis  in terms of an electromagnetic model
of matter, and introduced length transformation and the concept of
a `local time'.  Poincar\'{e} called for the development of a new
mechanics to replace the Newtonian mechanics (see in Schr\"{o}der
\cite{Schroder}). He showed that the Maxwell equations {\it in
vacuo} are invariant under the Lorentz transformations. He further
showed that Lorentz transformations form a group\cite{Schroder}.
The inhomogeneous Lorentz space-time $x^{a}\rightarrow
x^{a^{\prime }}$ coordinate transformations , also called the
Poincar\'{e} group, are given by:
\begin{equation}
x^{a^{\prime }}=\Lambda _{p}^{a}\,x^{p}+d^{a},  \label{LorTrEq}
\end{equation}
where constants $d^{a}$ denote space and time translations, while
$\Lambda_{b}^{a}$ represent rotations and uniform motion of the
origins of inertial frames. The matrix $\Lambda$ must satisfy the
following condition:
\begin{equation}
\Lambda _{a}^{p\,}\Lambda _{b}^{q} \eta _{pq}=\eta _{ab},
\label{STReq4}
\end{equation}
with
\begin{equation}
\eta _{pq}=\eta ^{pq}=diag(1,-1,-1,-1).  \label{STReq5}
\end{equation}
 The so-called \emph{proper} Lorentz group is obtained by imposing the
 following additional conditions:
\begin{equation}\label{STReq5a}
 \Lambda _{0}^{0\,}\geq 1,\mbox{   }det(\Lambda )=1.
\end{equation}
If there are no translations, i.e. $d^a=0$, we get the homogeneous
Lorentz group.

The incompatibility of mechanics with electrodynamics was finally
resolved by Albert Einstein who proposed a replacement of the
Galilean transformations by the Lorentz transformations.
Einstein's theory of special relativity (STR) is based on the
following postulates:
\begin{itemize}
  \item \textbf{The principle of relativity.} The laws of physics
  are the same in all inertial reference frames.
  \item \textbf{The constancy of the speed of light.} The speed of
  light in vacuum is the same for all inertial observers
  irrespective of the motion of the source.
\end{itemize}
Following Einstein's approach, then the above Lorentz
transformations $(\ref{LorTrEq})$ become a consequence of the
constancy of the speed of light. As already mentioned, in special
relativity physical laws are formulated in four-dimensional
Minkowski space-time. The interval $I_{AB}$ between two events
$A$ and $B$, with coordinates $\left( t_{A},x_{A},y_{A},z_{A}\right) $%
\ and $\left( t_{B},x_{B},y_{B},z_{B}\right) $, given by:
\begin{equation}
I_{AB}=(x_{A}^{0}-x_{B}^{0})^{2}-(x_{A}^{1}-x_{B}^{1})^{2}-
(x_{A}^{2}-x_{B}^{2})^{2}-(x_{A}^{3}-x_{B}^{3})^{2},
\label{STReq6}
\end{equation}
\begin{figure*}
\centerline{ \epsfxsize 15cm
   \epsfbox{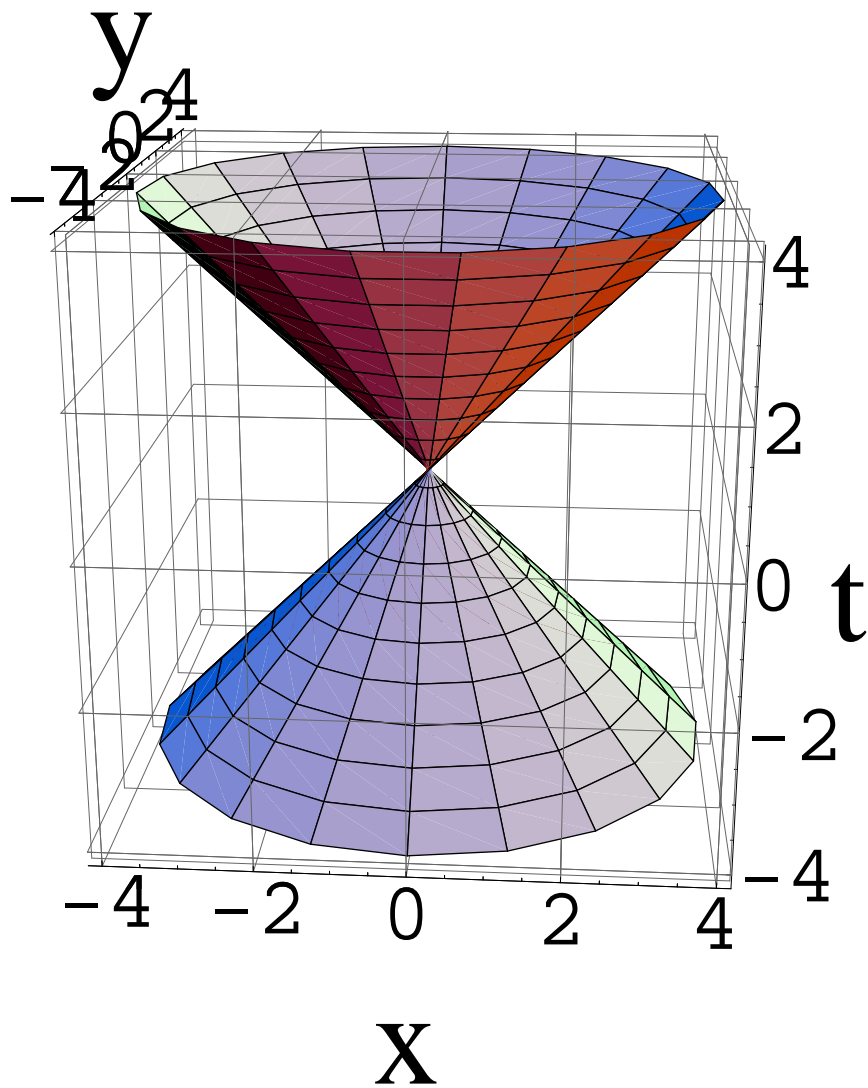}}
 \caption[ ]
   {$t>0$ and $t<0$ represent forward and backward null cones, respectively.}
\label{fig1}
\end{figure*}
is an invariant under Lorentz transformations. (Note that $x^0=ct$
and that $c=1$). $I_{AB}$ is said to be \textit{timelike} if
$I_{AB}>0,$ \textit{spacelike} if $I_{AB}<0,$ and
\textit{lightlike} or \textit{null} if $I_{AB}=0$. For a timelike
interval of two events, it is possible to do a Lorentz
transformation to an inertial frame where the two events occur at
the same point, but there is no inertial frame where they occur at
the same time. Therefore, for any two timelike events if
$x_{A}^{0}>x_{B}^{0}$ then $A$ is said to be in the absolute
future of $B$, while if $x_{A}^{0}<x_{B}^{0}$ then $A$ is said to
be in the absolute past of $B$. In a spacelike interval for any
two events it is always possible to do a Lorentz transformation to
an inertial frame where the two events occur at the same time, but
it is impossible to choose a frame where they occur at the same
point.

In Minkowski space-time the motion of a mass point is described by
the world line $x^i=x^i(\lambda)$, where $\lambda$ is parameter
determining motion. Now, by observing that the proper time $d\tau $, given by:
\begin{equation}
d\tau ^{2}=\eta _{pq}\,dx^{p}\,dx^{q},  \label{STReq7}
\end{equation}
is an invariant under Lorentz coordinate transformations,the
velocity four-vector $u^i$ and the acceleration four-vector are
defined with respect to the proper time as:
\begin{equation}\label{VelEq}
  u^i=\frac{dx^i}{d\tau},
\end{equation}
and
\begin{equation}\label{AccEq}
  \dot{u^i}=\frac{du^i}{d\tau}.
\end{equation}
Newton's equations, in their original form, are no longer
invariant under the Lorentz transformations, but using the above
definitions the four-vector relativistic force of Newton's second
law takes the form:
\begin{equation}
f^{a}=m\frac{d^{2}x^{a}}{d\tau ^{2}},  \label{STReq8}
\end{equation}
or in terms of the relativistic energy-momentum:
\begin{equation}
p^{a}=m\frac{dx^{a}}{d\tau },  \label{STReq9}
\end{equation}
the relativistic force may be expressed by:
\begin{equation}
f^{a}=\frac{dp^{a}}{d\tau },  \label{STReq10}
\end{equation}
where $p^{a}$\ is also a four-vector. $f^a$ is commonly referred
to as the Minkowski force. The spatial components of the
four-vector Minkowski force $f^a=(f^0, \mathbf{f})$ are related to
the Newtonian force $\mathbf{F}$ by the following:
\begin{equation}\label{}
  \mathbf{f}=\gamma \mathbf{F},
\end{equation}
and its time component is given by
\begin{equation}\label{}
  f^0=\gamma \mathbf{F\cdot v},
\end{equation}
where ${\bf v:=}\frac{d{\bf x}}{dt}$ and $\gamma :=\left( 1-{\bf v}%
^{2}\right) ^{-1/2}$. The time component of $p^{a}$ is the energy:
\begin{equation}
p^{0}=E=m\gamma , \label{STReq11}
\end{equation}
and the space components $p^{\alpha }$ form the momentum vector:
\begin{equation}
{\bf p}=m\gamma {\bf v}.  \label{STReq12}
\end{equation}
Thus the four-vector $p^a$ is called the energy-momentum vector.
This energy-momentum vector is conserved in all inertial reference
frames related by Lorentz transformations. For a particle with
rest mass $m>0$ we have that:
\begin{equation}\label{STReq12a}
  p_a p^a = m^2,
\end{equation}
which gives the following equation connecting the energy and
momentum\footnote{If we now make use of the quantum mechanical
correspondence between numbers and operators: $\mathbf{p}\mapsto
-i \mathbf{\nabla}, \mbox{   } E\mapsto -i
\frac{\partial}{\partial t}$ in $(\ref{EnMoEq})$ then one obtains
the Klein-Gordon wave equation
\begin{equation}\label{}
  (\Box + m^2)\phi (x)=0,
\end{equation}
for a scalar field $\phi$ with mass $m$.}:
\begin{equation}\label{EnMoEq}
  E^2-\mathbf{p}^2 = m^2.
\end{equation}
Taking the ratio of equations $(\ref{STReq11})$ and
$(\ref{STReq12})$ we get the following useful expression:
\begin{equation}
\frac{{\bf p}}{E}={\bf v}.  \label{STReq13}
\end{equation}

The Maxwell equations of electrodynamics in vacuum:
\begin{eqnarray}
\nabla \cdot{\bf E} &=&4\pi\rho ,  \label{MxlEq1} \\
\nabla \times {\bf B}-\frac{\partial {\bf E}}{\partial t} &=&4\pi
{\bf j},
\label{MxlEq2} \\
\nabla \cdot {\bf B} &=&0,  \label{MxlEq3} \\
\nabla \times {\bf E}+\frac{\partial {\bf B}}{\partial t} &=&{\bf
0}, \label{MxlEq4}
\end{eqnarray}
where $\mathbf{E}$ and $\mathbf{B}$ are respectively the electric
and magnetic fields, are invariant under Lorentz coordinate
transformations. This may be illustrated by expressing the above
equations in a covariant form in terms of an anti-symmetric
electromagnetic field tensor $F_{ab}$ which may be defined as:
\begin{equation}
F^{ab}:=\left[
\begin{array}{cccc}
0 & E_{x} & E_{y} & E_{z} \\
-E_{x} & 0 & B_{z} & -B_{y} \\
-E_{y} & -B_{z} & 0 & B_{x} \\
-E_{z} & B_{y} & -B_{x} & 0
\end{array}
\right]   \label{MxlTeq}
\end{equation}
Then using Eq. $(\ref{MxlTeq})$ the Maxwell equations $(\ref
{MxlEq1},\ref{MxlEq2},\ref{MxlEq3},\ref{MxlEq4})$ may now be
reduced  to the following two tensorial equations:
\begin{eqnarray}
\frac{\partial F^{ab}}{\partial x^{b}} &=&-4\pi j^{a},  \label{MxlCovEq1} \\
\frac{\partial ^*F^{ab}}{\partial x^{b}} &=&0,  \label{MxlCovEq2}
\end{eqnarray}
where $ ^*F^{ab}$ is the dual tensor of $F^{ab}$, $j^a$ is the
four-vector of current density with the components:
\begin{equation}\label{CurDeq}
  j^a=(\rho,\mathbf{\rho \mathbf{v}}),
\end{equation}
with $\rho$ being the charge density, and $\mathbf{v}$ the
velocity of charge. The equation of continuity for the current
density:
\begin{equation}\label{ContEq}
  \frac{\partial j^{a}}{\partial x^{a}}=0,
\end{equation}
now follows directly from Eq.$(\ref{MxlCovEq1})$.
Eq.$(\ref{MxlCovEq2})$ can also be expressed as:
\begin{equation}\label{MxlCovEq2b}
  \frac{\partial }{\partial x^{a}}F_{bc}+\frac{\partial }{\partial
  x^{b}}F_{ca}+\frac{\partial }{\partial x^{c}}F_{ab}=0.
\end{equation}
The {\em relativistic electromagnetic force} on a charged particle
may be written in terms of the Maxwell tensor as:
\begin{equation}
f^{a}=eF_{~p}^{a}\frac{dx^{p}}{d\tau },  \label{EMFeq}
\end{equation}
where $e$ is the charge parameter.

\section{The Energy-momentum tensor}
The energy-momentum of the electromagnetic field in free-space may
be obtained by considering the action integral formed with the
Lagrangian density:
\begin{equation}\label{LagEMeq}
  \mathcal{L}_\mathbf{F}=-\frac{1}{16\pi}F_{ab}F^{ab},
\end{equation}
which gives the following contravariant components of the
energy-momentum tensor:
\begin{equation}\label{EMTeq1a}
  T^{ab}=-\frac{1}{4\pi}\frac{\partial A^p}{\partial
  x_a}F^b_{~p}+\frac{1}{16\pi}\eta^{ab}F_{pq}F^{pq},
\end{equation}
where $A^a$ is the four-vector of the potentials. This tensor is
then symmetrized by adding the quantity $\frac{1}{4\pi}
\frac{\partial A^a}{\partial x_p}F^b_{~p}$ and using the fact that
in the absence of charges then Eq.$(\ref{MxlCovEq1})$ takes the
form:\[\frac{\partial F^{ab}}{\partial x^{b}}=0,\] which  finally
leads to the following symmetric energy-momentum tensor
expression:
\begin{equation}\label{EMTeq1b}
  T^{ab}=\frac{1}{4\pi}(-F^{ap}F^b_{~p}+\frac{1}{4}\eta^{ab}F_{pq}F^{pq}).
\end{equation}
This tensor $(\ref{EMTeq1b})$ is gauge invariant and has a
vanishing trace, $T^a_{~a}=0$ and most importantly it satisfies:
\begin{equation}\label{DivLeq}
  T^a_{~b,a}=0 .
\end{equation}
The fact that $T^a_{~b}$ is divergeless  indicates that the
four-vector:
\begin{equation}\label{EMVeq1}
P^a=\int_\sigma T^{ab}dS_b ,
\end{equation}
is conserved, where $\sigma$ is an arbitrary spacelike
hypersurface. Now since the integral $(\ref{EMVeq1})$ is
independent of the hypersurface $\sigma$, choosing the
hypersurface $x^0=constant$, i.e. a three-dimensional space, then:
\begin{equation}\label{EMVeq2}
P^a=\int T^{a0}d^3x ,
\end{equation}
may be identified as the total energy-momentum four-vector of the
field. The angular momentum tensor:
\begin{equation}\label{AngMTeq1}
  M^{abc}=T^{ab}x^c-T^{ac}x^b,
\end{equation}
defined in terms of $T^{ab}$, satisfies:\[M^{abc}_{~~~,a}=0,\]
which shows that the total field angular momentum
\begin{equation}\label{AngMTeq2}
  J^{ab}=\int M^{0ab}d^3x,
\end{equation}
is also conserved.

The energy-momentum tensor $(\ref{EMTeq1b})$ is only defined for
charge-free fields. In the presence of charged objects then we
should not only consider contributions from the electromagnetic
field but also from the charged objects themselves. Thus the
energy-momentum tensor of field together with charges is:
\[T^{ab}=T^{ab}_\mathbf{F}+T^{ab}_\mathbf{M},\]
where subscripts $\mathbf{F}$ and $\mathbf{M}$ indicate field and
matter contributions, respectively. Now we should have that:
\begin{equation}\label{EMconsEq}
  \frac{\partial}{\partial x^{a}} (T^{ab}_\mathbf{F}+T^{ab}_\mathbf{M})=0,
\end{equation}
where $T^{ab}_\mathbf{F}$ is given by $(\ref{EMTeq1b})$. Now
differentiating equation $(\ref{EMTeq1b})$ and simplifying, and
making use of the inhomogeneous Maxwell equations
$(\ref{MxlCovEq1})$ together with $(\ref{MxlCovEq2})$ we get
\begin{equation}\label{EFeq}
  \frac{\partial}{\partial x^{a}} T^{ab}_{\mathbf{F}}=-
  F^{bp}j_p .
\end{equation}
For a system of non-interacting particles, the energy momentum
tensor is given by
\begin{equation}\label{}
  T^{ab}_\mathbf{M}=\mu u^a u^b \frac{ds}{dt},
\end{equation}
where $\mu$  defined as:
\begin{equation}\label{}
  \mu=\mathrel{\mathop{\sum }\limits_{\mbox{n}}}m_{\mbox{n}}
  \delta(\mathbf{r-r_{\mbox{n}}}),
\end{equation}
may be termed the `mass density' since it indicates the continuous
mass distribution in space. Noting that mass should be conserved
for non-interacting particles, we have:
\begin{equation}\label{}
  \frac{\partial}{\partial x^{a}}(\mu \frac{ds}{dt})=0,
\end{equation}
and making use of  Eq. $(\ref{EMFeq})$ we get
\begin{equation}\label{EMMeq}
  \frac{\partial}{\partial x^{a}} T^{ab}_{\mathbf{M}}=
  F^{bp}j_p .
\end{equation}
Hence equations $(\ref{EFeq})$ and $(\ref{EMMeq})$ show that
equation $(\ref{EMconsEq})$ is indeed satisfied. From this
discussion it is clear that electromagnetic fields are determined
by the motion of charges, while on the other hand the motion of
charges is determined by the fields, i.e. the two systems are
interdependent on each other.

\section{The Einstein field equations}

In this section we give a derivation of the Einstein field
equations for gravitation from a variational principle, but before
that we give both the equivalence principle and the principle of
general covariance which guided Einstein in his search for
equations of gravitation. The equivalence principle is based on
the equality of the gravitational mass and the inertial mass.
Several experiments have been performed, starting from Galileo,
Newton, Bessel, E\"{o}tv\"{o}s, and many others, to investigate
whether or not there is a difference between the gravitational and
inertial masses of objects. The E\"{o}tv\"{o}s torsion balance
experiment showed with a high degree of accuracy the equality of
the inertial and gravitational mass. The principle of equivalence
 may be formulated as$\cite{Weinberg}$: {\em %
At every spacetime point in an arbitrary gravitational field it is
possible to choose a ``locally inertial coordinate system'' such
that, within a sufficiently small region of the point in question,
the laws of nature take the same form as in a Minkowski spacetime}
, i.e. the gravitational field may be eliminated locally by the
use of a freely falling coordinate system.

The principle of general covariance may be stated in one of the
following forms$\cite{Carmeli}$, namely
\begin{enumerate}
\item  All coordinate systems are equally good for stating the laws of
physics.
\item  The equations that describe the laws of physics should have tensorial
forms and be expressed in four-dimensional Riemannian spacetime.
\item  The equations describing the laws of physics should have the same
form in all coordinate systems.
\end{enumerate}
One considers the Ricci scalar curvature $%
\,R\,=g^{pq}R_{pq}$\ as our Lagrangian to derive the Einstein
field equations. The action integral for the gravitational field
is then given by:
\begin{equation}
\int \sqrt{-g}\,R\,d^{4}x ,  \label{Efeq1}
\end{equation}
where the integration is taken over all space and over the time component $%
x^{0}$ between two given values. To encompass non-gravitational
fields in a physical system we include another Lagrangian $L_{F}$
for all the other fields, so that our expression for the action
integral is given by:
\begin{equation}
I=\int \sqrt{-g}\,(R-2\,\kappa \,L_{F})\,d^{4}x ,  \label{Efeq2}
\end{equation}
where $\kappa $\ is Einstein's gravitational constant. We require
that the variation of the above be equal to zero,
\begin{equation}
\delta I=0.  \label{Efeq3}
\end{equation}
The variation of the first part of I is:
\begin{equation}
\delta \int \sqrt{-g}\,R\,d^{4}x=\int \sqrt{-g}\,g^{pq}\,\delta
R_{pq}\,d^{4}x+\int R_{pq}\,\delta \left( \sqrt{-g}\,g^{pq}\right)
\,d^{4}x .\label{Efeq4}
\end{equation}
By simplifying and applying Gauss theorem it can then be shown
that:
\begin{equation}
\int \sqrt{-g}\,g^{pq}\,\delta R_{pq}\,d^{4}x=0 . \label{Efeq5}
\end{equation}
Therefore
\begin{eqnarray}
\delta \int \sqrt{-g}\,R\,d^{4}x &=&\int \sqrt{-g}\,R_{pq}\,\delta
g^{pq}\,d^{4}x+\int R\,\delta \sqrt{-g}^{pq}\,d^{4}x  \nonumber \\
&=&\int \sqrt{-g}\,\left( R_{pq}-\frac{1}{2}g_{pg}\,R\,\right)
\delta g^{pq}\,d^{4}x .  \label{Efeq6}
\end{eqnarray}
Now the second part of action integral:
\begin{eqnarray}
\delta \int \sqrt{-g}\,L_{F}\,d^{4}x &=&\int \left[ \frac{\partial
\left( \sqrt{-g}\,L_{F}\right) }{\partial g^{pq}}\delta
g^{pq}-\frac{\partial
\left( \sqrt{-g}\,L_{F}\right) }{\partial g_{,a}^{pq}}\delta g_{,a}^{pq}%
\right] \,d^{4}x  \nonumber \\
&=&\frac{1}{2}\int \sqrt{-g}\,T_{pq}\,\delta g^{pq}\,d^{4}x ,
\label{Efeq7}
\end{eqnarray}
where $\,T_{pq}$ is the energy-momentum tensor defined as:
\begin{equation}
\,T_{pq}=\frac{2}{\sqrt{-g}}\left[ \frac{\partial \left( \sqrt{-g}%
\,L_{F}\right) }{\partial g^{pq}}-\frac{\partial }{\partial
x^{a}}\left\{
\frac{\partial \left( \sqrt{-g}\,L_{F}\right) }{\partial g_{,a}^{pq}}%
\right\} \right] .  \label{Efeq8}
\end{equation}
Now using results of $(\ref{Efeq6})$\ and $(\ref{Efeq7})$\ then
the variation of $(\ref{Efeq2})$\ is:
\begin{equation}
\delta I=\int \sqrt{-g}\,\left( R_{pq}-\frac{1}{2}g_{pg}\,R-\kappa
\,T_{pq}\,\right) \delta g^{pq}\,d^{4}x , \label{Efeq9}
\end{equation}
and since $\delta g^{pq}$\ is an arbitrary variation this gives us
the Einstein field equations:
\begin{equation}
R_{pq}-\frac{1}{2}g_{pg}\,R=\kappa \,T_{pq}.  \label{Efeq10}
\end{equation}

\section{Maxwell equations  in presence of gravitation}

In the absence of gravitation, the Maxwell field equations are
expressed by   $(\ref{MxlCovEq1})$ and $(\ref{MxlCovEq2})$.
However, in the presence of gravitation these equations are
generalized by the following equations by replacing partial
derivatives to covariant derivatives. Therefore one has now
\begin{equation}\label{MxlCovGeq1a}
  F_{ab;c}+F_{bc;a}+F_{ca;b}=0,
\end{equation}
and
\begin{equation}\label{MaxCovGeq2}
F^{ap}_{~~;p}=-4\pi j^a.
\end{equation}
The above two equations can  easily be expressed as
\begin{equation}\label{MxlCovGeq1b}
  F_{ab,c}+F_{bc,a}+F_{ca,b}=0,
\end{equation}
and
\begin{equation}\label{MaxCovGeq2b}
\frac{\partial}{\partial x^p} \left[\sqrt{-g} F^{ap}\right]=-4\pi \sqrt{-g} j^a.
\end{equation}
The continuity equation $(\ref{ContEq})$ now takes the form:
\begin{equation}\label{ContGEq}
j^{a}_{~;a}=0,
\end{equation}
and the equation of motion of a charged particle in combined
electromagnetic and gravitational fields, given as:
\begin{equation}\label{GMotEq}
m\left(\frac{du^a}{d\tau}+\Gamma^a_{bc}u^b u^c\right)=eF^{ab}_b,
\end{equation}
is obtained from $(\ref{EMFeq})$ by replacing ${du^a}/{d\tau}$
with the intrinsic derivative ${Du^a}/{d\tau}$. Equations
$(\ref{MxlCovGeq1a})$ and $(\ref{MaxCovGeq2})$ are the Maxwell
equations in the presence of gravitation. We now consider the
Einstein field equations:
\begin{equation}
R_{pq}-\frac{1}{2}g_{pg}\,R=\kappa \,T_{pq},  \label{EGeq9}
\end{equation}
in the presence of electromagnetic field, with the energy-momentum
tensor $T_{pq}$\ for the electromagnetic field given by
\begin{equation}
\,T_{pq}=\frac{2}{\sqrt{-g}}\left[ \frac{\partial \left( \sqrt{-g}%
\,L_{F}\right) }{\partial g^{pq}}-\frac{\partial }{\partial
x^{a}}\left\{
\frac{\partial \left( \sqrt{-g}\,L_{F}\right) }{\partial g_{,a}^{pq}}%
\right\} \right].   \label{EGeq10}
\end{equation}
The Lagrangian $\,L_{F}$ is given by:
\begin{equation}
L_{F}=-\frac{1}{16\pi }\sqrt{-g}\,g^{ap}g^{bq}F_{ab}F_{pq},
\label{EGeq11}
\end{equation}
with the indices of Maxwell tensor components (see
$(\ref{MaxCovGeq2})$ for
definition) $F_{ab}$ being raised or lowered by using metric components $%
g^{ab}$ or $g_{ab}$. Using the above, the energy-momentum tensor
for electromagnetic field may be simplified to:
\begin{equation}
T_{ab}=\frac{1}{4\pi }\left(-F_{ap}F_{b}^{p}+ \frac{1}{4}g_{ab}F_{pq}F^{pq}%
\right).   \label{EGeq12}
\end{equation}
As  this tensor is traceless,  the Ricci scalar
curvature $R$\  vanishes. Therefore the Einstein field equations
in the presence of electromagnetic field are given by:
\begin{equation}
R_{ab}=\kappa T_{ab}.  \label{EGeq13}
\end{equation}

\section{Klein-Gordon equation in curved space-time background}

In the absence of gravitation the Lagrangian density for a massive
scalar field $\phi $\ is given by
\[
L=\frac{1}{2}\left( \eta ^{pq}\phi _{,p}\phi _{,q}-m^{2}\phi
^{2}\right) .
\]
The field equation obtained from above by using variation
procedure is then given by
\[
\frac{\partial L}{\partial \phi }-\frac{\partial }{\partial x^{p}}\frac{%
\partial L}{\partial \left( \partial \phi /\partial x^{p}\right)
} = 0 ,
\]
which simplifies to
\begin{equation}
\Box \phi +m^{2}\phi =0 ,
 \label{KGeq1}
\end{equation}
where the differential operator $\Box :=\eta ^{pq}\partial
^{2}/\partial x^{p}\partial x^{q}$. This gives the familiar
Klein-Gordon equation in a flat space-time. To obtain the
Klein-Gordon equation in curved space-time we invoke the principle
of equivalence and general covariance. In the presence of
gravitation the above equation $(\ref{KGeq1})$ can be written as:
\begin{equation}
\phi _{\quad ;\,p}^{;\,p}+m^{2}\phi =0.
 \label{KGeq2}
\end{equation}
This is the Klein-Gordon equation in curved space-time.
\section{Conclusion}
In this introductory chapter we gave a list formulas which will be
required in our later work. In the next chapter we will  discuss the
conservation laws and  the dilemma associated
with the definition of gravitational energy in general relativity.
The lack of a generally accepted definition of gravitational
energy has lead to doubts concerning energy localization in GR.
The large number of available pseudotensorial
expressions used for computing energy and momentum distributions
has even lead to suspicions that these nontensorial quantities
would give different energy distributions for a given space-time.
However, the pioneering work of Virbhadra on energy localization,
on particular space-time manifolds, has consistently shown this to
be fallacious. In this work we are extending Virbhadra's work by
considering further space-times and showing that different
energy-momentum complexes give the same energy distribution in a
given space-time. Hence energy-momentum complexes are useful
expressions for computing energy distributions in GR.
\newpage

\chapter{Energy Localization}

\label{chapter-TWO}

\section{Introduction}   %

The concept of total  energy and momentum in asymptotically flat
space-time is unanimously accepted; however, the localization of
these physical quantities still remains an elusive problem when
one includes the gravitational field. In the special theory of
relativity the energy-momentum conservation laws of matter plus
non-gravitational fields are given by:
\begin{equation}
T_{i,k}^{k}=0 ,\label{Leq1}
\end{equation}
where $T_{i}^{k}$ denotes the symmetric energy-momentum tensor in
an inertial frame, whereas  general relativity  leads to the following
generalization of Eq.$(\ref{Leq1})$:
\begin{equation}
T_{i;k}^{k}=0. \label{Leq2}
\end{equation}
In this form  Eq.$(\ref{Leq2})$ does not give rise to
any\footnote{ However, Eq.$(\ref{Leq2})$ is the statement of the
energy-momentum conservation laws in Special relativity in
non-Cartesian and/or non-inertial coordinate systems.} integral
conservation law whatsoever. In fact, if this equation is written
as:
\begin{equation}
\frac{\partial (\sqrt{-g}T_{i}^{k})}{\partial x^{k}}=K_{i},
\label{Leq3}
\end{equation}
with
\begin{equation}
K_{i}=\frac{1}{2}\sqrt{-g}\frac{\partial g_{kp}}{\partial
x^{i}}\,T^{kp}, \label{Leq4}
\end{equation}
then it is clear that the quantity $K_{i}$\ is not a general  four-vector,
therefore in a local system of inertia \ we can always make
$K_{i}$ to vanish in a given spacetime point and, in this case
Eq.$(\ref{Leq3})$\ simply reduces to Eq.$(\ref {Leq1})$.\ In
general, $K_{i}\neq 0$\ and for $i=0$ then Eq.$(\ref{Leq3})$
expresses the fact that matter energy is not conserved.

Einstein formulated the conservation law in the form of a
divergence to include contribution from gravitational field energy
by introducing the energy-momentum pseudotensor $t_{i}^{k}$, so
that:
\begin{equation}
\frac{\partial }{\partial x^{k}}\left( \sqrt{-g}\left(
T_{i}^{k}+t_{i}^{k}\right) \right) =0.  \label{Eeq11}
\end{equation}
The quantity $t_{i}^{k}$\ is homogeneous quadratic in the first
derivatives of the metric tensor and thus it is obviously not a
tensor. With a suitable choice of a coordinates system
$t_{i}^{k}$\ can be made to vanish at a particular point. It can
also be shown that if we form the integral $\int t_{0}^{0}d^{3}x$
in a flat spacetime using quasi-Cartesian coordinates,then its
value is zero while if we transform to spherical coordinates the
value of this integral is infinite (Bauer, 1918). Furthermore, it
is possible to find a coordinate system for the Schwarzschild
solution such that the pseudotensor vanishes everywhere outside
the Schwarzschild radius (Schr\"{o}dinger, 1918). Einstein
ascribed these shortcomings to the coordinates used. However, the
difficulties associated with Einstein's nontensorial quantities
posed serious problems concerning the localizability of energy in
general relativity.

The problem of energy-momentum localization has been a subject of
many research activities dating back to the very onset of the
theory of general relativity but it still remains an open
question. The numerous attempts aimed at finding a more suitable
quantity for describing distribution of energy-momentum due to
matter, non-gravitational and gravitational fields resulted in
more energy-momentum complexes, notably those proposed by Landau
and Lifshitz, Papapetrou, M\o ller, and Weinberg. The physical
meaning of these nontensorial (under general coordinate
transformations) complexes have been questioned by some
researchers (see references in Chandrasekhar and Ferrari
\cite{ChaFer}). There are suspicions that different
energy-momentum complexes could give different energy
distributions in a given space-time. The problems associated with
energy-momentum complexes resulted in some researchers even
doubting the concept of energy-momentum localization in GR.
According to Misner, Thorne and Wheeler \cite{MTW} the energy is
localizable only for spherical systems. However, Cooperstock and
Sarracino \cite{CoopSar} refuted this viewpoint and stated that if
the energy is localizable in spherical systems then it is also
localizable for all systems. Bondi \cite{Bondi} wrote `` {\em In
relativity a non-localizable form of energy is inadmissible,
because any form of energy contributes to gravitation and so its
location can in principle be found}.'' It is rather unfortunate
that the controversy surrounding energy localization which first
appeared in electromagnetism,{\em where there is an ambiguity
concerning the choice of the Pointing vector}, is also present in
the most beautiful theory of general relativity (see in Feynmann,
Leighton and Sands\cite{Feynmann}). The \emph{ambiguity} in
electromagnetism is not nearly as great a problem as in general
relativity because in the former, we are dealing with truly
tensorial quantities.

Over the past two decades considerable effort has been put in
trying to define an alternative concept of energy, the so-called
quasilocal energy. The idea in this case is to determine the
effective energy of a source by measurements on a two-surface.
These masses are obtained over a two-surface as opposed to an
integral spanning over a three-surface of a local density as is
the case for pseudocomplexes. A large number of definitions of
quasi-local mass have been proposed, notable those by Hawkins,
Penrose, and many others (see in Brown and York \cite{BroYor},
Hayward\cite{Hayward}). Although, these quasi-local masses are
conceptually very important (as Penrose emphasized) they still
have serious problems. Bergqvist \cite{Bergqv} furnished
computations with seven different definitions of quasi-local
masses for the Reissner-Nordstr\"{o}m and Kerr space-times and
came to the conclusion that no two of these definitions gave the
same result. Moreover, the seminal quasi-local mass definition of
Penrose is not adequate to handle even the Kerr metric (Beinsten
and Tod \cite{BeinstTod}). On the contrary, several authors
studied energy-momentum complexes and obtained stimulating
results. The leading contributions of Virbhadra and his
collaborators (Rosen, Parikh, Chamorro, and Aguirregabiria) have
demonstrated with several examples that for a given spacetime,
different energy-momentum complexes show a high degree of
consistency in giving the same and acceptable energy and momentum
distribution. In the rest of this chapter we present a brief
introduction to each of the following : the Einstein, Landau and
Lifshitz, M\o ller, Papapetrou, and Weinberg energy-momentum
complexes which are going to be used in our later work.
\section{Einstein energy-momentum complex} %
In order to arrive at Einstein's conservation laws for a system consisting
of both matter and gravitational field we start with the gravitational field
equations
\begin{equation}
R^{ik}-\frac{1}{2}g^{ik}R=8\pi T^{ik},  \label{Eeq1}
\end{equation}
and then using the contracted Bianchi identity $\left( R^{ik}-\frac{1}{2}%
g^{ik}R\right) _{;\,k}=0$, Eq. $(\ref{Leq2})$ becomes a consequence of the
field equations. Eq. $(\ref{Leq4})$ can be written as
\begin{equation}
K_{i}=-\frac{1}{2}\sqrt{-g}g_{~~\,,i}^{pq}\,T_{pq},  \label{Eeq2}
\end{equation}
where $g_{,i}^{pq}=\frac{\partial g^{pq}}{\partial x^{i}}$.\ Using the field
equations $(\ref{Eeq1})$ we eliminate $T_{ik}$ from $(\ref{Eeq2})$ and write
$K_{i}$\ as
\begin{eqnarray}
K_{i} &=&-\frac{1}{16\pi }\sqrt{-g}g_{~~\,,i}^{pq}\,\left[ R_{pq}-\frac{1}{2}%
g_{pq}R\right]   \nonumber \\
&=&-\frac{1}{16\pi }\left[ \frac{\partial L}{\partial g^{pq}}g_{~~\,,i}^{pq}-%
\frac{\partial }{\partial x^{k}}\left( \frac{\partial L}{\partial
g_{~~,k}^{pq}}\right) g_{~~\,,i}^{pq}\right]   \nonumber \\
&=&-\frac{\partial (\sqrt{-g}t_{i}^{~k})}{\partial x^{k}},
\label{Eeq3}
\end{eqnarray}
where
\begin{equation}
\sqrt{-g}t_{i}^{~k}=\frac{1}{16\pi }\left( \delta _{i}^{k}L-\frac{\partial L%
}{\partial g_{~~,k}^{pq}}g_{~~,i}^{pq}\right) ,  \label{Eeq4}
\end{equation}
and $L$ is the Lagrangian density
\begin{equation}
L=\sqrt{-g}g^{ik}\left( \Gamma _{ik}^{p}\Gamma _{pq}^{q}-\Gamma
_{iq}^{p}\Gamma _{kp}^{q}\right) .  \label{Eeq5}
\end{equation}
Obviously $t_{i}^{~k}$\ is a function of the metric tensor and its first
derivatives. Now combining $(\ref{Eeq3})$ with$\ (\ref{Leq3})$ we get the
following equation expressing Einstein's conservation law:
\begin{equation}
\frac{\partial \theta _{i}^{~k}}{\partial x^{k}}=0,  \label{Eeq6}
\end{equation}
where
\begin{eqnarray}
\theta _{i}^{k} &=&\sqrt{-g}\left( T_{i}^{~k}+t_{i}^{~k}\right)
\label{Eeq7a} \\
&=&\Im _{i}^{~k}+\upsilon _{i}^{~k},  \label{Eeq7b}
\end{eqnarray}
is the total energy-momentum complex for the combined matter plus
gravitational field, while $\Im _{i}^{~k}=\sqrt{-g}T_{i}^{~k}$ and $\upsilon
_{i}^{~k}=\sqrt{-g}t_{i}^{~k}$.\ By introducing a local system of inertia,
the ``gravitational'' part $\upsilon _{i}^{~k}$ can always be reduced to
zero for any given space-time point. In general $\Im _{i}^{~k}$\ is a
function of matter and gravitational tensor, and hence the division of $%
\theta _{i}^{~k}$\ into ``matter'' part and ``gravitational'' part is highly
arbitrary. The matter part may even be eliminated entirely from $(\ref{Eeq7a}%
)$\ and the $\theta _{i}^{~k}$ expressed only as a function of the metric
tensor together with its first and second derivatives, as:
\begin{equation}
\theta _{i}^{~k}=\frac{\partial S_{i}^{~kp}}{\partial x^{p}},  \label{Eeq8}
\end{equation}
with $S_{i}^{~kp}$ given in Tolman\cite{Tolman34} as
\begin{equation}
S_{i}^{~kl}=\frac{1}{8\pi }\frac{\partial L}{\partial
g_{~~,l}^{ip}}g^{kp}. \label{Eeq9}
\end{equation}
M\o ller \cite{Moller58} suggested a more useful expression for $\theta
_{i}^{~k}$. A quantity $\theta _{i}^{~k}$\ which satisfies Eq. $(\ref{Eeq6})$%
\ identically, must be writable in the form:
\begin{equation}
\theta _{i}^{k}=\frac{1}{16\pi }\frac{\partial h_{i}^{~kp}}{\partial x^{p}},
\label{Eeq10}
\end{equation}
where $h_{i}^{~kp}=-h_{i}^{~pk}$ and $h_{i}^{~kp}$\ is a function of the
metric tensor and its first derivatives. It is easy to verify that this is
the case for
\begin{equation}
h_{i}^{~kl}=\frac{g_{in}}{\sqrt{-g}}\left[ \left( -g\right) \left(
g^{kn}g^{lm}-g^{l\,n}g^{km}\right) \right] _{,m}.  \label{Eeq11}
\end{equation}
If the physical system under consideration is such that we can introduce
quasi-Cartesian coordinates $x^{a}$ for which the $g_{ik}$ converge
sufficiently rapidly towards the constant values $\eta _{ik}$ where:
\begin{equation}
\eta _{ik}=diag(1,-1,-1,-1)  \label{Eeq12}
\end{equation}
then it follows from Eq. $(\ref{Eeq6})$\ that the quantities:
\begin{equation}
P_{i}=\int \int \int \theta _{i}^{~0}dx^{1}dx^{2}dx^{3}  \label{Eeq13}
\end{equation}
are constant in time, provided that $\theta _{i}^{~k}$ are everywhere
regular. The integral in Eq. $(\ref{Eeq13})$\ is extended over all space for
$x^{0}=const.$ Further, Gauss's theorem furnishes\
\begin{equation}
P_{i}=\frac{1}{16\pi }\int \int h_{i}^{~0\alpha }\mu _{\alpha }dS
\label{Eeq14}
\end{equation}
where $\mu _{\alpha }=\frac{x_{i}}{r}$ is the outward unit normal vector
over an infinitesimal surface element $dS$

The main problem with interpreting the integrand $\theta _{0}^{~0}$\ in $%
P_{0}$ as the energy density is that it does not behave like a three-scalar
density under purely spatial transformations. It can be shown that if we
form the integral $\int t_{0}^{~0}d^{3}x$ then its value is zero in a flat
spacetime using quasi-Cartesian coordinates, while if we transform to
spherical coordinates the value of this integral is infinite \cite{Moller58}%
. Furthermore, Schr\"{o}dinger \cite{Moller58} showed that there
exists a coordinate system for the Schwarzschild solution such
that the pseudotensor vanishes everywhere outside the
Schwarzschild\ radius.\ Many other prominent scientists, including
Weyl, Pauli, and Eddington, questioned the nontensorial nature of
$t_{k}^{~i}$ because with a suitable choice of a coordinate system
it can be made to vanish at any point in spacetime, (for details
see in Chandrasekhar and Ferrari \cite{ChaFer}). Einstein (see in
Goldberg \cite{Goldberg}) pointed out that these effects were
artifacts of the coordinates used and that they were not related
to the physical system used. Einstein showed that for a spacetime
that approaches the Minkowski spacetime at spatial infinity then
the energy-momentum $P_{i}$\ transforms as a four-vector under all
linear transformations. Consider the following example of a closed
system at rest whose coordinates $x^{a}$ are chosen so that at
large distances the line element is given by
\begin{equation}
ds^{2}=\left( 1-\alpha /r\right) \left( dx^{0}\right) ^{2}-\left( 1+\alpha
/r\right) \left( \left( dx^{1}\right) ^{2}+\left( dx^{2}\right) ^{2}+\left(
dx^{3}\right) ^{2}\right)   \label{Exq1}
\end{equation}
with $r^{2}=\left( \left( x^{1}\right) ^{2}+\left( x^{2}\right) ^{2}+\left(
x^{3}\right) ^{2}\right) $, and where $\alpha =2M_{0}$ could be regarded as
a constant connected to the total Newtonian gravitational mass $M_{0}$.
Using Eq. $(\ref{Eeq14})$\ the total energy-momentum components $P_{i}$\ in
this system will be given by
\begin{equation}
P_{i}=-\delta _{i}^{o}\frac{\alpha }{2}=-\delta _{i}^{o}M_{0}.  \label{Exq2}
\end{equation}
Now using a Lorentz transformation of the form
\begin{equation}
x^{0^{\prime }}=\frac{x^{0}+vx^{1}}{\sqrt{1-v^{2}}},\qquad x^{1^{\prime }}=%
\frac{x^{1}+vx^{0}}{\sqrt{1-v^{2}}},\qquad x^{2^{\prime }}=x^{2},\qquad
x^{3^{\prime }}=x^{3}.  \label{Exq3}
\end{equation}
the total energy-momentum components $P_{i^{\prime }}$\ in the new
coordinates system $x^{a^{\prime }}$\ have values
\begin{equation}
P_{i^{\prime }}=\left\{ -\frac{M_{0}v}{\sqrt{1-v^{2}}},\qquad \frac{M_{0}v}{%
\sqrt{1-v^{2}}},\qquad 0,\qquad 0\right\}   \label{Exq4}
\end{equation}
in other words, in an inertial system the total energy-momentum components
have the same values as the components of the four-momentum of a particle of
proper mass $M_{0}$ moving with velocity $v$ along the $x$-axis. Another
important result given by Einstein is that any two systems of
quasi-Cartesian coordinates $S$ and $S\prime $ which coincide at spatial
infinity, but differ arbitrarily elsewhere will have $P^{i}=P^{i^{\prime }}$%
.\ Although Einstein was able to show that the energy-momentum
pseudo-complex $\theta _{i}^{~k}$ provides satisfactory expressions for the
total energy and momentum of closed system in the form of three-dimensional
integrals $(\ref{Eeq13})$, to get meaningful values for these integrals one
is restricted to the use of quasi-Cartesian coordinates.

An alternative form of Einstein's pseudo-complex, which we found useful in
some of our calculations, is given by Tolman \cite{Tolman34} as:
\begin{equation}
\theta _{i}^{~k}=\frac{1}{8\pi }\left[ -\varrho ^{kp}\frac{\partial L}{%
\partial \varrho _{~~m}^{ip}}+\frac{1}{2}\delta _{i}^{k}\varrho ^{pq}\frac{%
\partial L}{\partial \varrho _{~~m}^{pq}}\right] _{,m},  \label{Eeq15}
\end{equation}
where $L$ is the Lagrangian given by $(\ref{Eeq5})$ whereas $\varrho ^{ab}:=%
\sqrt{-g}g^{ab}$ while $\varrho
_{~~c}^{ab}:=\sqrt{-g}g_{~~,c}^{ab}$ , so that:
\begin{equation}
\frac{\partial L}{\varrho _{~~c}^{ab}}=-\Gamma _{ab}^{c}+\frac{1}{2}\delta
_{a}^{c}\Gamma _{bp}^{p}+\frac{1}{2}\delta _{b}^{c}\Gamma _{ap}^{p}.
\label{Eeq16}
\end{equation}
Another useful expression, from Tolman \cite{Tolman34}, for obtaining energy
for a static or quasi-static system using quasi-Cartesian type of
coordinates is:
\begin{equation}
E=\int \int \int \left( \Im _{0}^{~0}-\Im _{1}^{~1}-\Im _{2}^{~2}-\Im
_{3}^{~3}\right) dx^{1}dx^{2}dx^{3}.  \label{Eeq17}
\end{equation}
The main advantage derived from using this expression is that it can be
evaluated by integrating only over the region actually occupied by matter or
electromagnetic energy, since $\Im _{b}^{~a}$\ vanishes in empty space.

\section{Landau-Lifshitz energy-momentum complex}  %
One of the main objections to Einstein's energy-momentum complex was that it
is not even symmetric in its indices, so cannot be used to define
conservation laws of angular momentum. In this section we discuss an
energy-momentum complex which satisfies this requirement. In deriving the
conserved total four-momentum for a gravitational field plus matter and
all non-gravitational fields, Landau and Lifshitz \cite{LL}\ introduced a
geodesic coordinate system at some particular point in spacetime in which
all the first derivatives of the metric tensor $g_{ik}$\ vanish. Then at
this point Eq. $(\ref{Leq2})$\ can be reduced into a form:
\begin{equation}
\frac{\partial T^{ik}}{\partial x^{k}}=0 , \label{LLeq1}
\end{equation}
similar to Eq. $(\ref{Leq1})$.\ Further, it can be shown that quantities $%
T^{ik}$, which satisfy Eq. $(\ref{LLeq1})$\ identically, can be expressed in
terms of the following
\begin{equation}
T^{ik}=\frac{\partial S^{ikl}}{\partial x^{l}},  \label{LLeq2}
\end{equation}
where the quantities $S^{ikl}$\ are antisymmetric in their last two indices $%
k$\ and $l$. At the point under consideration the Ricci tensor may be
written as
\begin{equation}
R^{ik}=\frac{1}{2}g^{ip}g^{kq}g^{rs}\left(
g_{pr,qs}+g_{qs,pr}-g_{pq,rs}-g_{rs,pq}\right) ,  \label{LLeq2a}
\end{equation}
since the Christoffel symbols vanish. Now using the gravitational
field
equations $(\ref{Eeq1})$ we may deduce that the energy-momentum tensor $%
T^{ik}$ can indeed be expressed as
\begin{equation}
T^{ik}=\frac{\partial }{\partial x^{l}}\left\{ \frac{1}{16\pi }\frac{1}{%
\left( -g\right) }\frac{\partial }{\partial x^{m}}\left[ \left(
-g\right) \left( g^{ik}g^{lm}-g^{il}g^{km}\right) \right] \right\}
,
\end{equation}
where the expression inside curly brackets can be associated with
$S^{ikl}$. Defining the quantities:
\begin{equation}
h^{ikl}=\frac{1}{16\pi }\frac{\partial }{\partial x^{m}}\left[ \left(
-g\right) \left( g^{ik}g^{lm}-g^{il}g^{km}\right) \right] ,  \label{LLeq4}
\end{equation}
then obviously $h^{ikl}=-h^{ilk}$and since all the first derivatives of the
metric tensor\ vanish, Eq. $(\ref{LLeq2})$\ may be written as:
\begin{equation}
\frac{\partial h^{ikl}}{\partial x^{l}}-\left( -g\right) T^{ik}=0.
\label{LLeq5}
\end{equation}
Eq. $(\ref{LLeq5})$\ will only hold in spacetime at some particular point of
a special coordinates system in which all the first derivatives of the
metric tensor $g_{ik}$\ vanish. In an arbitrary coordinate system, the
difference $\partial h^{ikl}/\partial x^{l}-\left( -g\right) T^{ik}$ will no
longer be zero. We denote this difference by: $\left( -g\right) t^{ik}$, and
thus in general Eq. $(\ref{LLeq5})$\ will be of the form:
\begin{equation}
\left( -g\right) \left( T^{ik}+t^{ik}\right) =\frac{\partial h^{ikl}}{%
\partial x^{l}},  \label{LLeq6}
\end{equation}
where the quantities $t^{ik}$\ are symmetric in their indices since both $%
T^{ik}$\ and $\partial h^{ikl}/\partial x^{l}$\ are symmetric in the indices
$i$\ and $k$. It is obvious that $t^{ik}$\ is not a tensor quantity. $T^{ik}$
can be eliminated from Eq. $(\ref{LLeq6})$ by making use of the Einstein
gravitational field equations , the above equation may then be written as
\begin{equation}
\left( -g\right) t^{ik}=\frac{g}{8\pi }\left( R^{ik}-\frac{1}{2}%
g^{ik}R\right) +h_{\quad ,l}^{ikl} . \label{LLeq7}
\end{equation}
Now using the expression of $h^{ikl}$ in Eq.$\ (\ref{LLeq4})$ and that of
the Ricci tensor, and after a lot of simplifications the expression of $%
t^{ik}$ reduces to the following
\begin{eqnarray*}
16\pi t^{ik} &=&\left\{ \left( g^{ip}g^{kq}-g^{ik}g^{pq}\right) \left(
2\Gamma _{pq}^{a}\Gamma _{ab}^{b}-\Gamma _{pb}^{a}\Gamma _{qa}^{b}-\Gamma
_{pa}^{a}\Gamma _{qb}^{b}\right) \right.  \\
&&+\left. g^{ip}g^{qr}\left( \Gamma _{pa}^{k}\Gamma _{qr}^{a}+\Gamma
_{qr}^{k}\Gamma _{pa}^{a}-\Gamma _{ra}^{k}\Gamma _{pq}^{a}-\Gamma
_{pq}^{k}\Gamma _{ra}^{a}\right) \right.  \\
&&+\left. g^{kp}g^{qr}\left( \Gamma _{pa}^{i}\Gamma _{qr}^{a}+\Gamma
_{qr}^{i}\Gamma _{pa}^{a}-\Gamma _{ra}^{i}\Gamma _{pq}^{a}-\Gamma
_{pq}^{i}\Gamma _{ra}^{a}\right) \right.  \\
&&+\left. g^{kp}g^{qr}\left( \Gamma _{pa}^{i}\Gamma _{qr}^{a}+\Gamma
_{qr}^{i}\Gamma _{pa}^{a}-\Gamma _{ra}^{i}\Gamma _{pq}^{a}-\Gamma
_{pq}^{i}\Gamma _{ra}^{a}\right) \right.  \\
&&+\left. g^{pq}g^{rs}\left( \Gamma _{pr}^{i}\Gamma
_{qs}^{k}-\Gamma _{pq}^{i}\Gamma _{rs}^{k}\right) \right\}.
\end{eqnarray*}
\ Also, since $h^{ikl}$\ is antisymmetric in indices $k$\ and $l$, it
follows from equation $(\ref{LLeq6})$ that
\begin{equation}
\frac{\partial }{\partial x^{k}}\left[ \left( -g\right) \left(
T^{ik}+t^{ik}\right) \right] =0,
\end{equation}
which means that there is a conservation law for the quantities

\begin{equation}
P^{i}=\int \left( -g\right) \left( T^{ik}+t^{ik}\right) dS_{k} .
\label{LLeq8}
\end{equation}
where the integration may be taken over any infinite hypersurface including
all of three dimensional space. In the absence of gravitation, in
quasi-Cartesian coordinates system, the set of quantities $t^{ik}$ vanishes
and $P^{i}$\ reduces to $\int \left( -g\right) T^{ik}dS_{k}$\ which is the
four-momentum of the physical system without gravitation. Therefore $P^{i}$\
in $(\ref{LLeq8})$ is identified with the total four-momentum of the whole
physical system including gravitation. So we refer to $t^{ik}$ as the
energy-momentum pseudo-tensor and to
\begin{equation}
L ^{ik}=\left( -g\right) \left( T^{ik}+t^{ik}\right)
\end{equation}
as the energy-momentum complex. Now choosing the hypersurface $x^{0}=const$,
then $P^{i}$\ can be written in the form of a three dimensional \ space
integral

\begin{equation}
P^{i}=\int \int \int L ^{i0}dx^{1}dx^{2}dx^{3} . \label{LLeq10}
\end{equation}
Hence we might interpret the quantity $L ^{00}$ as representing
the energy density of the whole physical system including
gravitation, and interpret the quantity $L ^{0k}$ as representing
the components of the total momentum density.

Unlike the Einstein energy-momentum complex $\theta _{i}^{k}$, the
main advantage with the Landau-Lifshitz energy-momentum complex $L
^{ik}$ is that it is symmetric with respect to its indices, and
therefore it can be used to define a conservation law for the
angular momentum. We define it as
\begin{eqnarray}
M^{ik} &=&\int \left( x^{i}dP^{k}-x^{k}dP^{i}\right)  \nonumber \\
&=&\int \left( x^{i}L ^{mk}-x^{k}L ^{mi}\right) dS_{m}.
\label{LLeq11}
\end{eqnarray}
By using an argument similar to one used by Einstein, it can be
shown that:
\begin{enumerate}
\item  For asymptotically flat spacetime the quantities $P^{i}$\ are
constant in time.
\item  For any two systems of quasi-Cartesian coordinates $S$ and $S\prime $
which coincide at spatial infinity, but differ arbitrarily elsewhere we have
$P^{i}=P^{i^{\prime }}$.
\item  $P^{i}$ transforms like contravariant components of a four-vector
under all linear transformations, including Lorentz transformations.
Therefore using $(\ref{LLeq6})$ the energy-momentum densities $(\ref{LLeq8})$
may be written as
\begin{equation}
P^{i}=\int h_{\quad ,m}^{ikm}dS_{k}=\frac{1}{2}\int \left( h_{\quad
,m}^{ikm}dS_{k}-h_{\quad ,k}^{ikm}dS_{m}\right)  \label{LLeq12}
\end{equation}
\ so that the above integral can then be written as an integral over an
ordinary surface giving
\begin{equation}
P^{i}=\oint h^{ikm}d^{\ast }{\it n}_{km},  \label{LLeq13}
\end{equation}
where $d^{\ast }{\it n}_{km}$\ is the normal to the surface element related
to tangential element $d{\it n}^{km}$\ by $d^{\ast }{\it n}_{ik}=\frac{1}{2}%
\epsilon _{ikmq}d{\it n}^{mq}$.\ Choosing the hypersurface $x^{0}=const$ for
the surface of integration in $(\ref{LLeq8})$ then the surface of
integration in $(\ref{LLeq13})$ becomes an ordinary space, thus we obtain
\begin{equation}
P^{i}=\oint h^{i0m}d{\it n}_{m},  \label{LLeq14}
\end{equation}
where $d{\it n}_{m}=d^{\ast }{\it n}_{0m}$ is a three-dimensional
element of an ordinary space. Similarly, an analogous formula for
angular momentum is given by
\[
M^{ik}=\int \left( x^{i}h^{k0m}-x^{k}h^{i0m}+\lambda ^{i0mk}\right) d{\it n}%
_{m}.
\]
\end{enumerate}

\section{M\o ller energy-momentum complex}      %
M\o ller \cite{Moller58} argued that although the Einstein
energy-momentum complex provides useful expressions for the total
energy and momentum of closed physical systems, the singling out
of quasi-Cartesian coordinates is somehow unsatisfactory from the
general relativity viewpoint. Most of the criticism of Einstein's
prescription centred around the nontensorial nature of the
quantity $t_{i}^{~k}$. A mere change of a coordinates system from
quasi-Cartesian into spherical polar coordinates creates energy in
vacuum. So M\o ller searched for an expression of energy and
momentum which is not dependent on any particular coordinates
system. If $\theta _{i}^{~k}$\ is Einstein's energy-momentum complex and $%
S_{i}^{~k}$ is another quantity with an identically vanishing
divergence, then their sum $\theta _{i}^{~k}+S_{i}^{~k}$ will also
vanish identically. Therefore the energy-momentum complex  is not
uniquely determined by the condition that its divergence vanishes.
M\o ller \cite{Moller58} exploited this freedom by searching for a
quantity $S_{i}^{~k}$\ that can be added to $\theta _{i}^{~k}$ so
that it transforms as tensor for spatial transformations. In order
to retain the satisfactory features of Einstein's theory
$S_{i}^{~k}$\ had to be chosen in such a way that it was form
invariant function which depends on the metric tensor and on its
first and second derivatives. Under linear transformations it had
to behave like a tensor density satisfying the following
conditions:
\begin{enumerate}
\item  $S_{i~,k}^{~k}=0$\ identically, therefore it must be expressible in
terms of $\Psi _{i~~,p}^{~kp}$\ where $\Psi _{i}^{~kp}=-\Psi _{i}^{~pk}$is
an affine tensor of rank $3$.
\item  $\int S_{i}^{~0}d^{3}x=0$ over total three-space for a closed system
if we use quasi-Cartesian coordinates.
\item  $\theta _{0}^{~k}+S_{0}^{~k}$ behaves like a four-vector density
under all transformations of the type:
\begin{equation}
x^{0^{\prime }}\rightarrow x^{0},\ \ x^{\alpha ^{\prime
}}=f^{\alpha ^{\prime }}(x^{\beta })  .\label{Meq1}
\end{equation}
\end{enumerate}
Thus for
\begin{equation}
\Im _{i}^{~k}=\theta _{i}^{~k}+S_{i}^{~k},  \label{Meq2}
\end{equation}
condition (2) implies that
\begin{equation}
\int \int \int \Im _{i}^{~0}dx^{1}dx^{2}dx^{3}=\int \int \int
\theta _{i}^{~0}dx^{1}dx^{2}dx^{3},  \label{Meq3}
\end{equation}
for a closed physical system. In order to find an $S_{i}^{~k}$ satisfying
the above conditions M\o ller \cite{Moller58} first investigated
transformation properties of $\theta _{0}^{~0}$\ under arbitrary
infinitesimal transformations of the type Eq. $(\ref{Meq1})$ so as to
establish the deviation of the variation of $\theta _{0}^{~0}$\ from a
scalar density. Following this procedure he finally arrived at:
\begin{equation}
\Im _{i}^{~k}=\frac{1}{8\pi }\chi _{i~~,p}^{~kp} , \label{Meq4}
\end{equation}
where
\begin{eqnarray}
\chi _{i}^{~kl} &=&2h_{i}^{~kl}-\delta _{i}^{k}h_{p}^{~pl}+\delta
_{i}^{l}h_{p}^{~pk}  \label{Meq5} \\
&=&\sqrt{-g}\left( \frac{\partial g_{ip}}{x^{q}}-\frac{\partial g_{iq}}{x^{p}%
}\right) g^{kq}g^{lp} , \label{Meq6}
\end{eqnarray}
Because of the antisymmetry, $\chi _{i}^{~kl}=-\chi _{i}^{~lk}$, it now
follows that:
\begin{equation}
\frac{\partial \Im _{i}^{~k}}{x^{k}}=0.  \label{Meq7}
\end{equation}
M\o ller was able to show that the quantities:
\begin{equation}
P_{i}=\int \int \int \Im _{i}^{~0}dx^{1}dx^{2}dx^{3},
\label{Meq8}
\end{equation}
for a closed system at rest in quasi-Cartesian coordinates coincide with the
corresponding ones for the Einstein pseudo-complex, and hence equation $(\ref
{Meq3})$ is satisfied and then following an argument similar to Einstein's
he then deduced that $P_{i}$ transform like four-vectors under Lorentz
transformations, and thus the integrals have the same values as the
corresponding integrals of Einstein in all cases where the latter are
meaningful at all. Finally, he showed that $\Im _{0}^{k}$ transforms like a
four-vector under the transformations of the form $(\ref{Meq1})$. Using
Gauss's theorem the total energy-momentum components are given by\
\begin{equation}
P^{k}=\frac{1}{8\pi }\int \int \chi _{i}^{~0\alpha }\mu _{\alpha
}dS, \label{Meq9}
\end{equation}
where $\mu _{\alpha }$ is the outward unit normal vector over an
infinitesimal surface element $dS$

M\o ller's coordinate independent prescription appeared to have finally
solved the problem of energy localization until three years later when M\o %
ller \cite{Moller61} performed a Lorentz transformation for a
closed system at rest using the line element $(\ref{Exq1})$. As
shown above $(\ref{Exq4})$, the Einstein pseudo-complex exhibited
the correct transformation properties. M\o ller \cite{Moller61}
first obtained the total energy-momentum components $P_{i}$ such
that
\begin{equation}
P_{i}=-\delta _{i}^{o}M_{0},  \label{Meq10}
\end{equation}
in the coordinates system $x^{a}$, using Eq. $(\ref{Meq8})$, which
agrees with those obtained in Eq. $(\ref{Exq2})$\ using Einstein's
prescription. Now using a Lorentz transformation, Eq.
$(\ref{Exq3})$, to obtain the total energy-momentum
components $P_{i^{\prime }}$\ in the new coordinates system $x^{a^{\prime }}$%
\ he got the following values
\begin{equation}
P_{i^{\prime }}=\left\{ -\frac{M_{0}}{\sqrt{1-v^{2}}}\left( 1+\frac{2}{3}%
v^{2}\right) ,\qquad \frac{5}{3}\frac{M_{0}v}{\sqrt{1-v^{2}}},\qquad
0,\qquad 0\right\}   \label{Meq11}
\end{equation}
which did not seem to give expected transformation properties under the
Lorentz transformation, and hence showing that $P_{i}$ does not transform
like a four-vector under Lorentz transformations. After a critical analysis
of M\o ller's result, Kovacs \cite{Kovacs} claimed to have found a defect in
M\o ller's calculation. However, Novotny \cite{Novotny} has shown that M\o %
ller\cite{Moller61} was right in concluding that $P_{i}$ does not
transform like a four-vector under Lorentz transformations.
Lessner \cite{Lessner} finally showed that the problem lies with
the interpretation of the result from a special relativistic point
of view instead of a general relativistic point of view. According
to Lessner \cite{Lessner}: {\em The energy-momentum four-vector
can only transform according to special relativity \ only if it is
transformed to a reference system with an everywhere constant
velocity. This cannot be achieved by a global Lorentz
transformation.} He concludes by stating that M\o ller's
energy-momentum complex is a powerful expression of energy and
momentum in general relativity.

\section{Papapetrou energy-momentum complex}    %
Amongst the five energy-momentum complexes under discussion, the
Papapetrou energy-momentum complex is the least known and as a
result it has been rediscovered several times, first in 1948 by
Papapetrou. Gupta, a high energy physicist well-known for the
Gupta-Bluer formalism in quantizing the electromagnetic field, in
1954 also obtained this complex using a slightly different method
(see Gupta \cite{Gupta}). This is the reason Misner refers to this
complex as the Papapetrou-Gupta energy-momentum complex. In 1994,
a renowned particle physicist, R. Jackiw of MIT - USA, with his
collaborators re-obtained the same energy-momentum complex using
the same method used by Papapetrou (see Bak, Cangemi, and Jackiw
\cite{BakCanJac}.) This fact was pointed out to them by Virbhadra
and then they sent an errata to PRD. Papapetrou \cite{Papap}
formulated this conservation law of general relativity by
explicitly introducing in calculations and in the final formulae
the flat-space metric tensor $\eta _{ik}$. The formula was
obtained following the generalized Belinfante method. First, from
Eq. $(\ref{Eeq12})$ Einstein's pseudocomplex may be written as:
\begin{equation}
\theta _{i}^{~k}=\frac{1}{8\pi }\frac{\partial }{\partial x^{p}}R_{i}^{~kp},
\label{Peq1}
\end{equation}
where
\begin{equation}
R_{i}^{~km}:=\left[ -\varrho ^{kp}\frac{\partial L}{\partial \varrho
_{~~m}^{ip}}+\frac{1}{2}\delta _{i}^{k}\varrho ^{pq}\frac{\partial L}{%
\partial \varrho _{~~m}^{pq}}\right],   \label{Peq2}
\end{equation}
and $L$ is the Lagrangian given by $(\ref{Eeq5})$, $\varrho
^{kp}$and $\varrho _{~~m}^{ip}$\ are the same as defined
previously. Now to symmetrize the above total energy-momentum
complex using Belinfante method we start by assuming the existence
of a quantity $\Omega ^{ik}$ so that $\Omega ^{ik}=\Omega ^{ki}$\
which differs from $\eta ^{ip}\theta _{p}^{~k}$\ only by a
divergence:
\begin{equation}
\Omega ^{ik}=\eta ^{ip}\theta _{p}^{~k}+\frac{\partial }{\partial x^{p}}%
B^{ikp}  \label{Peq3}
\end{equation}
so that its divergence vanishes:
\begin{equation}
\frac{\partial }{\partial x^{k}}\Omega ^{ik}=0.  \label{Peq4}
\end{equation}
The above equation will only be satisfied if $B^{ikl}$\ \ is antisymmetric
in its last two indices, i.e. if $B^{ikl}=-B^{ilk}$.\ By making use of the
Belinfante method $B^{ikl}$\ may be expressed by the relation:
\begin{equation}
B^{ikl}=-\frac{1}{2}\left( f^{ikl}+f^{lik}+f^{lki}\right)   \label{Peq5}
\end{equation}
where $f^{ikl}$\ is the spin density of the field given by:
\begin{equation}
f^{ikl}=\frac{1}{8\pi }\frac{\partial L}{\partial \varrho _{~~l}^{ab}}\left(
\varrho ^{ia}\eta ^{kb}-\varrho ^{ka}\eta ^{ib}\right) ,  \label{Peq6}
\end{equation}
which, using Eq. $(\ref{Peq2})$, may also be written as
\begin{equation}
f^{ikl}=\frac{1}{8\pi }\left( R_{p}^{~li}\eta
^{kp}-R_{p}^{~lk}\eta ^{ik}\right),   \label{Peq7}
\end{equation}
Now using this in Eq. $(\ref{Peq3})$ and $(\ref{Peq5})$, we get:
\begin{equation}
\Omega ^{ik}=\frac{1}{16\pi }\frac{\partial }{\partial
x^{p}}\left[ \eta ^{ia}\left( R_{a}^{~pk}+R_{a}^{~kp}\right) +\eta
^{ka}\left( R_{a}^{~pi}+R_{a}^{~ip}\right) -\eta ^{pa}\left(
R_{a}^{~ik}+R_{a}^{~ki}\right) \right]   \label{Peq8}
\end{equation}
then using $(\ref{Eeq16})$\ in $(\ref{Peq2})$\ we get
\[
R_{a}^{~bc}+R_{a}^{~cb}=\left( \delta _{a}^{p}\varrho ^{bc}-\frac{1}{2}%
\delta _{a}^{c}\varrho ^{bp}-\frac{1}{2}\delta _{a}^{b}\varrho ^{cp}\right)
_{,p}
\]
which is used to simplify $(\ref{Peq8})$ to
\begin{equation}
\Omega ^{ik}=\frac{1}{16\pi }{\cal {N}}_{\quad \quad ,ab}^{ikab}
\label{Peq9}
\end{equation}
where
\begin{equation}
{\cal {N}}^{ikab}=\sqrt{-g}\left( g^{ik}\eta ^{ab}-g^{ia}\eta
^{kb}+g^{ab}\eta ^{ik}-g^{kb}\eta ^{ia}\right) .  \label{ePq10}
\end{equation}
Note that the quantities ${\cal {N}}^{ikab}$\ are symmetric with
respect to the first two indices $i$ and $k$.\ The energy-momentum
complex $\Omega ^{ik} $ of Papapetrou satisfies the local
conservation laws $(\ref{Peq4})$. This locally conserved quantity
$\Omega ^{ik}$ contains contributions from the
matter, non-gravitational and gravitational fields. $\Omega ^{00}$ \ and $%
\Omega ^{\alpha 0}$ \ are the energy and momentum (energy current)
density components. The energy and momentum components are given
by
\begin{equation}
P^{i}=\int \int \int \Omega ^{i0}dx^{1}dx^{2}dx^{3}  \label{Peq12}
\end{equation}
and for the time-independent metrics Gauss's theorem furnishes
\[
P^{i}=\frac{1}{16\pi }\int \int {\cal {N}}_{\quad \quad ,\beta }^{i0\alpha
\beta }\,n_{\alpha }\,dS
\]
where $n_{\alpha }$ is the outward unit normal vector over an infinitesimal
surface element $dS$.

The energy-momentum density $\Omega ^{ik}$\ is symmetric with
respect to the two indices $i$ and $k$, therefore it can be used
to define angular momentum density
\[
M^{ikl}=r^{i}\,\Omega ^{kl}-r^{k}\,\Omega ^{il}.
\]
\section{Weinberg energy-momentum complex}     %
Weinberg\cite{Weinberg} obtained an energy-momentum complex by
considering a quasi-Minkowskian coordinate system\footnote{We
refer to a coordinate system as being quasi-Minkowskian if the
metric $g_{ab}$ approaches the Minkowski metric $\eta _{ab}$ far
away from a given finite material system.}. In this
quasi-Minkowskian coordinate system  $g_{ab}$ may be viewed as the
sum of the components of the Minkowski metric $\eta _{ab}$ and
that part $h_{ab}$ which vanishes at infinity, i.e.
\begin{equation}
g_{ab}=\eta _{ab}+h_{ab}.  \label{Weq1}
\end{equation}
$h_{ab}$ is only assumed to vanish at infinity but may take
arbitrarily large values elsewhere. Using this notation the Ricci
tensor may then be written as:
\begin{equation}
R_{ab}=R_{~~~ab}^{(1)}+R_{~~~ab}^{(2)}+\bigcirc(h^3),
\label{Weq1a}
\end{equation}
where the linear part $R_{~~~ab}^{(1)}$ in $h_{ab}$ is given by:
\begin{equation}
R_{~~~ab}^{(1)}=\frac{1}{2}\left( \frac{\partial
^{2}h_{p}^{p}}{\partial x^{a}\partial x^{b}}-\frac{\partial
^{2}h_{a}^{p}}{\partial x^{p}\partial
x^{b}}-\frac{\partial ^{2}h_{b}^{p}}{\partial x^{p}\partial x^{a}}+\frac{%
\partial ^{2}h_{ab}}{\partial x^{p}\partial x_{p}}\right) , \label{Weq2}
\end{equation}
and the second-order part $R_{~~~ab}^{(2)}$ in $h_{ab}$ is given
by:
\begin{eqnarray*}
R_{~~~ab}^{(2)}=&-&\frac{1}{2}h^{pq}\left[\frac{\partial^2
h_{pq}}{\partial x^{b}\partial x^{a}}-\frac{\partial^2
h_{aq}}{\partial x^{b}\partial x^{p}}-\frac{\partial^2
h_{pb}}{\partial x^{q}\partial x^{a}}+\frac{\partial^2
h_{ab}}{\partial x^{q}\partial x^{p}}\right]\\
&+&\frac{1}{4}\left[\frac{\partial h^{q}_{~a}}{\partial
x^{b}}+\frac{\partial h^{q}_{~b}}{\partial x^{a}}-\frac{\partial
h_{ab}}{\partial x_{q}}\right]\left[2\frac{\partial
h^{p}_{~q}}{\partial x^{p}}-\frac{\partial h^{p}_{~p}}{\partial
x^{q}}\right]\\
&-&\frac{1}{4}\left[\frac{\partial h^{p}_{~a}}{\partial
x_{q}}+\frac{\partial h^{pq}}{\partial x^{a}}-\frac{\partial
h^{q}_{~a}}{\partial x_{p}}\right]\left[\frac{\partial
h_{pb}}{\partial x^{q}}+\frac{\partial h_{pq}}{\partial
x^{b}}-\frac{\partial h_{qb}}{\partial x^{p}}\right].
\label{Weq2b}
\end{eqnarray*}
In the above expressions of $R_{~~~ab}^{(1)}$ and
$R_{~~~ab}^{(2)}$, indices on $h_{ik}$  and $\partial/\partial
x^{i}$ are raised and lowered with $\eta$'s, whereas indices on
true tensors such as $R_{ik}$\ are raised and lowered with $g$'s
as usual. Using the  above notation, Einstein's field equations
may now be written as:
\begin{equation}
R_{~~~ab}^{(1)}-\frac{1}{2}\eta_{ab}R_{~~~~p}^{(1)p}=-8\pi \left[
T_{ab}+t_{ab}\right] , \label{Weq3}
\end{equation}
where
\begin{equation}
t_{ab}=\frac{1}{8\pi }\left[ R_{ab}-\frac{1}{2}g_{ab}R_{~p}^{p}-R_{~~~ab}^{(1)}-%
\frac{1}{2}\eta _{ab}R_{~~~~p}^{(1)p}\right]. \label{Weq4}
\end{equation}
Note that the left hand side of Eq. $(\ref{Weq3})$\ may also be
written as
\begin{equation}
R^{(1)ik}-\frac{1}{2}\eta ^{ik}R_{\text{ \ \ \ }p}^{(1)p}={\omega
}_{\quad ,p}^{pik}  \label{Weq3a}
\end{equation}
where
\begin{equation}
{\omega }^{pik}=\frac{\partial h_{a}^{a}}{\partial x_{i}}\eta ^{pk}-\frac{%
\partial h_{a}^{a}}{\partial x_{p}}\eta ^{ik}-\frac{\partial h^{ai}}{%
\partial x^{a}}\eta ^{pk}+\frac{\partial h^{ap}}{\partial x^{a}}\eta ^{ik}+%
\frac{\partial h^{ik}}{\partial x_{p}}-\frac{\partial
h^{pk}}{\partial x_{i}} \label{Weq3b}
\end{equation}
is antisymmetric in its first two indices $i$\ and $p$, that is, ${\omega }%
^{pik}=-{\omega }^{ipk}$\ so that it now follows that:
\begin{equation}
\frac{\partial }{\partial x^{p}}\left[ R_{\text{
}}^{(1)pb}-\frac{1}{2}\eta ^{pb}R_{\text{ \ \ \ }q}^{(1)q}\right]
=0.  \label{Weq5}
\end{equation}
After analyzing Eq.$(\ref{Weq3})$\ and comparing its form with
that of the wave equation of a field spin of 2, and further noting
that, since the quantities $R_{~~~ab}^{(1)}$ obey the linearized
Bianchi identities Eq.$(\ref{Weq5})$\ Weinberg concluded that the
quantity $W^{ab}$\ given by:
\begin{equation}
W ^{ab}=\eta ^{ap}\eta ^{bq}\left[ T_{pq}+t_{pq}\right],
\label{Weq6}
\end{equation}
which by Eqs. $(\ref{Weq3})$ and $(\ref{Weq5})$ satisfies the
following local conservation law:
\begin{equation}
\frac{\partial }{\partial x^{p}}W ^{pb}=0,  \label{Weq7}
\end{equation}
may be interpreted as consisting of the total energy-momentum
pseudocomplex of matter and gravitation, with $t_{ik}$\ indicating
the energy-momentum pseudotensor of gravitation. Therefore for any
finite system of volume $V$ bounded by the surface $S$, we have:
\begin{equation}
\frac{d}{dt}\int_{V}W ^{0i}d^{3}x=-\int_{S}W ^{\alpha i}n_{\alpha
}dS ,\label{Weq8}
\end{equation}
where {\bf n}\ is the unit outward normal to the surface.
Analogously, the quantities $P^{i}$ given by:
\begin{equation}
P^{i}=\int \int\int W ^{0i}dx^1 dx^2 dx^3 ,\label{Weq9}
\end{equation}
may therefore be interpreted as representing the total
energy-momentum components of the system including matter,
electromagnetism, and gravitation.

Since $h_{ab}\rightarrow 0$ as $r\rightarrow \infty $, the
energy-momentum
tensor of matter plus non-gravitational field $%
T_{ab}$ also vanishes at infinity. From
\begin{eqnarray}
h_{ab} &=&{\Large O}\left( \frac{1}{r}\right) ,  \nonumber \\
h_{ab,c} &=&{\Large O}\left( \frac{1}{r^{2}}\right) ,  \nonumber \\
h_{ab,cd} &=&{\Large O}\left( \frac{1}{r^{3}}\right) ,
\label{Weq15}
\end{eqnarray}
then using Eqs. $(\ref{Weq1a},\ref{Weq2},\ref{Weq4})$ and noting
that the quantity $t_{ab}$ is of the second order in $h$, it
follows that
\begin{equation}
t_{ik}={\Large O}\left( \frac{1}{r^{4}}\right),   \label{Weq16}
\end{equation}
approaches zero at infinity. This shows that the source term on
the left hand side of $(\ref{Weq3})$\ is effectively confined to a
finite region. Therefore the quantities $P^{i}$\ in
$(\ref{Weq9})$\ that give the total energy and momentum converge.

Weinberg further justifies his choice of the quantities $P^{i}$ by
showing that these quantities are four-vectors and are additive.
We illustrate below that $P^{i}$  are invariant under any
transformation that reduces to an identity transformation at
infinity
\begin{enumerate}
\item First consider the transformation of the form
\begin{equation}
x^{a^{\prime }}=x^{a}+f^{a}(x)  \label{Weq17}
\end{equation}
where $f^{a}(x)\rightarrow 0$ as $r\rightarrow \infty $. Then to
first order
in both $f^{a}$ and $h_{ab}$, the metric tensor $g^{a^{\prime }b^{\prime }}$%
\ in the new coordinate system $x^{a^{\prime }}$\ will then be
given by
\begin{equation}
g^{a^{\prime }b^{\prime }}=\eta ^{ab}-h^{a^{\prime }b^{\prime }}
\label{Weq18}
\end{equation}
where
\begin{equation}
h^{a^{\prime }b^{\prime }}=h^{ab}-\frac{\partial f^{a}}{\partial x_{b}}-%
\frac{\partial f^{b}}{\partial x_{a}}  \label{Weq19}
\end{equation}
since as $r\rightarrow \infty $ both $f^{a}$ and $h_{ab}$\ are
small. This change in coordinate transformation will lead to the
following change in quantity ${\omega }^{pik}$ defined by
$(\ref{Weq3b})$\
\begin{equation}
\Delta \omega ^{bik}=D_{\quad \,\,,a}^{abik}  \label{Weq20}
\end{equation}
where
\begin{equation}
D^{abik}=\frac{1}{2}\left\{ -\frac{\partial f^{a}}{\partial x_{i}}\eta ^{bk}+%
\frac{\partial f^{a}}{\partial x_{b}}\eta ^{ik}+\frac{\partial f^{i}}{%
\partial x_{a}}\eta ^{bk}-\frac{\partial f^{b}}{\partial x_{a}}\eta ^{ik}-%
\frac{\partial f^{i}}{\partial x_{b}}\eta ^{ak}+\frac{\partial f^{b}}{%
\partial x_{i}}\eta ^{ak}\right\} .  \label{Weq21}
\end{equation}
Let us express $P^{k}$ in terms of ${\omega }^{pik}$\ as
\begin{equation}
P^{k}=+\frac{1}{8\pi }\int \int \int {\omega }_{\quad
,p}^{p0k}dx^{1}dx^{2}dx^{3}  \label{Weq22}
\end{equation}
which, using Gauss's theorem, gives
\begin{equation}
P^{k}=+\frac{1}{8\pi }\int \int \omega ^{\alpha 0k}\mu _{\alpha
}dS \label{Weq23}
\end{equation}
where $\mu _{\alpha }=\frac{x_{i}}{r}$ is the outward unit normal
vector
over an infinitesimal surface element $dS=r^{2}\sin \theta d\theta d\varphi $%
. Now noting that $D^{abik}$ is totally antisymmetric with respect
to its first three indices $a,b$\ and $i$, then the above change
$\Delta \omega ^{bik}$ will result in the following change in the
surface integral
\begin{eqnarray}
\Delta P^{k} &=&+\frac{1}{8\pi }\int \int D_{\quad
\,\,,a}^{a\alpha 0k}\mu
_{\alpha }dS  \label{Weq24} \\
&=&+\frac{1}{8\pi }\int \int D_{\quad \,\,,\alpha }^{\alpha \beta
0k}\mu _{\beta }dS  \nonumber
\end{eqnarray}
which, using Gauss's theorem, gives
\begin{equation}
\Delta P^{k}=+\frac{1}{8\pi }\int \int \int D_{\quad ,\alpha \beta
}^{\alpha \beta 0k}\mu _{\alpha }dx^{1}dx^{2}dx^{3}=0,
\label{Weq25}
\end{equation}
thus showing that $P^{i}$ is invariant under any transformation
that reduces to an identity transformation at infinity. An
important consequence of this result is that $P^{i}$\ transforms
as a four-vector under any transformation that leaves the
Minkowski metric $\eta _{ab}$ at infinity unchanged, because any
such transformation can be expressed as the product of a Lorentz
transformation under which $P^{i}$\ transforms as a four-vector.\

\item  We now show the additive property of  $P^{i}$. Dividing the
matter in our system into distant subsystems $S_{(n)}$,
we can approximate the gravitational field $h_{ab}$ as the sum of $%
h_{ab}^{(n)}$ 's that would be produced by each subsystem acting
alone. Thus, from the above calculation of $P^{i}$,\ it follows
that the total energy and momentum are equal to the sum of the
values $P_{(n)}^{i}$\ for each subsystem alone.
\end{enumerate}
Now from the above, it has been shown that $P^{i}$\ is conserved,
is a four-vector, and is additive. In addition to these
properties, the total energy-momentum complex $W^{ik}$ is
conserved and symmetric in its indices, therefore we can use it to
define a conservation law for angular momentum
\begin{equation}
M_{\,\,,p}^{pik}=0  \label{Weq26}
\end{equation}
where
\[
M^{pik}=x^{i}W^{pk}-x^{k}W^{pi}
\]
so that $M^{0ik}$\ and $M^{\alpha ik}$\ can be taken as
representing the density and flux of a total angular momentum
\[
J^{ik}=\int \int \int M^{0ik}dx^{1}dx^{2}dx^{3}
\]
\ which is constant if $M^{\alpha ik}$\ vanishes on the surface of
the volume of integration. Both $M^{pik}$\ and $J^{ik}$\ are
antisymmetric with respect to indices $i$\ and $k$.\ As above, the
total angular momentum complex can be written in terms of ${\omega
}^{pik}$\ as
\begin{equation}
J^{ik}=+\frac{1}{8\pi }\int \int \int \left( x^{i}{\omega }_{\quad
,\alpha }^{\alpha 0k}-x^{k}{\omega }_{\quad ,\alpha }^{\alpha
0i}\right) \,dx^{1}dx^{2}dx^{3}
\end{equation}
which, using Gauss's theorem, gives
\begin{eqnarray}
16\pi J^{\alpha \beta } &=&+\int \int \left\{ -x_{\alpha
}\frac{\partial
h_{o\beta }}{\partial x^{\gamma }}+x_{\beta }\frac{\partial h_{o\gamma }}{%
\partial x^{\alpha }}-x_{\beta }\frac{\partial h_{\alpha \gamma }}{\partial t%
}\,\right. \nonumber\\
&&+\left. x_{\alpha }\frac{\partial h_{\beta \gamma }}{\partial
t}+h_{0\beta }\delta _{\alpha \gamma }-h_{0\alpha }\delta _{\beta
\gamma }\right\} \mu _{\alpha }dS.
\end{eqnarray}
We only give the physically meaningful components of $J^{ik}$\
which are the three independent space-space components
$J^{1}=J^{23}$\ , $J^{2}=J^{31}$\ and $J^{3}=J^{12}$. In order to
calculate the total momentum, energy, and angular momentum of an
arbitrary system, one only needs to know the
asymptotic behavior of $h_{ab}$\ at great distances. Though the quantities $%
t_{ab},$ $W^{ab}$ and $M^{pik}$ are not tensors, they are at least
Lorentz covariant. Thus for a closed system $P^{i}$\ and $J^{ik}$\
will not only be constant but Lorentz-covariant.

\section{Present study of energy localization}  %
Rosen and Virbhadra \cite{RosVir} investigated the energy and momentum of
the Einstein-Rosen metric using the Einstein energy-momentum complex. The
Einstein-Rosen metric is a non-static vacuum solution of Einstein's field
equations that describes the gravitational field of cylindrical
gravitational waves given in cylindrical polar coordinates ($\rho ,\,\phi
\,,z)$ by the line element
\begin{equation}
ds^{2}=e^{2\gamma -2\Psi }(dt^{2}-d\rho ^{2})-e^{-2\Psi }\rho
^{2}d\phi ^{2}-e^{2\Psi }dz^{2}, \label{Veq1}
\end{equation}
where $\gamma =\gamma \left( \rho ,t\right) $ and $\Psi =\Psi \left( \rho
,t\right) $ and
\begin{eqnarray*}
\Psi _{,tt}-\Psi _{,\rho \rho }-\frac{1}{\rho }\Psi _{,\rho } &=&0, \\
\gamma _{,t} &=&2\rho \Psi _{,\rho }\Psi _{,t}, \\
\gamma _{,t} &=&\rho \left( \Psi _{,\rho }^{2}+\Psi _{,t}^{2}\right) .
\end{eqnarray*}
These authors carried out their calculations in quasi-Cartesian
coordinates, and reported that the energy and momentum density
components are non-vanishing and reasonable. Rosen (see in
\cite{Rosen58}) had earlier computed, in cylindrical polar
coordinates, the energy and momentum components in this metric
using the energy-momentum complexes of Einstein and Landau and
Lifshitz. For both prescriptions the energy and momentum density
components vanished. Initially, the vanishing of these components
seemed to confirm Scheidegger's conjecture that a physical system
cannot radiate gravitational energy. However, two years later
Rosen \cite{Rosen58} realized his mistake and recalculated energy
and momentum density components in quasi-Cartesian coordinates and
found finite and reasonable results, which were later reported by
himself and Virbhadra \cite{RosVir}. Virbhadra \cite{KSV95} showed
that the energy-momentum complexes of Einstein and Landau and
Lifshitz give the same energy and momentum densities when
calculations are carried out in quasi-Cartesian coordinates. The
energy density of the cylindrical gravitational waves was found to
be finite and positive definite. The momentum density components
was found to reflect the symmetry of the spacetime.

In a recent paper Virbhadra \cite{KSV99} showed that the
energy-momentum complexes of Einstein, Landau and Lifshitz,
Papapetrou, and Weinberg, and the Penrose quasi-local definition
give the same result for a general nonstatic spherically symmetric
metric of the Kerr-Schild class. The well-known spacetimes of the
Kerr-Schild class are for example the Schwarzschild,
Reissner-Nordstr\"{o}m, Kerr, Kerr-Newman, Vaidya, Dybney
\textit{et al.}, Kinnersley, Bonnor-Vaidya and Vaidya-Patel. These
spacetimes are defined in terms of the following metrics:
\begin{equation}
g_{ab}=\eta _{ab}-Hk_{a}k_{b},\label{Veq2}
\end{equation}
where $\eta _{ab}$ is the Minkowski metric, $H$ is a scalar field, and $%
k_{a} $ is a null, geodesic and shear free vector field in the Minkowski
spacetime. Each of these are given by
\begin{eqnarray}
\eta ^{pq}k_{p}k_{q} &=&0, \label{Veq3}\\
\eta ^{pq}k_{i,p}k_{q} &=&0,\label{Veq4} \\
 \left( k_{p,q}+k_{q,p}\right) k_{\,\,,r}^{p}\eta
^{qr}-\left( k_{\,\,,p}^{p}\right) ^{2} &=&0.\label{Veq5}
\end{eqnarray}
The vector $k_{a}$ of the Kerr-Schild class metric $g_{ab}$
remains null, geodesic and shear free with the metric $g_{ab}$ .
Eqs. $(\ref{Veq3})$-$(\ref{Veq5})$ lead to
\begin{eqnarray*}
g^{pq}k_{p}k_{q} &=&0, \\
g^{pq}k_{i;p}k_{q} &=&0, \\
\left( k_{p;q}+k_{q;p}\right) k_{\,\,;r}^{p}g^{qr}-\left(
k_{\,\,;p}^{p}\right) ^{2} &=&0.
\end{eqnarray*}
Aguirregabiria et. al.\cite{ACV96} obtained the following results
\begin{eqnarray}
\theta _{i}^{k} &=&\frac{1}{16\pi }\eta _{ir}\Lambda _{\qquad ,pq}^{rkpq} \label{Veq8}\\
\Omega^{ik} &=&L^{ik}=W^{ik}=\frac{1}{16\pi }\Lambda _{\qquad
,pq}^{ikpq}\label{Veq9}
\end{eqnarray}
where
\[
\Lambda ^{abcd}=H\left( \eta ^{ab}k^{c}k^{d}+\eta
^{cd}k^{a}k^{b}-\eta ^{ac}k^{b}k^{d}-\eta ^{bd}k^{a}k^{c}\right)
\]
therefore the energy-momentum complexes of Einstein $\theta
_{i}^{k}$, Landau and Lifshitz $L^{ik}$, Papapetrou $\Omega^{ik}$,
and Weinberg $W^{ik}$ `coincide' for any Kerr-Schild class metric.
Only the null conditions of equations $ (\ref{Veq3})$ -
$(\ref{Veq5}) $ was used to obtain the above results in terms of
the scalar function $H$ and the vector $k_{a}$\ for the Landau and
Lifshitz, Papapetrou, and Weinberg complexes was used, while for
the Einstein complex the null as well as geodesic conditions were
used. These energy-momentum complexes `coincide' for a class of
solutions more general than the Kerr-Schild class since the
shear-free conditions were not required to obtain the above
equations. The energy and momentum components are given by
\[
P^{i}=\frac{1}{16\pi }\int \int \Lambda _{\qquad ,q}^{i0\alpha
q}n_{\alpha }dS.
\]
Since the energy-momentum complexes of Landau and Lifshitz, Papapetrou, and
Weinberg are symmetric in their indices the corresponding spatial components
of angular momentum are defined as
\[
J^{\alpha \beta }=\frac{1}{16\pi }\int \int \left( x^{\alpha
}\Lambda _{\qquad ,q}^{\beta 0\gamma q}-x^{\beta }\Lambda _{\qquad
,q}^{\alpha 0\gamma q}+\Lambda ^{\alpha 0\gamma \beta }\right)
n_{\gamma }dS.
\]

\section{Conclusion}%
In this chapter we elaborated on the problem of energy
localization in general relativity, a concept which still remains
a puzzle.  There have been many attempts at finding an appropriate
method for obtaining energy-momentum distribution in a curved
space-time, which resulted in various energy-momentum complexes.
These complexes are restricted to the use particular coordinates.
The problem associated with energy-momentum complexes resulted in
some researchers doubting the concept of energy localization.
However, the leading contributions of Virbhadra and his
collaborators have demonstrated with several examples that some of
these complexes consistently give the same and acceptable energy
and momentum distribution for a particular spacetime. We
elaborated on only a few energy-momentum complexes used in our
work. We also highlighted alternative attempts aimed at finding
the alternative concept of quasilocal energy-momentum. The
coordinate-independent quasilocal mass definitions  are important
conceptually; however, serious problems have been found  with
these.
 Chang, Nester and Chen\cite{Changetal} have also shown
that the Einstein, Landau and Lifshitz, M\o ller, Papapetrou, and
Weinberg energy-momentum complexes may each be associated with a
legitimate Hamiltonian boundary term, and because quasilocal
energy-momentum are obtainable from a Hamiltonian then each of
these complexes may also said to be quasilocal.

\newpage


\chapter{Energy Distribution of Charged Dilaton Black Hole Spacetime}

\label{chapter-THREE}
\section{Charged dilaton black hole spacetime}  %

Properties of the well-known Reissner-Norstr\"{o}m black holes
have been studied extensively.  If one couples the dilaton field
to the Maxwell field of the black holes many of these properties
are known to change (see in Holzhey and Wilczek \cite{HW92}; Horne
and Horowitz\cite{HH92}). Charged dilaton black holes have been a
subject of study in many recent investigations
\cite{HW92}-\cite{SSAD}. Garfinkle, Horowitz and Strominger (GHS)
\cite{GHS91} considered the action
\begin{equation}
S=\int \sqrt{-g}\left[ -R+2\left( \nabla \Phi \right) ^{2}+e^{-2\gamma \Phi
}F^{2}\right] d^{4}x  \label{Action1}
\end{equation}
where $R$ is the Ricci scalar, $\Phi $\ is the dilaton field,
$F^2=F_{ab}F^{ab}$ where $F_{ab}$\ is the electromagnetic field
tensor, and $\gamma $ is an arbitrary parameter which governs the
strength of the coupling between the dilaton and the Maxwell
fields. We shall consider only nonnegative values of $\gamma $
since a change in the sign of the parameter $\gamma $ will have
the same effect as
a change in the sign of the dilaton field $\Phi $. The action $(\ref{Action1}%
)$ reduces to the Einstein-Maxwell scalar theory for $\gamma =0$. When $%
\gamma =1$ then $(\ref{Action1})$ gives an action which is part of the
low-energy action of string theory. For $\gamma =\sqrt{3}$ we get the action
for the Kaluza-Klein theory. By varying the action $(\ref{Action1})$ we
obtain the following equations of motion:
\begin{eqnarray}
&&\nabla _{i}\left( e^{-2\gamma \Phi }F^{ik}\right) =0,  \nonumber \\
&&\nabla ^{2}\Phi +\frac{\gamma }{2}e^{-2\gamma \Phi }F^{2}=0,  \nonumber \\
R_{ik} &=&2\nabla _{i}\Phi \nabla _{k}\Phi +2e^{-2\gamma \Phi
}F_{ip}F_{k}^{p}-\frac{1}{2}g_{ik}e^{-2\gamma \Phi }F^{2}.
\end{eqnarray}
Garfinkle, Horowitz and Strominger \cite{GHS91} obtained static spherically
symmetric asymptotically flat black-hole solution described by the line
element
\begin{eqnarray}
ds^{2} &=&(1-\frac{r_{+}}{r})(1-\frac{r_{-}}{r})^{\sigma }dt^{2}-(1-\frac{%
r_{+}}{r})^{-1}(1-\frac{r_{-}}{r})^{-\sigma }dr^{2}  \nonumber \\
&&-(1-\frac{r_{-}}{r})^{1-\sigma }r^{2}(d\theta ^{2}+sin^{2}\theta
d\phi ^{2})  \label{eqn3.1}
\end{eqnarray}
with the dilaton field $\Phi $ given by
\begin{equation}
e^{2\Phi }=\left( 1-\frac{r_{-}}{r}\right) ^{\frac{1-\sigma }{\gamma }},
\end{equation}
and the electromagnetic field tensor component
\begin{equation}
F_{tr}=\frac{Q}{r^{2}}
\end{equation}
where
\begin{equation}
\sigma =\frac{1-{\gamma }^{2}}{1+{\gamma }^{2}}.  \label{eqn3.3}
\end{equation}
$r_{-}$ and $r_{+}$ are related to mass $M$ and charge $Q$ parameters as
follows:
\begin{eqnarray}
M &=&\frac{r_{+}+\sigma r_{-}}{2}  \nonumber \\
Q^{2} &=&\frac{r_{+}r_{-}}{(1+\gamma ^{2})}.
\end{eqnarray}
Virbhadra\cite{KSV97} proved that for $Q=0$ the GHS solution
yields the Janis-Newman-Winicour solution\cite{JNW68} to the
Einstein massless scalar equations. Virbhadra, Jhingan and
Joshi\cite{Viretal97} showed that the Janis-Newman-Winicour
solution has a globally naked strong curvature singularity.

For $\gamma =0$ the solution yields the standard
Reissner-Nordstr\"{o}m of the Einstein-Maxwell theory, but for
$\gamma \neq 0$ the solution is qualitatively different.

Certain qualitative features of the solutions of non-rotating
charged dilaton black-holes are independent of the parameter
$\gamma $. For instance, the surface $r=r_{+}$ is an event horizon
for all values of $\gamma $. A number of interesting properties of
charged dilaton black holes critically depend on the dimensionless
parameter $\gamma $ which controls the coupling between the
dilaton and the Maxwell fields. The maximum charge, for a given
mass, that can be carried by a charged dilaton black-hole depends
on $\gamma $ \cite{HW92}. When $\gamma \neq 0$, the surface
$r=r_{-}$ is a curvature singularity while at $\gamma =0$ the
surface $r=r_{-} $ is a nonsingular inner horizon \cite{HH92}.
Both the entropy and temperature of these black holes depend on
$\gamma $ \cite{HW92}. The gyromagnetic ratio, i.e. the ratio of
the magnetic dipole moment to the angular momentum, for charged
slowly rotating dilaton black holes depends on parameter $\gamma $
\cite{HW92}. Chamorro and Virbhadra \cite{ChaVir96} showed, using
Einstein's prescription, that the energy distribution of charged
dilaton black holes depends on the value of $\gamma $.

\section{Energy distribution in charged dilaton black holes}%
Virbhadra and Parikh\cite{VirPar93} calculated, using the energy-momentum
complex of Einstein, the energy distribution with stringy charged black
holes ($\gamma = 1$) and found that
\[
E=M.
\]
Thus the entire energy of a charged black-hole in low-energy string theory
is confined to the interior of the black-hole. We\cite{Xulu98a} found the
same result using the Tolman definition. For the Reissner-Nordstr\"{o}m
metric, several definitions of energy give
\[
E=M-\frac{Q^{2}}{2r}
\]
(Tod \cite{Tod}; Hayward \cite{Hayward}; Aguirregabiria {\it et
al.}\ \cite {ACV96}). Thus the energy is both in its interior and
exterior. Chamorro and Virbhadra \cite{ChaVir96} studied, using
the energy-momentum complex of Einstein, the energy distribution
associated with static spherically symmetric charged dilaton black
holes for an arbitrary value of the coupling parameter $\gamma $
which controls the strength of the dilaton to the Maxwell field.
We \cite{Xulu98b} investigated the energy distribution in the same
spacetime in Tolman's prescription and got the same result as
obtained by Chamorro and Virbhadra. The energy distribution of
charged dilaton black holes depends on the value of $\gamma $ and
the total energy is independent of this parameter. In this chapter
we present the computations of the energy distribution for the
Garfinkle-Horowitz-Strominger spacetime performed using the Tolman
energy-momentum complex\footnote{Virbhadra\cite{KSV99} pointed out
that though the Tolman energy-momentum complex differs in form
from the energy-momentum complex obtained by Einstein, both are
equivalent in import. Therefore, the Tolman energy-momentum
complex should be correctly referred to as the Tolman form of
Einstein's energy-momentum complex. The author was not aware of
this fact at the time of writing papers \cite{Xulu98a,Xulu98b}.}.

\section{Tolman energy distribution}%

We start by transforming the line element $(\ref{eqn3.1})$ to
quasi-Cartesian coordinates:
\begin{eqnarray}
ds^{2} &=&(1-\frac{r_{+}}{r})(1-\frac{r_{-}}{r})^{\sigma }dt^{2}-(1-\frac{%
r_{-}}{r})^{1-\sigma }(dx^{2}+dy^{2}+dz^{2})  \nonumber \\
&-&\frac{(1-\frac{r_{+}}{r})^{-1}(1-\frac{r_{-}}{r})^{-\sigma }-(1-\frac{%
r_{-}}{r})^{1-\sigma }}{r^{2}}(xdx+ydy+zdz)^{2},  \label{eqn3.5}
\end{eqnarray}
according to
\begin{eqnarray}
r &=&\sqrt{x^{2}+y^{2}+z^{2}},  \nonumber \\
\theta &=&\cos ^{-1}\left( \frac{z}{\sqrt{x^{2}+y^{2}+z^{2}}}\right) ,
\label{eqn3.6} \\
\phi &=&\tan ^{-1}\left( \frac{y}{x}\right) .  \nonumber
\end{eqnarray}

Tolman's\cite{Tolman34} energy-momentum complex is
\begin{equation}
{\cal {T}}_{k}{}^{i}=\frac{1}{8\pi }U_{k,j}^{ij},  \label{eqn3.7}
\end{equation}
where
\begin{eqnarray}
U_{k}^{ij} &=&\sqrt{-g}\left[ -g^{pi}(-\Gamma _{kp}^{j}+\frac{1}{2}%
g_{k}^{j}\Gamma _{ap}^{a}+\frac{1}{2}g_{p}^{j}\Gamma _{ak}^{a})\right.
\nonumber \\
&&+\frac{1}{2}g_{k}^{i}g^{pm}\left. (-\Gamma _{pm}^{j}+\frac{1}{2}%
g_{p}^{j}\Gamma _{am}^{a}+\frac{1}{2}g_{m}^{j}\Gamma
_{ap}^{a})\right] , \label{eqn3.8}
\end{eqnarray}
${\cal {T}}_{0}^{0}$ is the energy density, ${\cal {T}}_{0}^{\alpha }$ are
the components of energy current density, ${\cal {T}}_{\alpha }^{0}$ are the
momentum density components. Therefore, energy $E$ for a stationary metric
is given by the expression
\begin{equation}
E=\frac{1}{8\pi }\int \int \int U_{0,\ \alpha }^{0\alpha }\
dxdydz.
\end{equation}
After applying the Gauss theorem one has
\begin{equation}
E=\frac{1}{8\pi }\int \int U_{0}^{0\alpha }\mu _{\alpha
}dS,\label{eqn3.9}
\end{equation}
where $\mu _{\alpha }=(x/r,y/r,z/r)$ are the three components of a normal
vector over an infinitesimal surface element $dS=r^{2}sin\theta d\theta
d\phi $.

The determinant of the metric tensor is given by
\begin{equation}
g=-\left( \frac{r}{r-r_{-}}\right) ^{2\left( \sigma -1\right) }
\end{equation}
and its non-vanishing contravariant components are:
\begin{eqnarray}
g^{00} &=&\frac{r^{\sigma +1}}{\left( r-r_{-}\right) ^{\sigma }\left(
r-r_{+}\right) },\nonumber \\
g^{11} &=&\frac{\left( r-r_{-}\right) ^{\sigma -1}}{r^{\sigma +3}}\left[
x^{2}\left( rr_{+}+rr_{-}-r_{+}r_{-}\right) -r^{4}\right] ,\nonumber \\
g^{12} &=&\frac{xy\left( r-r_{-}\right) ^{\sigma -1}}{r^{\sigma +3}}\left[
r\left( r_{+}+r_{-}\right) -r_{+}r_{-}\right] ,\nonumber \\
g^{22} &=&\frac{\left( r-r_{-}\right) ^{\sigma -1}}{r^{\sigma +3}}\left[
y^{2}\left( rr_{+}+rr_{-}-r_{+}r_{-}\right) -r^{4}\right] ,\nonumber \\
g^{23} &=&\frac{yz\left( r-r_{-}\right) ^{\sigma -1}}{r^{\sigma +3}}\left[
r\left( r_{+}+r_{-}\right) -r_{+}r_{-}\right] , \nonumber\\
g^{33} &=&\frac{\left( r-r_{-}\right) ^{\sigma -1}}{r^{\sigma +3}}\left[
z^{2}\left( rr_{+}+rr_{-}-r_{+}r_{-}\right) -r^{4}\right] , \nonumber\\
g^{31} &=&\frac{zx\left( r-r_{-}\right) ^{\sigma -1}}{r^{\sigma
+3}}\left[ r\left( r_{+}+r_{-}\right) -r_{+}r_{-}\right]
\label{eqn3.10}.
\end{eqnarray}
To compute the energy distribution using Eq. $(\ref{eqn3.9})$ we
also require the following list of nonvanishing components of the
Christoffel symbol of the second kind.
\begin{eqnarray}
\Gamma _{11}^{1} &=&x(c_{1}+c_{2}x^{2}),  \nonumber \\
\Gamma _{22}^{2} &=&y(c_{1}+c_{2}y^{2}),  \nonumber \\
\Gamma _{33}^{3} &=&z(c_{1}+c_{2}z^{2}),  \nonumber \\
\Gamma _{22}^{1} &=&x(c_{3}+c_{2}y^{2}),  \nonumber \\
\Gamma _{11}^{2} &=&y(c_{3}+c_{2}x^{2}),  \nonumber \\
\Gamma _{33}^{1} &=&x(c_{3}+c_{2}z^{2}),  \nonumber \\
\Gamma _{11}^{3} &=&z(a_{3}+a_{2}x^{2}),  \nonumber \\
\Gamma _{33}^{2} &=&y(c_{3}+c_{2}z^{2}),  \nonumber \\
\Gamma _{22}^{3} &=&z(c_{3}+c_{2}y^{2}),  \nonumber \\
\Gamma _{12}^{1} &=&y(c_{4}+c_{2}x^{2}),  \nonumber \\
\Gamma _{13}^{1} &=&z(c_{4}+c_{2}x^{2}),  \nonumber \\
\Gamma _{21}^{2} &=&x(c_{4}+c_{2}y^{2}),  \nonumber \\
\Gamma _{23}^{2} &=&z(c_{4}+c_{2}y^{2}),  \nonumber \\
\Gamma _{31}^{3} &=&x(c_{4}+c_{2}z^{2}),  \nonumber \\
\Gamma _{32}^{3} &=&y(c_{4}+c_{2}z^{2}),  \nonumber \\
\Gamma _{00}^{1} &=&xc_{5},  \nonumber \\
\Gamma _{00}^{2} &=&yc_{5},  \nonumber \\
\Gamma _{00}^{3} &=&zc_{5},  \nonumber \\
\Gamma _{01}^{0} &=&xc_{6},  \nonumber \\
\Gamma _{02}^{0} &=&yc_{6},  \nonumber \\
\Gamma _{03}^{0} &=&zc_{6},  \nonumber \\
\Gamma _{23}^{1}~~ &=&~~~\Gamma _{13}^{2}~~~=~~~\Gamma
_{12}^{3}~~~=~~~c_{2}xyz,  \label{eqn3.11}
\end{eqnarray}
where\newline
\begin{eqnarray}
c_{1} &=&\frac{1}{2r^{4}(r-r_{-})}[2r_{+}r^{2}+3r_{-}r^{2}-3r_{-}r_{+}r
\nonumber \\
&&+{r_{-}}^{2}r_{+}-{r_{-}}^{2}r+(r_{-}r_{+}-r_{-}r-r_{+}r-r^{2})r_{-}\sigma
],  \nonumber \\
c_{2} &=&\frac{1}{2r^{4}}\left[ \frac{2{r_{+}}%
^{2}r+6r_{-}r_{+}r-3r_{-}r^{2}-3r_{+}r^{2}-3r_{-}{r_{+}}^{2}}{r-r_{+}}\right.
\nonumber \\
&&+\left. \frac{(2r_{-}-r_{-}\sigma )(r_{-}r_{+}-r_{-}r-r_{+}r)}{r-r_{-}}%
\right] ,  \nonumber \\
c_{3} &=&\frac{1}{2r^{4}}[r_{-}r+2r_{+}r-r_{-}r_{+}+(r-r_{+})r_{-}\sigma ],
\nonumber \\
c_{4} &=&\frac{1}{2r^{2}}\left[ \frac{r_{-}-r_{-}\sigma }{r-r_{-}}\right] ,
\nonumber \\
c_{5} &=&\frac{(r-r_{-})^{2\sigma -1}}{2r^{2\sigma +4}}%
(r-r_{+})[(r-r_{+})r_{+}+(r-r_{+})r_{-}\sigma ],  \nonumber \\
c_{6} &=&\frac{1}{2r^{2}}\left[ \frac{r_{+}}{r-r_{+}}+\frac{r_{-}\sigma }{%
r-r_{-}}\right] .  \label{eqn3.12}
\end{eqnarray}
Using Eqs. $(\ref{eqn3.8})$ and $(\ref{eqn3.11})$ we obtain
required components of $U_{k}^{ij}$. These are
\begin{eqnarray}
U_{0}^{01} &=&\frac{x}{r^{4}}\left[ r(\sigma r_{-}+r_{+})-\sigma r_{-}r_{+}%
\right] ,  \nonumber \\
U_{0}^{02} &=&\frac{y}{r^{4}}\left[ r(\sigma r_{-}+r_{+})-\sigma r_{-}r_{+}%
\right] ,  \label{eqn3.13} \\
U_{0}^{03} &=&\frac{z}{r^{4}}\left[ r(\sigma r_{-}+r_{+})-\sigma r_{-}r_{+}%
\right] .  \nonumber
\end{eqnarray}
Now using $(\ref{eqn3.13})$ with $(\ref{eqn3.3})$ in
$(\ref{eqn3.9})$ we get
\begin{equation}
E(r)=M-\frac{Q^{2}}{2r}(1-\gamma ^{2}).  \label{eqn3.14}
\end{equation}
\begin{figure}[hbt]
\centering
 \mbox{\epsfig{figure=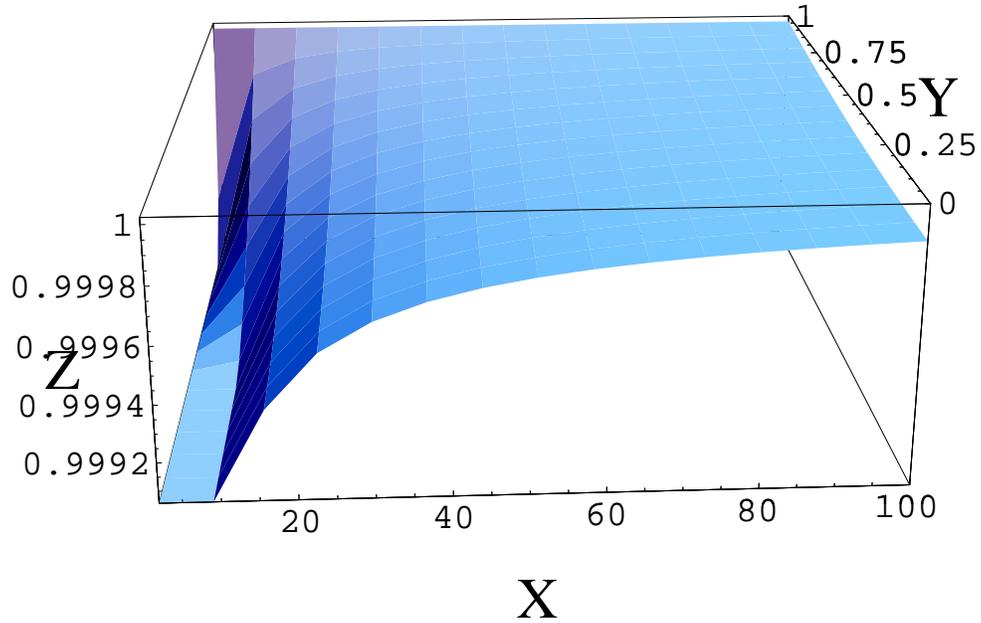,width=\linewidth}}
\caption{ $E/M$  on Z-axis is plotted against  $r/M$ on X-axis
          and $\gamma$ on Y-axis for $Q/M = 0.1$.}  \label{fig3.3}
\end{figure}

Thus, we get
the same result as Chamorro and Virbhadra\cite{ChaVir96} obtained
using the Einstein energy-momentum complex. This is against the
``folklore'' that different energy-momentum complexes could give
different and hence unacceptable energy distribution in a given
spacetime. For the Reissner-Nordstr\"{o}m metric ($\gamma = 0$) one gets
$E=M-Q^{2}/2r$, which is the same as obtained by using several
other energy-momentum complexes(Aguirregabiria, Chamorro and
Virbhadra)\cite{ACV96} and definitions of Penrose as well as
Hayward \cite{Tod},\cite{Hayward}. $E(r)$, given by
$(\ref{eqn3.14})$, can be interpreted as the ``effective
gravitational mass'' that a neutral test particle ``feels'' in the
GHS spacetime. The ``effective gravitational mass'' becomes
negative at radial distances less than $Q^{2}(1-\gamma ^{2})/2M$.
\newpage


\chapter{Energy Distribution in Ernst Space-time}

\label{chapter-FOUR}
\section{Introduction} %

The well-known Melvin's magnetic universe\cite{Melvin1} is a
solution of the Einstein-Maxwell equations corresponding to a
collection of parallel magnetic lines of force held together by
mutual gravitation. Thorne\cite{Thorne} studied extensively the
physical structure of the magnetic universe and investigated its
dynamical behaviour under arbitrarily large radial perturbations.
He showed that no radial perturbation can cause the magnetic field
to undergo gravitational collapse to a space-time singularity or
electromagnetic explosion to infinite dispersion. We\cite{Xulu5}
investigated the energy distribution in Melvin's magnetic universe
and found encouraging results. The energy-momentum complexes of
Einstein, Landau and Lifshitz, and Papapetrou give the same and
acceptable energy distribution in Melvin's magnetic universe. A
discussion of the Melvin's magnetic universe together with its
energy distribution is given in sections $(\ref{sec:melvin})$ and
$(\ref{sec:melvinED})$, below.

Ernst\cite{Ernst} obtained the axially symmetric exact solution to
the Einstein-Maxwell equations representing the Schwarzschild
black hole immersed in the Melvin's uniform magnetic universe.
Virbhadra and Prasanna \cite{VirPra} studied the spin dynamics of
charged particles in the Ernst space-time. We\cite{Xulu3}
investigated energy distribution in the Ernst space-time and
calculated the energy distribution using the Einstein
energy-momentum complex. The first term of the energy expression
is the rest-mass energy of the Schwarzschild black hole, the
second term is the classical value for the energy of the uniform
magnetic field and the remaining terms in the expression are due
to the general relativistic effect. The presence of the magnetic
field is found to increase the energy of the system. Both the
Ernst solution and a discuscussion of energy associated with a
Schwarzschild black hole are given in sections $(\ref{sec:ernst})$
and $(\ref{sec:ernstED})$.

\section{Melvin's magnetic universe}

\label{sec:melvin}

The Einstein-Maxwell equations are
\begin{equation}
R_{i}^{\ k}-\frac{1}{2}\ g_{i}^{\ k}R=8\pi T_{i}^{\ k},
\label{EMeq1}
\end{equation}
\begin{equation}
\frac{1}{\sqrt{-g}}\left( \sqrt{-g}\ F^{ik}\right) _{,k}=4\pi
J^{i}\ , \label{EMeq2}
\end{equation}
\begin{equation}
F_{ij,k}+F_{jk,i}+F_{ki,j}=0,  \label{EMeq3}
\end{equation}
where the energy-momentum tensor of the electromagnetic field is
\begin{equation}
T_{i}^{\ k}=\frac{1}{4\pi }\left[ -F_{im}F^{km}+\frac{1}{4}\
g_{i}^{\ k}F_{mn}F^{mn}\right] .  \label{EMeq4}
\end{equation}
$R_{i}^{\ k}$ is the Ricci tensor and $J^{i}$ is the electric
current density vector.

Melvin \cite{Melvin1} obtained the electrovac solution ($ J^{i}=0
$) to these equations  which is expressed by the line element
\begin{equation}
ds^{2}=\Lambda ^{2}\left[ dt^{2}-dr^{2}-r^{2}d\theta ^{2}\right] -\Lambda
^{-2}r^{2}\sin ^{2}\theta d\phi ^{2}  \label{melvinle}
\end{equation}
and the Cartan components of the magnetic field are
\begin{eqnarray}
H_{r} &=&\Lambda ^{-2}B_{o}\cos \theta ,  \nonumber \\
H_{\theta } &=&-\Lambda ^{-2}B_{o}\sin \theta ,\label{Cartan}
\end{eqnarray}
where
\begin{equation}
\Lambda =1+\frac{1}{4}B_{o}^{2}r^{2}\sin ^{2}\theta .  \label{Lambda}
\end{equation}
$B_{o}$ ($\equiv B_{o}\sqrt{G}/c^{2}$) is the magnetic field parameter and
this is a constant in the solution given above.

The above solution had been obtained earlier by Misra and Radhakrishna%
\footnote{%
Melvin\cite{Melvin2} mentioned that this solution is contained implicitly as
a special case among the solutions obtained by Misra and Radhakrishna.} \cite
{Misra}. This space-time is invariant under rotation about, and translation
along, an axis of symmetry. This is also invariant under reflection in
planes comprising that axis or perpendicular to it. Wheeler\cite{Wheel}
demonstrated that a magnetic universe could also be obtained in Newton's
theory of gravitation and showed that it is unstable according to elementary
Newtonian analysis. However, Melvin\cite{Melvin2} showed his universe to be
stable against small radial perturbations and Thorne \cite{Thorne} proved
the stability of the magnetic universe against arbitrary large
perturbations. Thorne \cite{Thorne} further pointed out that the Melvin
magnetic universe might be of great value in understanding the nature of
extragalactic sources of radio waves and thus the Melvin solution to the
Einstein-Maxwell equations is of immense astrophysical interest. We \cite
{Xulu3}, \cite{Xulu5} computed energy distribution in Melvin's universe
using the definitions of Einstein, Landau and Lifshitz, and Papapetrou. For
this space-time we found that these definitions of energy give the same and
convincing results. The energy distribution obtained here is the same for
all these energy momentum complexes. In the next section we give the
computations of energy distribution for this spacetime performed using
energy-momentum complexes of Landau and Lifshitz, and Papapetrou.

\section{Energy distribution in Melvin magnetic universe}%
\label{sec:melvinED}

The non-zero components of the energy-momentum tensor are
\begin{eqnarray}
T_{1}^{\ 1} &=&-T_{2}^{\ 2}=\frac{B_{o}^{2}\left( 1-2\sin ^{2}\theta \right)
}{8\pi \Lambda ^{4}},  \nonumber \\
T_{0}^{\ 0} &=&-T_{3}^{\ 3}=\frac{B_{o}^{2}}{8\pi \Lambda ^{4}},  \nonumber
\\
T_{2}^{\ 1} &=&-T_{1}^{\ 2}=\frac{2B_{o}^{2}\sin \theta \cos \theta }{8\pi
\Lambda ^{4}}.
\end{eqnarray}
To get meaningful results using these energy-momentum complexes
one is compelled to use ``Cartesian'' coordinates (see
\cite{Moller58} and \cite {KSV99}). We use the following
transformation:
\begin{eqnarray}
r &=&\sqrt{x^{2}+y^{2}+z^{2}},  \nonumber \\
\theta &=&\cos ^{-1}\left(
\frac{z}{\sqrt{x^{2}+y^{2}+z^{2}}}\right) ,
\nonumber \\
\phi &=&\tan ^{-1}(\frac{y}{x}).  \label{eqn4.9}
\end{eqnarray}
The line element $(\ref{melvinle})$  in $t,x,y,z$ coordinates
becomes
\begin{equation}
ds^{2}=\Lambda ^{2}dt^{2}-\Lambda ^{2}(dx^{2}+dy^{2}+dz^{2})+\left( \Lambda
^{2}+\frac{1}{\Lambda ^{2}}\right) \frac{(xdy-ydx)^{2}}{x^{2}+y^{2}}\text{.}
\label{melvinlecart}
\end{equation}
The determinant of the metric tensor is given by
\begin{equation}
g=-\Lambda ^{4},
\end{equation}
and the non-zero contravariant components of the metric tensor are
\begin{eqnarray}
g^{_{00}} &=&\Lambda ^{-2},  \nonumber \\
g^{_{^{11}}} &=&-\,\frac{\Lambda ^{-2}x^{2}+\Lambda ^{2}y^{2}}{x^{2}+y^{2}},
\nonumber \\
g^{_{12}} &=&-\left( \Lambda ^{2}-\frac{1}{\Lambda ^{2}}\right) \frac{xy}{%
x^{2}+y^{2}},  \nonumber \\
g^{_{22}} &=&-\,\frac{\Lambda ^{-2}y^{2}+\Lambda ^{2}x^{2}}{x^{2}+y^{2}},
\nonumber \\
g^{_{33}} &=&-\Lambda ^{-2}.  \label{gupik}
\end{eqnarray}

\subsection{The Landau and Lifshitz energy-momentum complex}%

The symmetric energy-momentum complex of Landau and
Lifshitz\cite{LL} may be written as
\begin{equation}
L^{ij}=\frac{1}{16\pi }\ell_{\quad ,kl}^{ikjl}  ,\label{Lij}
\end{equation}
where
\begin{equation}
\ell^{ikjl}=-g(g^{ij}g^{kl}-g^{il}g^{kj}) . \label{Sijkl}
\end{equation}
$L^{00}$ is the energy density and $L^{0\alpha }$ are the momentum
(energy current) density components. $ \ell ^{mjnk} $ has
symmetries of the Riemann curvature tensor. The energy $E$ is
given by the expression
\begin{equation}
E_{LL}=\frac{1}{16\pi }\int \int \ell_{\quad ,\alpha }^{0\alpha
0\beta }\ \mu _{\beta }\ dS  ,\label{LLGauss}
\end{equation}
where $\mu _{\beta }$ is the outward unit normal vector over an
infinitesimal surface element $dS$. In order to calculate the energy
component for Melvin's universe expressed by the line element $(\ref
{melvinlecart})$ we need the following non-zero components of $\ell%
^{ikjl}$
\begin{eqnarray}
\ell^{0101} &=&-\frac{x^{2}+y^{2}\Lambda
^{4}}{x^{2}+y^{2}}\text{,}
\nonumber \\
\ell^{0102} &=&\frac{xy(\Lambda ^{4}-1)}{x^{2}+y^{2}}\text{,}
\nonumber
\\
\ell^{0202} &=&-\frac{y^{2}+x^{2}\Lambda
^{4}}{x^{2}+y^{2}}\text{,}
\nonumber \\
\ell^{0303} &=&-1\text{.}  \label{Scomponents}
\end{eqnarray}
Equation $(\ref{Lij})$ with equations $(\ref{Sijkl})$ and $(\ref{Scomponents}%
)$ gives
\begin{equation}
L^{00}=\frac{1}{8\pi }B^{2}\Lambda ^{3}\text{.}  \label{L00}
\end{equation}
For a surface given by parametric equations $x=r\sin \theta \cos
\phi ,$ $\ y=r\sin \theta \sin \phi ,$ $\ z=r\cos \theta $ (where
$r$ is constant) one has $\mu _{\beta }=\{x/r,$ $y/r,$ $z/r\}$ and
$dS=r^{2}sin\theta d\theta d\phi $. Using equations
$(\ref{Scomponents})$ in $(\ref{LLGauss})$ over a surface
$r=constant$, we obtain
\begin{equation}
E_{LL}=\frac{1}{6}B_{o}^{2}r^{3}+\frac{1}{20}B_{o}^{4}r^{5}+\frac{1}{140}%
B_{o}^{6}r^{7}+\frac{1}{2520}B_{o}^{8}r^{9}.  \label{ELL}
\end{equation}

\subsection{The Energy-momentum complex of Papapetrou}%

The Papapetrou energy-momentum complex\cite{Papap}:
\begin{equation}
\Omega ^{ij}=\frac{1}{16\pi }{\cal {N}}_{\quad ,kl}^{ijkl},
\label{Omega}
\end{equation}
where
\begin{equation}
{\cal {N}}^{ijkl}=\sqrt{-g}\left( g^{ij}\eta ^{kl}-g^{ik}\eta
^{jl}+g^{kl}\eta ^{ij}-g^{jl}\eta ^{ik}\right) , \label{Nijkl}
\end{equation}
is also symmetric in its indices, as discussed in chapter 2.
$\Omega^{i0}$  are the energy and momentum density components. The
energy $E$ for a stationary metric is given by the expression
\begin{equation}
E_{P}=\frac{1}{16\pi }\int \int {\cal N}_{\qquad ,\beta }^{00\alpha \beta }\
\ \mu _{\alpha }dS.  \label{PapGauss}
\end{equation}
To find the energy component of the line element $(\ref{melvinlecart})$, we
require the following non-zero components of $\ {\cal N}^{ijkl}$
\begin{eqnarray}
{\cal N}^{0011} &=&-(1+\frac{x^{2}+y^{2}\Lambda ^{4}}{x^{2}+y^{2}})\text{,}
\nonumber \\
{\cal N}^{0012} &=&\frac{xy(\Lambda ^{4}-1)}{x^{2}+y^{2}}\text{,}  \nonumber
\\
{\cal N}^{0022} &=&-(1+\frac{y^{2}+x^{2}\Lambda ^{4}}{x^{2}+y^{2}})\text{,}
\nonumber \\
{\cal N}^{0033} &=&-2\text{.}  \label{Ncomponents}
\end{eqnarray}
Equations $(\ref{Ncomponents})$ in equation $(\ref{Omega})$ give
the energy density component
\begin{equation}
\Omega ^{00}=\frac{1}{8\pi }B^{2}\Lambda ^{3}\text{.}  \label{Omega00}
\end{equation}
Thus we find the same energy density as we obtained in Section
$4.3.1$. We now use Eq. $(\ref{Ncomponents})$ in
$(\ref{PapGauss})$ over a 2-surface (as in the last Section) and
obtain
\begin{equation}
E_{P}=\frac{1}{6}B_{o}^{2}r^{3}+\frac{1}{20}B_{o}^{4}r^{5}+\frac{1}{140}%
B_{o}^{6}r^{7}+\frac{1}{2520}B_{o}^{8}r^{9}.  \label{EPap}
\end{equation}
This result is expressed in geometrized units ($G=1$ and $c=1$). In the
following we restore $G$ and $c$ and get
\begin{equation}
E_{P}=\frac{1}{6}B_{o}^{2}r^{3}+\frac{1}{20}\frac{G}{c^{4}}B_{o}^{4}r^{5}+%
\frac{1}{140}\frac{G^{2}}{c^{8}}B_{o}^{6}r^{7}+\frac{1}{2520}\frac{G^{3}}{%
c^{12}}B_{o}^{8}r^{9}.
\end{equation}
The first term $\frac{B_{o}^{2}r^{3}}{6}$ is the known classical value of
energy and the rest of the terms are general relativistic corrections. The
general relativistic terms increase the value of energy.

\subsection{Discussion of Results}%

In this section we obtained the energy distribution in Melvin's
magnetic universe. We used the energy-momentum complexes of Landau
and Lifshitz, and Papapetrou. Both definitions give the same
results ($L^{00}=\Omega ^{00},E_{LL}=E_{P}$). The first term in
the energy expression (see equations $(\ref{ELL})$ and
$(\ref{EPap})$) is the well-known classical value for the energy
of the uniform magnetic field and the other terms are general
relativistic corrections. The general relativistic corrections
increase the value of the energy.

\section{The Ernst solution}                                   %
\label{sec:ernst}

Ernst\cite{Ernst} obtained an axially symmetric electrovac
solution to the  Einstein-Maxwell equations $(\ref{EMeq1})$ to
$(\ref{EMeq4})$, describing the Schwarzschild black hole in
Melvin's magnetic universe. The space-time is
\begin{equation}
ds^{2}=\Lambda ^{2}\left[ (1-\frac{2M}{r})dt^{2}-(1-\frac{2M}{r}%
)^{-1}dr^{2}-r^{2}d\theta ^{2}\right] -\Lambda ^{-2}r^{2}\sin
^{2}\theta d\phi ^{2} , \label{Ernsteq1}
\end{equation}
with $\Lambda$ given by $(\ref{Lambda})$ and the only Cartan
component of the magnetic field which differs from those given in
$(\ref{Cartan})$ is
\begin{equation}
H_{\theta } =-\Lambda ^{-2}B_{o}\left( 1-2M/r\right) ^{1/2}\sin
\theta .
\end{equation}
$M$ and $B_{o}$ are constants in the solution.
The non-zero components of the energy-momentum tensor are
\begin{eqnarray}
T_{1}^{\ 1} &=&-T_{2}^{\ 2}=\frac{B_{o}^{2}\left( 2M\sin
^{2}\theta -2r\sin
^{2}\theta +r\right) }{8\pi \Lambda ^{4}r},  \nonumber \\
T_{3}^{\ 3} &=&-T_{0}^{\ 0}=\frac{B_{o}^{2}\left( 2M\sin
^{2}\theta
-r\right) }{8\pi \Lambda ^{4}r},  \nonumber \\
T_{1}^{\ 2} &=&-T_{2}^{\ 1}=\frac{2B_{o}^{2}\left( 2M-r\right)
\sin \theta \cos \theta }{8\pi \Lambda ^{4}r}.  \label{ErnstEMTeq}
\end{eqnarray}
The Ernst solution is a black hole solution ($r=2M$ is the event
horizon). For $B_{o}=0$ it gives the Schwarzschild solution and
for $M=0$ it gives the Melvin's magnetic universe. The magnetic
field has a constant value $B_{o}$ everywhere along the axis.
Ernst pointed out an interesting feature of this solution. Within
the region $2M<<r<<B_{0}^{-1}$, the space is approximately flat
and the magnetic field approximately uniform, when $|B_{0}M|<<1$.
If the magnetic field is strong, i.e. $|B_{0}M|$ is of the order
unity, then it tends to be more concentrated near the poles
$\theta =0$\ and $\theta =\pi $.

\section{The energy associated with Schwarzschild black hole in a magnetic
universe.}
\label{sec:ernstED}

In this section we obtain the energy distribution of this
spacetime using Einstein's energy-momentum complex. As in the case
of the energy-momentum complexes of Landau and Lifshitz, and
Papapetrou, to get meaningful results for energy distribution in
the prescription of Einstein one is compelled to use ``Cartesian''
coordinates.  The line element $(\ref{Ernsteq1})$ is easily
transformed to ``Cartesian'' coordinates $t,x,y,z$ using the
standard transformation $(\ref{eqn4.9})$. We get
\begin{eqnarray}
ds^{2} &=&\Lambda ^{2}(1-\frac{2M}{r})dt^{2}-\left[ \Lambda ^{2}\left( \frac{%
ax^{2}}{r^{2}}\right) +\Lambda ^{-2}\left( \frac{y^{2}}{x^{2}+y^{2}}\right) %
\right] dx^{2} \nonumber\\
&-&\left[ \Lambda ^{2}\left(
\frac{ay^{2}}{r^{2}}\right) +\Lambda ^{-2}\left(
\frac{x^{2}}{x^{2}+y^{2}}\right) \right] dy^{2}
\nonumber \\
&-&\Lambda ^{2}\left[ 1+\frac{2Mz^{2}}{r^{2}(r-2M)}\right]
dz^{2}-\left[
\Lambda ^{2}\left( \frac{2axy}{r^{2}}\right) +\Lambda ^{-2}\left( -\frac{2xy%
}{x^{2}+y^{2}}\right) \right] dxdy  \nonumber \\
&-&\Lambda ^{2}\left[ \frac{4Mxz}{r^{2}(r-2M)}\right] dxdz-\Lambda
^{2}\left[ \frac{4Myz}{r^{2}(r-2M)}\right] dydz,  \label{Ernsteq2}
\end{eqnarray}
where
\begin{equation}
a=\frac{2M}{r-2M}+\frac{r^{2}}{x^{2}+y^{2}}.  \label{eqn4.10}
\end{equation}
Using the Einstein energy-momentum complex
\begin{equation}
\Theta _{i}^{k}=\frac{1}{16\pi }h_{i,\ \ l}^{kl}  \label{eqn4.13}
\end{equation}
where
\begin{equation}
h_{i}^{kl}=-h_{i}^{lk}=\frac{g_{in}}{\sqrt{-g}}\left[ -g\left(
g^{kn}g^{lm}-g^{ln}g^{km}\right) \right] _{,m}  \label{eqn4.14}
\end{equation}
the energy $E$ for a stationary metric is given by the expression
\begin{equation}
E=\frac{1}{16\pi }\int \int \int h_{0,\alpha }^{0\alpha }dxdydz,
\label{eqn4.16}
\end{equation}
and after applying the Gauss theorem, one has
\begin{equation}
E=\frac{1}{16\pi }\int \int h_{0}^{0\alpha }\mu _{\alpha }dS.
\label{eqn4.17}
\end{equation}

\subsection{Calculations}%

The determinant of the metric tensor is given by
\begin{equation}
g=-\Lambda ^{4} . \label{eqn4.18}
\end{equation}
The non-zero contravariant components of the metric tensor are
\begin{eqnarray}
g^{00} &=&\Lambda ^{-2}\frac{r}{r-2M},  \nonumber \\
g^{11} &=&\Lambda ^{-2}\left[ \frac{2Mx^{2}}{r^{3}}-\frac{x^{2}}{x^{2}+y^{2}}%
\right] -\Lambda ^{2}\left[ \frac{y^{2}}{x^{2}+y^{2}}\right] ,  \nonumber \\
g^{12} &=&\Lambda ^{-2}\left[ \frac{2Mxy}{r^{3}}-\frac{xy}{x^{2}+y^{2}}%
\right] +\Lambda ^{2}\left[ \frac{xy}{x^{2}+y^{2}}\right] ,  \nonumber \\
g^{22} &=&-\Lambda ^{-2}\left[ \frac{2My^{2}}{r^{3}}-\frac{y^{2}}{x^{2}+y^{2}%
}\right] -\Lambda ^{2}\left[ \frac{x^{2}}{x^{2}+y^{2}}\right] ,  \nonumber \\
g^{33} &=&\Lambda ^{-2}\left[ \frac{2Mz^{2}}{r^{3}}-1\right] ,  \nonumber \\
g^{13} &=&\Lambda ^{-2}\left[ \frac{2Mxz}{r^{3}}\right] ,  \nonumber \\
g^{23} &=&\Lambda ^{-2}\left[ \frac{2Myz}{r^{3}}\right] .
\label{eqn4.19}
\end{eqnarray}
The only required components of $h_{k}^{ij}$ in the calculation of
energy are the following:
\begin{eqnarray}
h_{0}^{01} &=&\frac{4Mx}{r^{3}}+(\Lambda ^{4}-1)\left[ \frac{x}{x^{2}+y^{2}}%
\right] ,  \nonumber \\
h_{0}^{02} &=&\frac{4My}{r^{3}}+(\Lambda ^{4}-1)\left[ \frac{y}{x^{2}+y^{2}}%
\right] ,  \nonumber \\
h_{0}^{03} &=&\frac{4Mz}{r^{3}}.  \label{eqn4.20}
\end{eqnarray}
Now using $(\ref{eqn4.20})$ with $(\ref{eqn4.17})$ we obtain the
energy distribution in the Ernst space-time.
\begin{equation}
E=M+\ \frac{1}{16\pi }\int_{\theta =0}^{\pi }\int_{\phi =0}^{2\pi
}(\Lambda ^{4}-1)r\sin \theta d\theta d\phi  .\label{eqn4.21}
\end{equation}
We substitute the value of $\Lambda $ in the above and then
integrate. We get
\begin{equation}
E=M+\frac{1}{6}B_{o}^{2}r^{3}+\frac{1}{20}B_{o}^{4}r^{5}+\frac{1}{140}%
B_{o}^{6}r^{7}+\frac{1}{2520}B_{o}^{8}r^{9}.  \label{eqn4.22}
\end{equation}
The above result is expressed in geometrized units (gravitational constant $%
G=1$ and the speed of light in vacuum $c=1$). In the following we
restore $G$ and $c$ and get
\begin{equation}
E=Mc^{2}+\frac{1}{6}B_{o}^{2}r^{3}+\frac{1}{20}\frac{G}{c^{4}}B_{o}^{4}r^{5}+%
\frac{1}{140}\frac{G^{2}}{c^{8}}B_{o}^{6}r^{7}+\frac{1}{2520}\frac{G^{3}}{%
c^{12}}B_{o}^{8}r^{9}.  \label{eqn4.23}
\end{equation}
The first term $Mc^{2}$ is the rest-mass energy of the
Schwarzschild black hole, the second term
$\frac{1}{6}B_{o}^{2}r^{3}$ is the well-known classical value of
the energy of the magnetic field under consideration, and the
rest of the terms are general relativistic corrections. For very large $%
B_{o}r$, the general relativistic contribution dominates over the
classical value for the magnetic field energy. As mentioned in
Section $2$, the gravitational field is weak for
$2M<<r<<B_{o}^{-1}$ (in $G=1,c=1$ units). Thus in the weak
gravitational field we have $B_{0}r<<1$ ; therefore, the classical
value for the magnetic field energy will be greater than the
general relativistic correction in these cases.

\subsection{ Discussion of Results}%

In the above section we considered the Ernst space-time and
calculated the energy distribution using the Einstein
energy-momentum complex. It beautifully yields the expected
result: The first term is the Schwarzschild rest-mass energy, the
second term is the classical value for energy due to the uniform
magnetic field ($E=\frac{1}{8\pi }\int \int \int B_{o}^{2}dV$,
where $dV$ is the infinitesimal volume element, yields exactly the
same value as the second term of $(\ref{eqn4.23})$ ), and the rest
of the terms are general relativistic corrections. The general
relativistic terms increase the value of the energy.

\section{Conclusion}%
In this chapter we calculated energy distributions in Melvin's
magnetic universe and Ernst space-time using prescriptions of
Landau and Lifshitz,  Papapetrou, and Einstein. We got encouraging
results for the asymptotically non-flat space-times considered
above. We also note that the results obtained indicate that the
energy-momentum complexes of Einstein, Landau and Lifshitz, and
Papapetrou give the same energy distribution for the Melvin
magnetic universe. It was believed that the results are meaningful
for energy distribution in the prescription of Einstein only when
the space-time studied is asymptotically Minkowskian. However,
recent investigations of Rosen and Virbhadra \cite{RosVir},
Virbhadra \cite{KSV95}, Aguirregabiria {\em et al.} \cite{ACV96},
and Xulu \cite{Xulu3}, \cite{Xulu5} showed that many
energy-momentum complexes can give the same and appealing results
even for asymptotically non-flat space-times. Aguirregabiria {\em
et al.} showed that many energy-momentum complexes give the same
results for any Kerr-Schild class metric. There are many known
solutions of the Kerr-Schild class which are asymptotically not
flat. For example, Schwarzschild metric with cosmological
constant. The general energy expression for any Kerr-Schild class
metric obtained by them immediately gives $E=M-(\lambda /3)r^{3}$
where $\lambda $ is the cosmological constant. This result is very
much convincing. $\lambda >0$ gives repulsive effect whereas
$\lambda <0$ gives attractive effect.

\newpage

\chapter{Total Energy of the Bianchi Type I Universes}

\label{chapter-FIVE}
\section{Introduction}   %
\label{sec:intro}A wide range of cosmological models may be
deduced from Einstein's field equations. The 1965 observation of
the cosmic microwave background radiation, by Penzias and Wilson,
was the most important cosmological discovery since Hubble's 1929
announcement that all galaxies recede from us at velocities
proportional to their distance from us. This discovery of
background radiation strongly support that some version of the big
bang theory is correct and it also resulted in some conjectures
regarding the total energy of the universe. Tryon\cite{Tryon},
assuming that our Universe appeared from nowhere about $10^{10}$
years ago, remarked that the conventional laws of physics need not
have been violated at the time of creation of the Universe. He
proposed that our Universe must have a zero net value for all
conserved quantities. The arguments he presented indicate that the
net energy of our Universe may be indeed zero. His big bang model
(in which our Universe is a fluctuation of the vacuum) predicted a
homogeneous, isotropic and closed Universe consisting of matter
and anti-matter equally. Tryon \cite {Tryon} also cited an elegant
topological argument by Bergmann that any closed universe has zero
energy.

Two decades later, the work of Cooperstock \cite{Coop94} and Rosen
\cite {Rosen} revived the interest in the investigations of the
energy of the Universe. Cooperstock \cite{Coop94} considered the
conformal Friedmann-Robertson-Walker (FRW) metric
\begin{equation}
ds^{2}=R(t)^{2}\left[ dt^{2}-\frac{dr^{2}}{1-kr^{2}}-r^{2}\left(
d\theta ^{2}+\sin ^{2}\theta \,d\varphi ^{2}\right) \right] .
\label{FRWeq1}
\end{equation}
By making use of calculations involving killing vectors,
Cooperstock and Israelit (see in \cite{Coop94}) were able to
express the covariant conservation laws $T_{\quad ;k}^{ik}=0,$ in
the form of an ordinary divergence:
\begin{equation}
\left[ \sqrt{-g}\left( T_{0}^{\,0}-\frac{3}{8\pi }\left\{ \frac{\dot{R}^{2}}{%
R^{4}}+\frac{k}{R^{2}}\right\} \right) \right] _{,0}=0.  \label{CoopEq}
\end{equation}
>From $\left( \ref{CoopEq}\right) $, Cooperstock \cite{Coop94}was able to
conclude that the total density of the universe is zero.

Rosen \cite{Rosen} considered a closed homogeneous isotropic
universe described by the Friedmann-Robertson-Walker (FRW) metric:
\begin{equation}
ds^{2}=dt^{2}-\frac{a(t)^{2}}{\left( 1+r^{2}/4\right) ^{2}}\left(
dr^{2}+r^{2}d\theta ^{2}+r^{2}\sin ^{2}\theta \,d\varphi ^{2}\right) .
\label{FRWeq2}
\end{equation}
Then using Einstein's prescription, he obtained the following
energy-momentum complex\footnote{%
To avoid any confusion we mention that we use the term energy-momentum
complex for one which satisfies the local conservation laws and gives the
contribution from the matter (including all non-gravitational fields) as
well as the gravitational field. Rosen \cite{Rosen} used the term
pseudo-tensor for this purpose. We reserve the term energy-momentum
pseudotensor for the part of the energy-momentum complex which comes due to
the gravitational field only.}
\begin{equation}
\Theta_{0}^{\,0}=\frac{a}{8\pi }\left[ \frac{3}{\left(
1+r^{2}/4\right) ^{2}}-\frac{r^{2}}{\left( 1+r^{2}/4\right)
^{3}}\right] . \label{RosEq}
\end{equation}
By integrating the above over all space, one finds that the total
energy $E$ of the universe is zero. These interesting results
fascinated some general relativists, for instance, Johri {\em et
al.} \cite{Johri}, Banerjee and Sen \cite{BanSen} and Xulu
\cite{Xulu4}. Johri {\em et al.} \cite{Johri}, using the Landau
and Lifshitz energy-momentum complex, showed that the total energy
of an FRW spatially closed universe is zero at all times
irrespective of equations of state of the cosmic fluid. They also
showed that the total energy enclosed within any finite volume of
the spatially flat FRW universe is zero at all times. In this
chapter we investigate the total energy of the Bianchi type I
universes.

\section{Bianchi type I space-times}  %

The Bianchi type I space-times are expressed by the line element
\begin{equation}
ds^{2}=dt^{2}-e^{2l}dx^{2}-e^{2m}dy^{2}-e^{2n}dz^{2},  \label{Ueqtn1}
\end{equation}
where $l$, $m$, $n$ are functions of $t$ alone. The nonvanishing components
of the energy-momentum tensor $T_{i}^{\ k}$ ( $\equiv \frac{1}{8\pi }%
G_{i}^{\ k}$, where $G_{i}^{\ k}$ is the Einstein tensor) are
\begin{eqnarray}
T_{0}^{\ 0} &=&\frac{1}{8\pi }\ \left( \ \dot{l}\dot{m}+\dot{m}\dot{n}+\dot{n%
}\dot{l}\ \right) ,  \nonumber \\
T_{1}^{\ 1} &=&\frac{1}{8\pi }\ \left( \ \dot{m}^{2}+\dot{n}^{2}+\dot{m}\dot{%
n}+\ddot{m}+\ddot{n}\ \right) ,  \nonumber \\
T_{2}^{\ 2} &=&\frac{1}{8\pi }\ \left( \ \dot{n}^{2}+\dot{l}^{2}+\dot{n}\dot{%
l}+\ddot{n}+\ddot{l}\ \right) ,  \nonumber \\
T_{3}^{\ 3} &=&\frac{1}{8\pi }\ \left( \ \dot{l}^{2}+\dot{m}^{2}+\dot{l}\dot{%
m}+\ddot{l}+\ddot{m}\ \right) .  \label{Ueqtn2}
\end{eqnarray}
The dot over $l,m,n$ stands for the derivative with respect to the
coordinate $t$. The metric given by Eq. ($\ref{Ueqtn1}$) reduces to the
spatially flat Friedmann-Robertson-Walker metric in a special case. With $%
l(t)=m(t)=n(t)$, defining $R\left( t\right) =e^{l\left( t\right) }$ and
transforming the line element ($\ref{Ueqtn1}$) to $t,x,y,z$ coordinates
according to $x=r\sin \theta \cos \phi ,\ y=r\sin \theta \sin \phi ,\
z=r\cos \theta $ gives
\begin{equation}
ds^{2}=dt^{2}-\left[ R\left( t\right) \right] ^{2}\left\{ dr^{2}+r^{2}\left(
d\theta ^{2}+sin^{2}\theta d\phi ^{2}\right) \right\} ,  \label{Ueqtn3}
\end{equation}
which describes the well-known spatially flat Friedmann-Robertson-Walker
space-time.

\section{Energy distribution in Bianchi type I space-times} %

The Bianchi type $I$ solutions, under a special case, reduce to
the spatially flat FRW solutions. Banerjee and Sen \cite{BanSen}
studied the Bianchi type $I$ solutions, using the Einstein
energy-momentum complex,  and found that the total (matter plus
field) energy density is zero everywhere. As the spatially flat
FRW solution is a special case of the Bianchi type $I$ solutions,
one observes that the energy-momentum complexes of Einstein and
Landau and Lifshitz give the same result for the spatially flat
FRW solutions. Because there is a perception that different
complexes could give different and hence meaningless results for a
given metric, we \cite{Xulu4} investigated energy-momentum
distribution in Bianchi type I space-times using energy-momentum
complexes of Landau and Lifshitz, Papapetrou, and Weinberg. The
results of these investigations are presented in this section.

\subsection{The Landau and Lifshitz energy-momentum complex}%

In order to calculate the energy and momentum density components of the line element $(%
\ref{Ueqtn1})$ using the symmetric energy-momentum complex of
Landau and Lifshitz \cite{LL}:
\begin{equation}
L^{ij}=\frac{1}{16\pi }{\cal S}_{\quad ,kl}^{ijkl}\text{,}  \label{Ueqtn4}
\end{equation}
where
\begin{equation}
{\cal S}^{ijkl}=-g(g^{ij}g^{kl}-g^{ik}g^{jl}), \label{Ueqtn5}
\end{equation}
the required nonvanishing components of ${\cal S}^{ijkl}$ are
\begin{eqnarray}
{\cal S}^{0101} &=&-e^{2m+2n}\text{,}  \nonumber \\
{\cal S}^{0110} &=&e^{2m+2n} \text{,}  \nonumber \\
{\cal S}^{0202} &=&-e^{2l+2n}\text{,}  \nonumber \\
{\cal S}^{0220} &=&e^{2l+2n} \text{,}  \nonumber \\
{\cal S}^{0303} &=&-e^{2l+2m} \text{,}  \nonumber \\
{\cal S}^{0330} &=&e^{2l+2m}\text{.}  \label{Ueqtn9}
\end{eqnarray}
Using the above results in $(\ref{Ueqtn4})$ and $(\ref{Ueqtn5})$
we obtain the energy density and energy current (momentum) density
components, respectively as:
\begin{equation}
L^{00}=L^{\alpha 0}=0 \text{.}  \label{Ueqtn10}
\end{equation}
Hence the energy and momentum components
\begin{equation}
P^{i}=\int \int \int L^{i0}dx^{1}dx^{2}dx^{3}  \label{Ueqtn8}
\end{equation}
vanish.

\subsection{The Energy-momentum complex of Papapetrou}  %

In order to calculate the energy and momentum density components of the line element $(%
\ref{Ueqtn1})$ using the symmetric energy-momentum complex of
Papapetrou \cite{Papap}:
\begin{equation}
\Omega ^{ij}=\frac{1}{16\pi }{\cal {N}}_{\quad ,kl}^{ijkl}  \label{Ueqtn11}
\end{equation}
where
\begin{equation}
{\cal {N}}^{ijkl}=\sqrt{-g}\left( g^{ij}\eta ^{kl}-g^{ik}\eta
^{jl}+g^{kl}\eta ^{ij}-g^{jl}\eta ^{ik}\right)   \label{Ueqtn12}
\end{equation}
and $ \eta ^{ik}$ is the Minkowski metric, we require the
following nonvanishing components of ${\cal {N}}^{ijkl}$:
\begin{eqnarray}
{\cal {N}}^{0011} &=&-(1+e^{-2l})e^{l+m+n}\text{,}  \nonumber \\
{\cal {N}}^{0110} &=&e^{-l+m+n}\text{,}  \nonumber \\
{\cal {N}}^{0022} &=&-(1+e^{-2m})e^{l+m+n}\text{,}  \nonumber \\
{\cal {N}}^{0220} &=&e^{l-m+n}\text{,}  \nonumber \\
{\cal {N}}^{0033} &=&-(1+e^{-2n})e^{l+m+n}\text{,}  \nonumber \\
{\cal {N}}^{0330} &=&e^{l+m-n}\text{,}  \nonumber \\
{\cal {N}}^{0101} &=&{\cal {N}}^{0202}={\cal
{N}}^{0303}=e^{l+m+n}\text{.} \label{Ueqtn16}
\end{eqnarray}
Using the above results in $(\ref{Ueqtn11})$ and $(\ref{Ueqtn12})$
we obtain the energy density component $\Omega ^{00}$ \ and
momentum (energy current) density components$\Omega ^{\alpha 0}$
as:
\begin{equation}
\Omega^{00}=\Omega^{\alpha 0}=0 \text{.}  \label{Ueqtn17}
\end{equation}
Again, we find that the energy and momentum components
\begin{equation}
P^{i}=\int \int \int \Omega ^{i0}dx^{1}dx^{2}dx^{3}  \label{Ueqtn15}
\end{equation}
vanish.

\subsection{The Weinberg energy-momentum complex} %

The energy and momentum density components of the line element $(%
\ref{Ueqtn1})$ were calculated using the symmetric energy-momentum
complex of Weinberg \cite{Weinberg}:
\begin{equation}
W^{ij}=\frac{1}{16\pi }{\Delta }_{\quad ,i}^{ijk}  \label{Ueqtn18}
\end{equation}
where
\begin{equation}
{\Delta }^{ijk}=\frac{\partial h_{a}^{a}}{\partial x_{i}}\eta ^{jk}-\frac{%
\partial h_{a}^{a}}{\partial x_{j}}\eta ^{ik}-\frac{\partial h^{ai}}{%
\partial x^{a}}\eta ^{jk}+\frac{\partial h^{aj}}{\partial x^{a}}\eta ^{ik}+%
\frac{\partial h^{ik}}{\partial x_{j}}-\frac{\partial h^{jk}}{\partial x_{i}}
\label{Ueqtn19}
\end{equation}
and
\begin{equation}
h_{ij}=g_{ij}-\eta _{ij}\text{.}  \label{Ueqtn20}
\end{equation}
$\eta _{ij}$ \ is the Minkowski metric. In this case, using the
equations $(\ref{Ueqtn1})$ and $(\ref{Ueqtn19})$, we find that all
the components of ${\Delta }^{ijk}$ vanish. Thus Eq.
$(\ref{Ueqtn18})$ yields
\begin{equation}
W^{ik}=0.  \label{Ueqtn23}
\end{equation}
Therefore the energy and momentum components
\begin{equation}
P^{i}=\int \int \int W^{i0}dx^{1}dx^{2}dx^{3} \label{Ueqtn24}
\end{equation}
also vanish.
\section{ Conclusion}%

In recent years some researchers showed interest in studying the
energy content of the universe in different models (see
Cooperstock \cite{Coop94} , Rosen \cite{Rosen}, Johri {\em et al.}
\cite{Johri}, Banerjee and Sen \cite {BanSen} . Cooperstock
\cite{Coop94} investigated energy density for conformal
Friedmann-Robertson-Walker metric and by making use of
calculations involving killing vectors he was able to deduce that
the total energy density is equal to zero. Rosen \cite{Rosen}
studied the total energy of a closed homogeneous isotropic
universe described by the FRW metric using the Einstein
energy-momentum complex,  and found that to be zero. Using the
Landau and Lifshitz prescription of energy and momentum Johri {\em
et al.} \cite {Johri} demonstrated that (a) the total energy of an
FRW spatially closed universe is zero at all times irrespective of
equations of state of the cosmic fluid and (b) the total energy
enclosed within any finite volume of the spatially flat FRW
universe is zero at all times. Banerjee and Sen \cite{BanSen}
showed that the energy and momentum density components vanish in
the Bianchi type I space-times (they used the energy-momentum
complex of Einstein).

It is usually suspected that different energy-momentum complexes
could give different results for a given geometry. Therefore,
we\cite{Xulu4} extended the investigations of Banerjee and Sen
with three more energy-momentum complexes (proposed by Landau and
Lifshitz, Papapetrou, and Weinberg) and found the same results
(see equations $(\ref{Ueqtn10})$, $(\ref{Ueqtn17})$ and
$(\ref{Ueqtn23})$) as reported by them. Note that the energy
density component of the energy-momentum tensor is not zero for
the Bianchi type I
solutions (see Eq. $(\ref{Ueqtn2})$); however, it is clear from equations $(%
\ref{Ueqtn10})$, $(\ref{Ueqtn17})$ and $(\ref{Ueqtn23})$ that the
total energy density (due to matter plus field, as given by the
energy-momentum complexes) vanishes everywhere . This is because
the energy contributions from the matter and field inside an
arbitrary two-surface in Bianchi type I space-times cancel each
other. These results illustrate the importance of energy-momentum
complexes (as opposed to the perception against them that
different complexes could give different and hence meaningless
results for a given metric) and also supports the viewpoint of
Tryon.


\newpage

\chapter{M\o ller Energy for the Kerr-Newman Metric} %
\label{chapter-SIX}
\section{Introduction}%
The investigations of Hawking, Israel, Carter, Robinson and others
on the properties of black holes built-up to the proof of the
so-called {\em ``No Hair"} theorem which shows that black holes
are completely described by only three quantities namely mass $M$,
charge $e$, and angular momentum $a$ (see in Israel
\cite{300years}). Hence, the stationary axially symmetric and
asymptotically flat Kerr-Newman solution which is parameterized by
mass $M$, charge $e$, and angular momentum $a$, is the most
general black hole solution to the Einstein-Maxwell equations.
This solution describes the exterior gravitational and
electromagnetic fields of a charged rotating object. When $e=0$,
it describes the Kerr family of axially symmetric  solutions that
give the geometry of space-time surrounding rotating uncharged
objects. When $a=0$, it describes the spherically symmetric
Reissner-Nordstr\"{o}m solution of charged non-rotating black
holes. For both $a=0$ and $e=0$ the solution reduces to the
spherically symmetric Schwarzschild solution of the simplest type
of black hole which is only characterized by the mass $M$. The
Kerr-Newman solution, as the most general black hole solution, is
therefore of vital importance in studying the geometry surrounding
compact objects. In this chapter we investigate energy
distribution in Kerr-Newman space-time.

The energy distribution in the Kerr-Newman (KN) space-time was
earlier computed by Cohen and de Felice \cite{CodeFe} using
Komar's prescription. Virbhadra (\cite{KSV90a},\cite{KSV90b})
showed that, up to the third order of the rotation parameter, the
energy-momentum complexes of Einstein and Landau-Lifshitz give the
same and reasonable energy distribution in the KN space-time when
calculations are carried out in Kerr-Schild Cartesian coordinates.
Cooperstock and Richardson \cite{CoopRi} extended the Virbhadra
energy calculations up to the seventh order of the rotation
parameter and found that these definitions give the same energy
distribution for the KN metric. Aguirregabiria {\it et al.}
\cite{ACV96} performed exact computations for the energy
distribution in KN space-time in Kerr-Schild Cartesian
coordinates. They showed that the energy distribution in the
prescriptions of Einstein, Landau-Lifshitz, Papapetrou, and
Weinberg (ELLPW) gave the same result. In a recent paper
Lessner\cite{Lessner}\ in his analysis of M\o ller's
energy-momentum expression concludes that it is a powerful concept
of energy and momentum in general relativity. We\cite{Xulu6}
evaluated the energy distribution in KN field using the M\o ller
energy-momentum prescription. The results of our investigation
\cite{Xulu6} are given below.

In a series of papers (\cite {Coop93},\cite {Coop99},\cite
{Coop2000}), Cooperstock has propounded a hypothesis which
essentially states that the energy and momentum in a curved
space-time are confined to the regions of non-vanishing
energy-momentum tensor $T_i^{\ k}$ of the matter and all
non-gravitational fields. It is of interest to investigate whether
or not the {\em Cooperstock hypothesis} holds good. Our results
(\cite{Xulu6},\cite{Xulu7}) and the recent results of Bringley
\cite{Bringley} support this hypothesis. In this chapter we use
the Kerr-Newman space-time for testing the Cooperstock hypothesis.
We first give the KN metric, followed by M\o ller energy
distribution and a discussion of results.

\section{The Kerr-Newman metric}%

The Kerr-Newman metric in Boyer-Lindquist coordinates $(t,\rho
,\theta ,\phi )$ is expressed by the line element:
\begin{equation}
ds^{2}=\frac{\Delta }{r_{0}^{2}}[dt-a\sin ^{2}\theta d\phi ]^{2}-\frac{ \sin
^{2}\theta }{r_{0}^{2}}[\left( \rho ^{2}+a^{2}\right) d\phi -a dt]^{2}-\frac{%
r_{0}^{2}}{\Delta }d\rho ^{2}-r_{0}^{2}d\theta ^{2},
\label{KNMetricBL}
\end{equation}
where $\Delta := \rho^2- 2M \rho + e^2+a^2$ and $r_{0}^{2} :=
\rho^{2}+a^{2}\cos^{2}\theta $. $M$, $e$ and $a$ are respectively {\em mass}%
, {\em electric charge} and {\em rotation} parameters and the
corresponding electromagnetic field tensor is:
\begin{equation}
\begin{split}
F &= e r_{0}^{-4}[(\rho^{2}-a^{2}\cos^{2}\theta) d\rho \wedge%
dv -2a^{2}\rho \cos \theta d\theta \wedge dv \\%
 &-a \sin^2 \theta (\rho^2 -a^2 \cos^2 \theta)%
d\rho \wedge d\phi +2a\rho(\rho^2 + a^2) \cos \theta \sin \theta%
d\theta \wedge d\phi ].%
\end{split} \label{Feq}%
\end{equation}%
(For details see in Carter\cite {Carter68}.) The KN space-time has
$ \{\rho = constant\}$ null hypersurfaces for $g^{\rho \rho}=0$,
which are given by
\begin{equation}
\rho _{\pm}=M \pm \sqrt{M^{2}-e^{2}-a^{2}} \text{\ .}  \label{nullhyp}
\end{equation}
There is a ring curvature singularity $\rho =0$ in the KN
space-time. This space-time has an event horizon at $\rho =
\rho_+$.  It describes a black hole if and only if $M^{2}\geq
e^{2}+a^{2}$.

The Boyer-Lindquist coordinates are singular at  $\rho
= \rho_{\pm}$. Therefore, to remove this coordinate singularity
$t$ is replaced with a null coordinate $v$, and $\phi$ with an
`untwisting' angular coordinate ${\varphi }$ using the following
transformation:
\begin{eqnarray}
dt &=& dv - \frac{\rho ^{2}+a^{2}}{\Delta }d\rho ,  \nonumber \\
d\phi &=& d\varphi - \frac{a}{\Delta }d\rho,
\end{eqnarray}
and thus we express the KN metric in advanced
Eddington-Finkelstein coordinates (Misner {\em et. al.} \cite{MTW}
refer to these as Kerr coordinates) $(v, \rho, \theta, \varphi)$
as:
\begin{eqnarray}
ds^{2} &=& \left( 1-\frac{2M\rho}{r_{0}^{2}}+\frac{e^{2}}{r_{0}^{2}}%
\right)dv^{2} -2dv\ d\rho +\frac{2a\sin^2 \theta} {r_{0}^{2}}\left(
2M\rho-e^{2}\right) dv\ d\varphi -r_{0}^{2}d\theta^{2}  \nonumber \\
&& + 2a \sin^{2} \theta d\rho d\varphi -\left[ \left(\rho^2+a^2\right)\sin^2%
\theta + \frac{2M \rho-e^2}{{r_o}^2} a^2 \sin^4\theta \right]
d\varphi^2 . \label {KNMetricAEF}
\end{eqnarray}
The energy-momentum complexes of Einstein, Landau-Lifshitz,
Papapetrou and Weinberg are coordinate-dependent and require the
use of quasi-Cartesian coordinates. Thus we transform the above to
Kerr-Schild Cartesian coordinates $(T,x,y,z)$ according to:
\begin{eqnarray}
T &=& v - \rho , \nonumber \\
x &=& \sin\theta \left(\rho \cos\varphi + a \sin\varphi\right),  \nonumber \\
y &=& \sin\theta \left(\rho \sin\varphi - a \cos\varphi\right),  \nonumber \\
z &=& \rho \cos\theta,
\end{eqnarray}
and one has the line element
\begin{eqnarray}
ds^2 &=& dT^2 - dx^2 -dy^2 -dz^2 - \frac{\left(2m\rho-e^2\right)\rho^2}{%
\rho^4+a^2 z^2} \times  \nonumber \\
&& \left( dT + \frac{\rho}{a^2+\rho^2} \left(x dx + y dy\right) + \frac{a}{%
a^2+\rho^2} \left(y dx - x dy\right) +\frac{z}{\rho} dz \right)^2
.\label{KNMetricKS}
\end{eqnarray}
The components of the energy-momentum tensor $T_a^{`b}$, in
quasi-Cartesian coordinates, are given by
\begin{equation}
T_{a}^{~b}=\frac{e^2}{8\pi r_0^6}
\begin{bmatrix}
     R^2+a^2  & 2ay            &-2ax     &    0\\
     -2ay     & -(R^2+a^2-2x^2) & 2xy    &    2xz\\
     2ax      & 2xy             & -(R^2+a^2-2y^2) & 2yz\\
     0        & 2xz             & 2yz     & -(R^2+a^2-2z^2)
\end{bmatrix}
\end{equation}
where
\begin{equation}
r_{0}^{4}=(R^{2}-a^2)^2+4a^2z^2
\end{equation}
(for details see in Cooperstock and Richardson\cite{CoopRi}).

\section{Energy distribution in Kerr-Newman metric.}%

In this Section we first give the energy distribution in the KN
space-time obtained by some authors and then using the M\o ller
energy-momentum complex we obtain the energy distribution for the
same space-time.
\subsection{Previous results}%

The energy distribution in Komar's prescription obtained by Cohen
and de Felice\cite{CodeFe}, using the KN metric ($\ref
{KNMetricBL}$) in Boyer-Lindquist coordinates, is given by
\begin{equation}
E_{{\rm K}} = M-\frac{e^{2}}{2\rho }\left[ 1+\frac{(a^{2}+\rho^{2})}{a\rho }
\arctan \left( \frac{a}{\rho }\right) \right] \text{.}  \label{EKomar}
\end{equation}
(The subscript K on the left hand side of the equation refers to
Komar.) Aguirregabiria {\it et al.}\cite{ACV96} studied the
energy-momentum complexes of Einstein, Landau-Lifshitz, Papapetrou
and Weinberg for the KN metric. They showed that these definitions
give the same results for the energy and energy current densities.
They used the KN metric ($\ref {KNMetricKS}$) in Kerr-Schild
Cartesian coordinates. They found that these definitions give the
same result for the energy distributon for the KN metric, which is
expressed as
\begin{equation}
E_{{\rm ELLPW}} = M-\frac{e^{2}}{4\rho }\left[ 1+\frac{(a^{2}+\rho^{2})}{%
a\rho }\arctan \left( \frac{a}{\rho }\right) \right] \text{.}  \label{EELLPW}
\end{equation}
(The subscript ELLPW on the left hand side of the above equation
refers to the Einstein, Landau-Lifshitz, Papapetrou and Weinberg
prescriptions.) It is obvious that the Komar definition gives a
different result for the Kerr-Newman metric as compared to those
obtained using energy-momentum complexes of ELLPW. However, for
the Kerr metric ($e=0$) all these definitions yield the same
results. These results obviously support the Cooperstock
hypothesis.

\subsection{The M\o ller energy distribution}%

In order to calculate the energy and momentum density components
of the Kerr-Newman metric using the M\o ller energy-momentum
complex \cite {Moller58} ${\Im }_{i}^{\ k}$\  given by
\begin{equation}
{\Im }_{i}^{\ k}=\frac{1}{8\pi }{\chi}_{i\ \ ,l}^{\ kl} \quad \text{,}
\label{MollerEMC}
\end{equation}
where the antisymmetric superpotential ${\chi}_{i}^{\ kl}$ is
\begin{equation}
{\chi}_{i}^{\ kl} = - {\chi}_{i}^{\ lk} = \sqrt{-g} \left[ g_{in,m}-g_{im,n}%
\right] g^{km}g^{nl},  \label{Chi}
\end{equation}
the only required non-vanishing component of ${\chi}_{i}^{\ kl}$
is
\begin{equation}
{\chi }_{0}^{\ 01}=\frac{-2(\rho ^{2}+a^{2})\sin \theta }{(\rho
^{2}+a^{2}\cos ^{2}\theta )^{2}}\left( Ma^{2}\cos ^{2}\theta
-M\rho ^{2}+e^{2}\rho \right).
\end{equation}
Using the above expression in
\begin{equation}
E = \frac{1}{8\pi } \int \int {\chi}_{0}^{\ 0\beta }\ \mu _{\beta
}\ dS \label{energy},
\end{equation}
for the energy $E$ of a stationary metric, (where $\mu _{\beta }$
is the outward unit normal vector over an infinitesimal surface
element $dS$), we then obtain the energy $E$ inside a surface with
$\{\rho =constant\}$\ given
\begin{equation}
E_{{\rm M\o l}}=M-\frac{e^{2}}{2\rho }\left[ 1+\frac{(a^{2}+\rho ^{2})}{%
a\rho }\arctan \left( \frac{a}{\rho }\right) \right] \text{.}  \label{EMol}
\end{equation}
(The subscript M\o l on the left hand side of this equation refers to M\o %
ller's prescription.)

\subsection{Discussion of Results}%

The above result (Eq. \ref{EMol}) , obtained using M\o ller's
complex, agrees with the energy distribution (Eq. \ref{EKomar})
obtained by Cohen and de Felice\cite{CodeFe} in Komar's
prescription. It differs by a factor of two in the second term of
the energy distribution  from that (Eq. \ref{EELLPW}) computed by
Aguirregabiria {\it et al.} using ELLPW complexes. However, in
both cases the energy is shared by both the interior and exterior
of the KN black hole. It is clear that the definitions of ELLPW,
Komar, and now that of M\o ller also upholds the Cooperstock
hypothesis for the KN metric. The total energy ($\rho \rightarrow
\infty $ in all these energy expressions) give the same result
$M$.
\begin{figure}[hbt]
\centering
 \mbox{\epsfig{figure=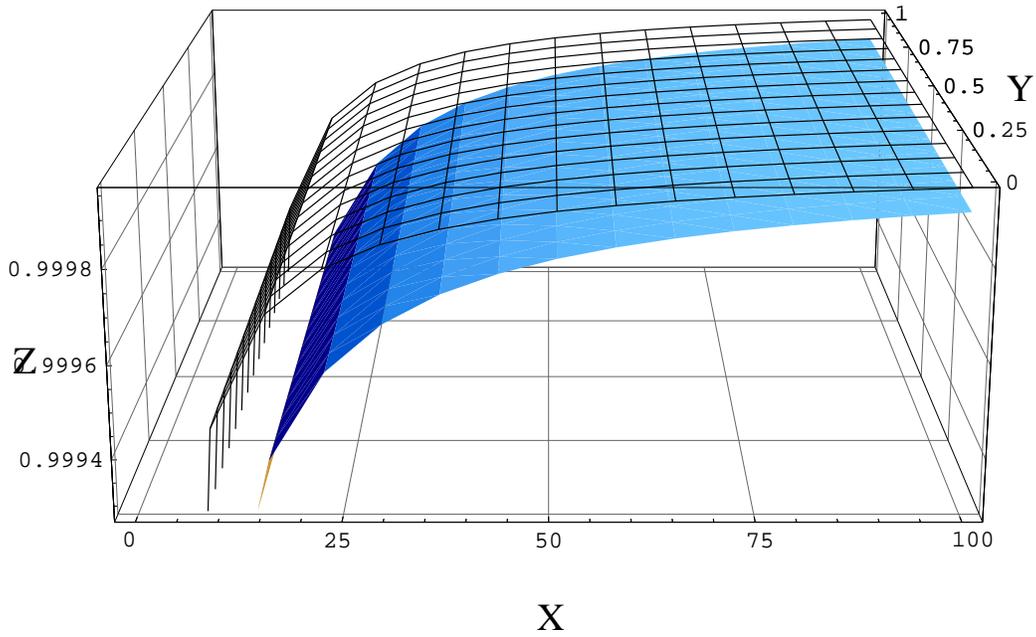,width=\linewidth}}
\caption{ ${\cal E}_{\rm ELLPW}$ and  ${\cal E}_{\rm KM}$ on
         Z-axis are plotted against  ${\cal R}$ on X-axis and ${\cal S}$ on
         Y-axis for ${\cal Q} = 0.1$. The upper (transparent one) and lower
         surfaces are for ${\cal E}_{\rm ELLPW}$ and  ${\cal E}_{\rm KM}$
         respectively.}
\label{fig6.1}
\end{figure}

Now defining
\begin{eqnarray}
&&{\cal E}_{{\rm ELLPW}}:=\frac{E_{{\rm ELLPW}}}{M},\quad {\cal E}_{{\rm %
ELLPW}}:=\frac{E_{{\rm ELLPW}}}{M},\quad {\cal E}_{{\rm KM}}:=\frac{E_{{\rm K%
}}}{M}=\frac{E_{{\rm M\o l}}}{M},  \nonumber \\
&&{\cal S}:=\frac{a}{M},\quad {\cal Q}:=\frac{e}{M},\quad {\cal R}:=\frac{%
\rho }{M}
\end{eqnarray}
the equations $(\ref{EKomar})$, $(\ref{EELLPW})$ and $(\ref{EMol})$ may be
expressed as
\begin{equation}
{\cal E}_{{\rm ELLPW}}=1-\frac{{\cal Q}^{2}}{4{\cal R}}\left[ 1+\left( \frac{%
{\cal S}}{{\cal R}}+\frac{{\cal R}}{{\cal S}}\right) \arctan \left( \frac{%
{\cal S}}{{\cal R}}\right) \right] ,  \label{CalEELLPW}
\end{equation}
and
\begin{equation}
{\cal E}_{{\rm KM}}=1-\frac{{\cal Q}^{2}}{2{\cal R}}\left[ 1+\left( \frac{%
{\cal S}}{{\cal R}}+\frac{{\cal R}}{{\cal S}}\right) \arctan \left( \frac{%
{\cal S}}{{\cal R}}\right) \right] \text{\ .}  \label{CalEKM}
\end{equation}
The ring curvature singularity in the KN metric is covered by the event
horizon for $\left( {\cal Q}^{2}+{\cal S}^{2}\right) \leq 1$ and is naked
for $\left( {\cal Q}^{2}+{\cal S}^{2}\right) >1$. In Fig. 1 we plot ${\cal E}%
_{{\rm ELLPW}}$ and ${\cal E}_{{\rm KM}}$ against ${\cal R}$ and
${\cal S}$ for ${\cal Q}=0.1$. As the value of ${\cal R}$
increases the two surfaces shown in the figure come closer.

\section{Conclusions}%

The prescriptions of Einstein, Landau and Lifshitz, Papapetrou,
and Weinberg used for calculating the energy-momentum distribution
in a general relativistic system restrict one to make calculations
in quasi-Cartesian coordinates. This shortcoming of singling out a
particular coordinate system prompted M\o ller\cite{Moller58} to
construct an expression which enables one to evaluate energy in
any coordinate system. According to M\o ller\cite{Moller58} this\
expression should give the same values for the total energy and
momentum as the Einstein's energy-momentum complex for a closed
system. However, M\o ller's energy-momentum complex was subjected
to some criticism (see in M\o ller\cite{Moller61},
Kovacs\cite{Kovacs}, Novotny\cite{Novotny}). Further
Komar\cite{Komar} formulated a new definition of energy in a
curved space-time. This prescription, though not restricted to the
use of ``Cartesian coordinates'', is applicable only to the
stationary space-times. The M\o ller energy-momentum complex is
neither restricted to the use of particular coordinates nor to the
stationary space-times. Recently, Lessner\cite{Lessner} pointed
out that the M\o ller definition is a powerful concept of energy
and momentum in general relativity. However, it is worth noting
that for the Reissner-Nordstr\"{o}m metric $E_{{\rm ELLPW}} = M -
e^2/(2 \rho)$ (the Penrose definition also gives the same
result\cite{KSV99} and this provides the weak field limit) whereas
$E_{{\rm KM\o l}} = M - e^2/\rho$ does not give the weak field
limit. This question must be investigated carefully. In the next
chapter we investigate the Cooperstock hypothesis for the
non-static space-times with M\o ller's energy-momentum complex.

\newpage

\chapter{Energy of the Nonstatic Spherically Symmetric Metrics}

\section{Introduction} \label{sec:intro}

In a recent paper, Virbhadra\cite{KSV99} investigated whether or
not the energy-momentum complexes of Einstein, Landau and Lifshitz
(LL) , Papapetrou, and Weinberg give the same energy distribution
for the most general nonstatic spherically symmetric metric and,
contrary to previous results of many asymptotically flat
spacetimes \cite
{ACV96,ChaVir96,ChaVir95,CoopRi,KSV90a,KSV90b,KSV91,KSV92,VirPar93,VirPar94,KSV95,KSV97,Xulu98a,Xulu98b}
 and asymptotically non-flat
spacetimes \cite{RosVir,KSV95,Xulu3,Xulu4}, he found that these
definitions disagree. He observed that the energy-momentum complex
of Einstein gave a consistent result for the Schwarzschild metric
whether one calculates in Kerr-Schild Cartesian coordinates or
Schwarzschild Cartesian coordinates. The prescriptions of LL,
Papapetrou and Weinberg furnish the same result as in the Einstein
prescription if computations are carried out in Kerr-Schild
Cartesian coordinates; however, they disagree with the Einstein
definition if computations are done in Schwarzschild Cartesian
coordinates. Thus, the definitions of LL, Papapetrou and Weinberg
do not furnish a consistent result. Based on this and some other
investigations (see also in Bergqvist\cite {Bergqv}, Bernstein and
Tod \cite{BeinstTod}, Virbhadra concluded that the Einstein method
seems to be the best among all known (including quasi-local mass
definitions) for energy distribution in a space-time.

In the previous chapter we highlighted Lessner's arguments
indicating the importance of the M\o ller energy-momentum
expression. So in the present chapter we wish to revisit the M\o
ller energy-momentum prescription by presenting the result of our
investigation \cite{Xulu7} of the energy distribution in the most
general nonstatic spherically symmetric space-time using M\o ller's energy-momentum complex. This result is compared with the
Virbhadra energy expression obtained by using the energy-momentum
complex of Einstein. We also discuss some examples of energy
distributions in different prescriptions. In the next section we
give the energy expression obtained by Virbhadra\cite{KSV99}.

\section{Virbhadra's  energy result}%

The most general nonstatic spherically symmetric space-time is
described by the line element
\begin{equation}
ds^{2}=\alpha (r,t)\,dt^{2}-\beta (r,t)\,dr^{2}-2\gamma (r,t)\,dt\,dr-\sigma
(r,t)\,r^{2}\,(d\theta ^{2}+\sin ^{2}\theta \,d\phi ^{2}) \text{.}
\label{Genmetric}
\end{equation}
This has, amongst others,  the following well-known space-times
 as special cases: The Schwarzschild metric, Reissner-Nordstr\"{o}m
metric, Vaidya metric,  Janis-Newman-Winicour metric,
 Garfinkle-Horowitz-Strominger metric, a general non-static spherically
symmetric metric of the Kerr-Schild class (discussed  in  Virbhadra's
 paper\cite{KSV99}).
Virbhadra\cite{KSV99} explored the energy distribution in the most
general nonstatic spherically symmetric space-time
($\ref{Genmetric}$) using the energy-momentum complex of Einstein
($\ref{Eeq10}$). To compute the energy $E$ ($=P_0$) using the
$\Theta_0{}^{0}$ component of the Einstein energy-momentum complex
Virbhadra transformed the line element $(\ref{Genmetric})$ to
``Cartesian coordinates'' $(t, x, y, z)$  using $x =  r \sin\theta
\cos\phi, y = r \sin\theta  \sin\phi,  z =  r \cos\theta$ and $t$
remaining the same. Then using ($\ref{Eeq14}$) he obtained the
energy distribution which is given below:
\begin{equation}
E_{\rm Einst} =  \frac{r\left[\alpha\left(\beta-\sigma-r \sigma_{,r}\right)
            - \gamma \left(r\sigma_{,t}-\gamma\right)\right]}
           {2 \sqrt{\alpha \beta+\gamma^2}}
\label{EEinst}
\end{equation}
where comma indicates partial differentiation. In the next Section
we obtain the energy distribution for the same metric in M\o ller's formulation.

\section{Energy distribution in M\o ller's formulation}%

In this Section we use the energy-momentum complex of M\o ller to
obtain energy distribution in the most general nonstatic
spherically symmetric metric given by the equation
$(\ref{Genmetric})$. Since the M\o ller complex is not restricted
to the use of ``Cartesian coordinates'' we perform the
computations in $t,r,\theta,\phi$ coordinates, because
computations in these coordinates are  easier compared to those in
$t,x,y,z$  coordinates.

The M\o ller  energy-momentum complex ${\Im }_{i}^{\ k}$ is given
by ($\ref{Meq4}$), with the anti-symmetric superpotential
${\chi}_{i}^{\ kl}$ given by ($\ref{Meq5}$). To compute the energy
distribution
\begin{equation}
E = \frac{1}{8\pi}\int\int {\chi}_{0}^{\ 0\beta}\mu_{\beta}dS,
\label{MEnergy}
\end{equation}
 for the line element $(\ref{Genmetric})$  under consideration
we calculate
\begin{equation}
{\chi}_{0}^{\ 01}=\frac{\left( \alpha _{,r}+\gamma _{,t}\right)
    \sigma r^{2}\sin\theta}{
\left( \alpha \beta +\gamma^{2}\right) ^{1/2}}
 \text{\  ,}
 \label{chi001}
\end{equation}
which is the only required component of ${\chi}_{i}^{kl}$ for our purpose.

Using the above expression in equation  $(\ref{energy})$ we obtain the energy
distribution
 \begin{equation}
E_{\rm M\o l} = \frac{\left( \alpha _{,r}+\gamma _{,t}\right)
\sigma r^2 }{2\left( \alpha \beta +\gamma ^{2}\right)^{1/2}}\text{
\  .} \label{EMol}
\end{equation}
It is evident that the energy distribution for the most general nonstatic
spherically symmetric metric  the definitions of Einstein and M\o ller
disagree in general (compare $(\ref{EEinst})$ with $(\ref{EMol})$).
However, these furnish the same results for some space-times, for instance,
the Schwarzschild and Vaidya space-times\cite{KSV92}.  In the next Section
we will compute energy distribution in a few  space-times using
$(\ref{EEinst})$ and  $(\ref{EMol})$.

\section{\protect\bigskip Examples}
In this Section we discuss  a  few examples of space-times in the
Einstein as well  as the M\o ller prescriptions. We also test the
Cooperstock hypothesis with these examples.

\begin{enumerate}
\item {\bf The Schwarzschild solution}\\
This solution is  expressed by the line element
\begin{equation}
ds^2 = \left(1-\frac{2M}{r}\right) dt^2
    - \left(1-\frac{2M}{r}\right)^{-1} dr^2
    -r^2 \left(d\theta +\sin^2\theta^2 d\phi^2\right) \text{\ .}
\end{equation}
 Equations $(\ref{EEinst})$ and
$(\ref{EMol})$ furnish (see also in \cite{Moller58,KSV97})
\begin{equation}
 E_{\rm Einst} = E_{\rm M\o l} = M
\end{equation}
showing that these two definitions of energy distribution agree
for the Schwarzschild space-time and the above results support the
Cooperstock hypothesis.

\item {\bf The Reissner-Nordstr\"{o}m  solution}\\
  The Reissner-Nordstr\"{o}m  solution is given by
\begin{equation}
ds^2 = \left(1-\frac{2M}{r}+\frac{e^2}{r^2}\right) dt^2
    - \left(1-\frac{2M}{r}+\frac{e^2}{r^2}\right)^{-1} dr^2
    -r^2 \left(d\theta^2 +\sin^2\theta d\phi^2\right) \text{\ ,}
\end{equation}
and the antisymmetric electromagnetic field tensor
\begin{equation}
F_{tr} = \frac{e}{r^2} \text{\ ,}
\end{equation}
where $M$  and $e$ are respectively the mass and electric charge
parameters.

For this space-time  equations $(\ref{EEinst})$ and  $(\ref{EMol})$ furnish
(see also in \cite{KSV90b,CoopRi})
\begin{equation}
 E_{\rm Einst}  = M - \frac{e^2}{2r}
\end{equation}
and
\begin{equation}
 E_{\rm M\o l} = M - \frac{e^2}{r} \text{\ .}
\end{equation}
Both of these results obviously support the Cooperstock
hypothesis.
\item {\bf The Janis-Newman-Winicour  solution}

This solution has been usually incorrectly referred to in the
literature as the Wyman solution. Virbhadra\cite{KSV97} proved
that the Wyman solution is the same as the Janis-Newman-Winicour
solution. As Janis, Newman and Winicour obtained this solution
much before Wyman, Virbhadra\cite{KSV99} rightly referred to this
as the Janis-Newman-Winicour solution. This solution is given by
\begin{equation}
ds^2 = \left(1-\frac{B}{r}\right)^{\mu} dt^2
      - \left(1-\frac{B}{r}\right)^{-\mu} dr^2
      - \left(1-\frac{B}{r}\right)^{1-\mu}
           r^2 \left(d\theta^2  +\sin^2\theta \  d\phi^2\right)
\label{JNWLE}
\end{equation}
and the scalar field
\begin{equation}
\Phi = \frac{q}{B\sqrt{4\pi}} \ln\left(1-\frac{B}{r}\right),
\label{JNWPHI}
\end{equation}
where
\begin{eqnarray}
\mu &=& \frac{2M}{B}, \nonumber\\
B &=& 2 \sqrt{M^2+q^2}.
\label{MUB}
\end{eqnarray}
$M$ and $q$ are the mass and scalar charge parameters respectively.
For $q=0$ this solution  furnishes the Schwarzschild solution.

Virbhadra\cite{KSV99} computed the energy expression for this
metric using Eq. $(\ref{EEinst})$. We do the same here using
equation  $(\ref{EMol})$. Thus we  find that
\begin{equation}
 E_{\rm Einst} = E_{\rm M\o l} = M \text{,}
\end{equation}
which shows  that these two definitions of energy distribution
agree for the Janis-Newman-Winicour space-time.

\item{\bf Garfinkle-Horowitz-Strominger  solution}

The Garfinkle-Horowitz-Strominger  static spherically symmetric
asymptotically flat solution (see in \cite {GHS91})  is described
by the line element ($\ref{eqn3.1}$). In order to compute the
energy distribution in Garfinkle-Horowitz-Strominger space-time
using the energy-momentum complex of Einstein, Chamorro and
Virbhadra\cite{ChaVir96} transformed ($\ref{eqn3.1}$) to
quasi-Cartesian coordinates ($\ref{eqn3.5}$) . Then, by making use
of ($\ref{Eeq10}$) they found the following expression in Einstein
prescription:
\begin{equation}
 E_{\rm Einst}  = M - \frac{e^2}{2r}\left(1-\lambda^2\right) \text{\ .}
\label{GHSEE}
\end{equation}
We compute the energy distribution for the Garfinkle-Horowitz-Strominger
space-time using equation $(\ref{EMol})$ and obtain
\begin{equation}
 E_{\rm M\o l}  = M - \frac{e^2}{r}\left(1-\lambda^2\right) \text{\ .}
\label{GHSME}
\end{equation}
Thus, these two definitions give different results (there is a
difference of a factor $2$ in the second term) but they obviously
support the Cooperstock hypothesis.

Now defining
\begin{equation}
{\cal E}_{\rm Einst} :=  \frac{E_{\rm Einst}}{M},  \quad
{\cal E}_{\rm  M\o l} :=  \frac{E_{\rm M\o l}}{M},  \quad
{\cal Q}  :=    \frac{e}{M},  \quad
{\cal R}  :=   \frac{r}{M}
\end{equation}
the equations  $(\ref{GHSEE})$ and  $(\ref{GHSME})$ may be expressed as
\begin{equation}
{\cal E}_{\rm Einst} =
    1-\frac{{\cal Q}^{2}}{2{\cal R}} \left(1-\lambda^2\right)
\label{CalGHSEE}
\end{equation}
and
\begin{equation}
{\cal E}_{\rm M\o l} =
         1-\frac{{\cal Q}^{2}}{{\cal R}} \left(1-\lambda^2\right)
\label{CalGHSME}
\end{equation}
For $\lambda^2 = 1, {\cal E}_{\rm Einst} = {\cal E}_{\rm M\o l} = M$;
however, they differ for any other values of $\lambda^2$. For any values of
$\lambda^2 < 1$,  ${\cal E}_{\rm Einst}$  as well as ${\cal E}_{\rm M\o l}$
decrease with an increase in ${\cal Q}^2$ and increase with increase in
${\cal R}$.
${\cal E}_{\rm Einst} > {\cal E}_{\rm M\o l}$ and they asymptotically
($\cal R \rightarrow \infty$) reach the value $1$.
The situation is just opposite for any values of $\lambda^2 > 1$ :
${\cal E}_{\rm Einst}$  as well as ${\cal E}_{\rm M\o l}$
increase with an increase in ${\cal Q}^2$ and decrease with increase in $R$.
${\cal E}_{\rm Einst} < {\cal E}_{\rm M\o l}$ and they asymptotically
($\cal R \rightarrow \infty$) reach the value $1$.

We plot the energy distributions ${\cal E}_{\rm Einst}$ and
${\cal E}_{\rm M\o l}$ for  $\lambda = 0$
(Reissner-Nordstr\"{o}m space-time) in the figure 1 and for $\lambda^2 =1.2$
in  figure  2.
\end{enumerate}
\begin{figure}[hbt]
\centering
 \mbox{\epsfig{figure=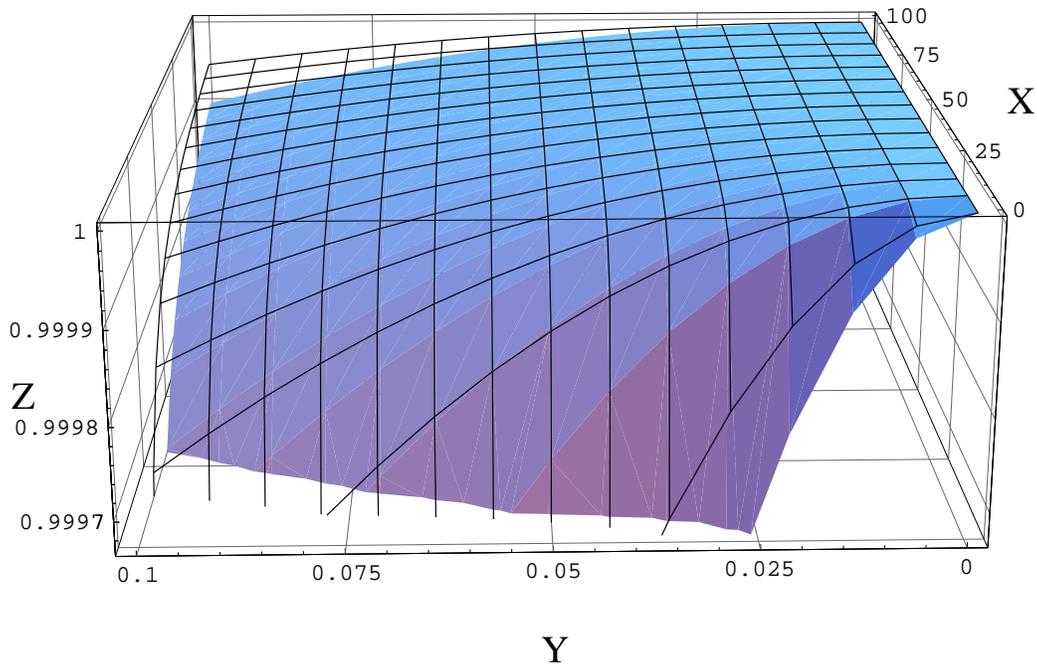,width=\linewidth}}
\caption{ ${\cal E}_{\rm Einst}$ and  ${\cal E}_{\rm M\o l}$ on
    Z-axis are plotted against  ${\cal R}$ on X-axis
    and ${\cal Q}$ on Y-axis for
   ${\lambda} = 0$ (Reissner-Nordstr\"{o}m metric).
  The upper (grid-like) and lower surfaces are for
  ${\cal E}_{\rm Einst}$ and  ${\cal E}_{\rm M\o l}$ respectively. }
  \label{fig7.1}
\end{figure}

\begin{figure}[hbt]
\centering
 \mbox{\epsfig{figure=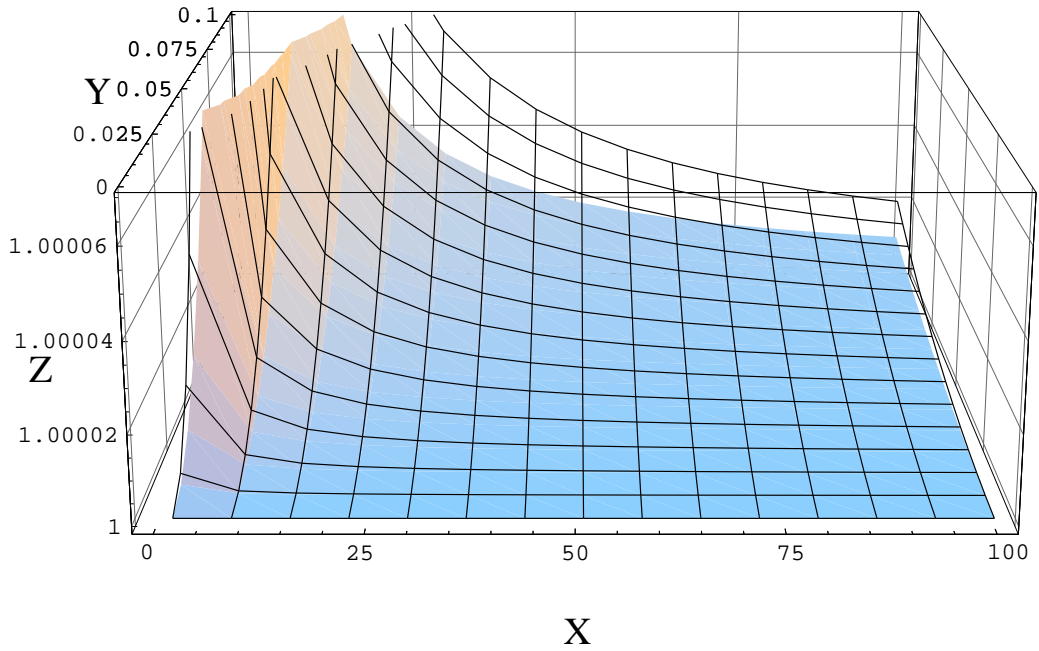,width=\linewidth}}
\caption{ ${\cal E}_{\rm Einst}$ and  ${\cal E}_{\rm M\o l}$ on
         Z-axis are plotted against  ${\cal R}$ on X-axis and
         ${\cal Q}$ on Y-axis for ${\lambda^2} = 1.2$.
         The upper (grid-like) and lower surfaces
         are for ${\cal E}_{\rm M\o l}$  and
         ${\cal E}_{\rm Einst}$ respectively.}
\label{fig7.2}
\end{figure}
\newpage


\section{\protect\bigskip Conclusion}

Based on some  analysis of the results known with many
prescriptions for energy distribution (including some  well-known
quasi-local mass definitions) in a given space-time
Virbhadra\cite{KSV99} remarked that the  formulation by Einstein
is still the best one. Lessner\cite{Lessner} argued that the M\o
ller energy-momentum expression is a powerful concept of energy
and momentum in general relativity, which motivated us to study
this further. We obtained the energy distribution for the most
general nonstatic spherically symmetric metric using  M\o ller's
definition. The result we found differs in general from that
obtained using the Einstein energy-momentum complex. However,
these agree for the Schwarzschild, Vaidya and
Janis-Newman-Winicour space-times. They disagree for the
Reissner-Nordstr\"{o}m space-time. For the Reissner-Nordstr\"{o}m
space-time $E_{\rm Einst} = M - e^2/(2 r)$ (the seminal Penrose
quasi-local mass definition also yields  the same result agreeing
with linear theory\cite{Tod}) whereas  $E_{\rm M\o l} = M -
e^2/r$. This question must be considered important. M\o ller's
energy- momentum complex is not constrained to the use of any
particular coordinates (unlike  the case of the Einstein complex);
however, as we have shown above, it does not furnish expected
result for the Reissner-Nordstr\"{o}m space-time.  We agree with
Virbhadra's conclusion  that the Einstein energy-momentum complex
is still the best tool for obtaining energy distribution in a
given space-time.
\newpage

\chapter{Summary and Conclusion}

An important feature of conserved quantities such as the energy,
momentum and angular momentum is that they play a very fundamental
role in any physical theory as they provide a first integral of
equations of motion (Nahmad-Achar and Schutz\cite{NahAchar}).
These help to solve, what would otherwise be, intractable
problems, for instance, collisions, stability properties of
physical systems etc. Conservation laws of energy-momentum,
together with the equivalence principle, played a significant role
in guiding Einstein's search for generally covariant field
equations. Evidently, it is desirable to incorporate conserved
quantities in general relativity. Energy-momentum is an important
conserved quantity whose definition has been a focus of many
investigations. Unfortunately, there is still no generally agreed
definition of energy and momentum in general relativity.

Einstein's formulation of energy-momentum conservation laws in the
form of a divergence to include contribution from gravitational
field involved the introduction of a pseudotensor quantity
$t_{j}{}^{i}$. Owing to the fact that $t_{j}{}^{i}$ is not a true
tensor (although covariant under linear transformations),
Levi-Civita, Schr\"{o}dinger, and Bauer expressed some doubts at
the validity of Einstein's energy-momentum conservation laws.
Although, Einstein defended the use of a pseudotensor quantity to
represent gravitational field and showed that his energy-momentum
pseudocomplex provides satisfactory expressions for the total
energy and momentum of closed systems, the problems associated
with Einstein's energy-momentum complex, used for calculating the
energy and momentum distribution in a general relativistic system,
was followed by many definitions, some of which are coordinate
dependent and others are not. The physical meaning of these was
questioned, and the large number of the definitions of
energy-momentum complexes only fuelled scepticism that different
energy-momentum complexes could give unacceptable different energy
distribution for a given space-time. The problems associated with
energy-momentum complexes resulted in some researchers even
doubting the concept of energy-momentum localization.

Misner \textit{et al} \cite{MTW} argued that to look for a local
energy-momentum is looking for the right answer to the wrong
question. They further argued that energy is only localizable for
spherical systems. Cooperstock and Sarracino \cite{CoopSar}
countered this point of view, arguing that if energy is
localizable in spherical systems then it is localizable in any space-times.
 Bondi\cite{Bondi}  noted that a nonlocalizable form of energy is not
admissible in general relativity. The viewpoints of Misner
\textit{et al} discouraged further study of energy localization
and on the other hand an alternative concept of energy, the
so-called quasi-local energy, was developed. To date, a large
number of definitions of quasi-local mass have been proposed. The
uses of quasi-local masses to obtain energy in a curved space-time
are not limited to a particular coordinates system whereas many
energy-momentum complexes are restricted to the use of ``Cartesian
coordinates.'' Penrose\cite {Penrose} emphasized that quasi-local
masses are conceptually very important. Nevertheless, the present
quasi-local mass definitions still have inadequacies. For
instance, Bergqvist\cite{Bergqv} considered quasi-local mass
definitions of Komar, Hawking, Penrose, Ludvigsen-Vickers,
Bergqvist-Ludvigsen, Kulkarni-Chellathurai-Dadhich, and
Dougan-Mason  and concluded that no two of these definitions give
agreed results for the Reissner-Nordstr\o m and Kerr space-times.
The shortcomings of the seminal quasi-local mass definition of
Penrose in handling the Kerr metric are discussed in Bernstein and
Tod\cite {BeinstTod}, and in Virbhadra\cite{KSV99}. On the
contrary, the remarkable work of Virbhadra, and some others, and
recent results of Chang, Nester and Chen have revived the interest
in various energy-momentum complexes.

Virbhadra, and co-workers considered many space-times and have
shown that several energy-momentum complexes give the same and
acceptable results  for a given space-time. Aguirregabiria {\em et
al.} \cite{ACV96} proved that several energy-momentum complexes
``coincide'' for any Kerr-Schild class metric. Virbhadra \cite
{KSV99} showed that for a general non-static spherically symmetric
metric of the Kerr-Schild class, the energy-momentum complexes of
Einstein, Landau and Lifshitz, Weinberg and Papapetrou furnish the
same result as Tod obtained using the Penrose quasi-local mass
definition. These are fascinating results. Recently, Chang, Nester
and Chen \cite{Changetal} demonstrated that by associating each of
the energy-momentum complexes of Einstein, Landau and Lifshitz,
M\o ller, Papapetrou, and Weinberg  with a legitimate Hamiltonian
boundary term, then each of these complexes may be said to be
quasi-local.  Quasi-local energy-momentum are obtainable from a
Hamiltonian. Hence energy-momentum complexes are useful
expressions for computing energy distributions.

Virbhadra and Parikh \cite{VirPar93} calculated, using the
energy-momentum complex of Einstein, the energy distribution for a
spherically symmetrically charged black hole in low-energy string
theory and found that the energy is confined to the interior of
the holes. Using Einstein's energy-momentum complex, Chamorro and
Virbhadra \cite{ChaVir96} studied the energy distribution
associated with static spherically symmetric charged dilaton black
holes for an arbitrary value of the coupling parameter which
controls the strength of the dilaton to the Maxwell field, and got
an acceptable result. We \cite{Xulu98a,Xulu98b}, (for a discussion
see chapter 3) computed energy distributions in these space-times
using the Tolman form of the Einstein's complex and confirmed both
of the above computations.  In the case of static spherically
symmetric charged dilaton black holes the energy distribution
depends on the value of the coupling parameter while the total
energy does not depend on this parameter.

Earlier investigations
\cite{ChaVir96,ChaVir95,CoopRi,KSV90a,KSV90b,KSV91,KSV92,VirPar93,VirPar94,KSV97}
with many asymptotically flat space-times indicated that several
energy-momentum complexes give the same and acceptable result for
a given space-time. Rosen and Virbhadra \cite{RosVir,KSV95b}
showed, using the Einstein-Rosen space-time, that even for an
asymptotically nonflat space-time many energy-momentum complexes
could give the same and persuading results. We\cite{Xulu3}
computed the energy distribution in the Ernst space-time, using
the Einstein energy-momentum complex and got encouraging results.
This prompted us to investigate energy distribution in Melvin's
magnetic universe (which is a special case of Ernst space-time)
using several different energy-momentum complexes. We\cite{Xulu5}
found that the energy-momentum complexes of Einstein, Landau and
Lifshitz, and Papapetrou give the same and acceptable energy
distribution in Melvin's magnetic universe(for a discussion see
chapter 4) . These results uphold the importance of
energy-momentum complexes.

The work of Rosen\cite{Rosen} and Cooperstock \cite {Coop94} on
the energy of the universe was followed by the investigations of
Johri \textit{et al} \cite{Johri}, and Banerjee and Sen
\cite{BanSen}. These researchers studied the energy content of the
universe using different models. Using the Einstein
energy-momentum complex, Rosen \cite{Rosen} studied the total
energy of a closed homogeneous isotropic universe described by the
Friedmann-Robertson-Walker (FRW) metric and found that  to be
zero. Cooperstock \cite {Coop94} concluded, after making use of
calculations involving killing vectors, that for a conformal
Friedmann-Robertson-Walker metric the total energy density is
equal to zero. Johri {\em et al.} \cite{Johri} showed, using the
Landau and Lifshitz definition, that the total energy of an FRW
spatially closed universe is zero at all times irrespective of
equations of state of the cosmic fluid. They also showed that the
total energy enclosed within any finite volume of the spatially
flat FRW universe is zero at all times. Using the energy-momentum
complex of Einstein, Banerjee and Sen \cite{BanSen} showed that
the energy and momentum density components vanish in the Bianchi
type I space-times. We\cite{Xulu4} extended the investigations of
Banerjee and Sen with the energy-momentum complexes of  Landau and
Lifshitz, Papapetrou, and Weinberg to check whether these
complexes could give different results for the Bianchi type I
space-times. We got the same results as obtained by Banerjee and
Sen \cite{BanSen} (for a discussion see chapter 5).

The Kerr-Newman (KN) solution is the most general black hole
solution to the Einstein-Maxwell equations. Cohen and de Felice
\cite {CodeFe} investigated energy distribution in this space-time
using Komar's prescription. This was followed by the
investigations of Virbhadra \cite{KSV90a,KSV90b} and Cooperstock
and Richardson \cite{CoopRi} who showed (up to the third order and
seventh order, respectively, of rotation parameter) that the
energy-momentum complexes of Einstein and Landau-Lifshitz give the
same and reasonable energy distribution in KN space-time.
Aguirregabiria {\it et al.} \cite{ACV96} performed exact
computations  for the KN energy distribution,in the prescriptions
of Einstein, Landau-Lifshitz, Papapetrou, and Weinberg (ELLPW) in
Kerr-Schild Cartesian coordinates. They showed that the ELLPW
complexes again gave the same energy distribution, but the second
term of their result differs by a factor of two from that obtained
by Cohen and de Felice. Seeing the results that the ELLPW all give
the same energy distribution, and Lessner's\cite{Lessner}
conclusion that M\o ller's energy-momentum expression \textit{is a
powerful representation of energy and momentum} we found it
tempting to obtain energy using this prescription. We first note
that Florides \cite{Florides} showed that for all static or
quasi-static space-times, the M\o ller's energy formula is
equivalent to the Tolman's energy formula (Eq. $ \ref{Eeq17}$).
We\cite{Xulu6} evaluated the energy distribution for the KN
space-time in M\o ller's prescription. We found that the energy
distribution in KN space-time computed using M\o ller
energy-momentum complex agrees with Komar mass obtained by Cohen
and de Felice\cite{CodeFe}. We also found that our results support
the Cooperstock hypothesis (for a discussion see chapter 6).

As already discussed, it has been shown with examples of many
space-times that several energy-momentum complexes give the same
and acceptable energy distribution for a given space-time.
Recently Virbhadra\cite{KSV99} investigated whether or not several
energy-momentum complexes give the same result for the most
general nonstatic spherically symmetric metric, and contrary to
previous results, he found that  the prescriptions of Einstein,
Landau and Lifshitz, Papapetrou, and Weinberg all give different
results. Based on some  analysis of the results known with many
prescriptions for energy distribution (including some  well-known
quasi-local mass definitions) in a given space-time
Virbhadra\cite{KSV99} \textit{concluded} that the  formulation by
Einstein is still the best one. In order to test the validity of
Virbhadra's conclusion and to further investigate the Cooperstock
hypothesis we\cite{Xulu7} used the M\o ller's energy-momentum
complex to study the energy distribution in the most general
nonstatic spherically symmetric space-time considered by Virbhadra
\cite{KSV99}. The M\o ller energy distribution differs in general
from that obtained using the Einstein energy-momentum complex.
Both prescriptions agree for the Schwarzschild, Vaidya and
Janis-Newman-Winicour space-times, but disagree for the
Reissner-Nordstr\"{o}m space-time. For the Reissner-Nordstr\"{o}m
space-time both the Einstein prescription and the Penrose
quasi-local mass definition yields the same result which agrees
with linear theory. This confirms Virbhadra's conclusion that
Einstein's prescription is still the best tool for finding energy
distribution in a given space-time.

The main weaknesses of energy-momentum complexes is that most of
these restrict one to make calculations in ``Cartesian
coordinates'', and the large number of these energy-momentum
complexes makes it difficult to decide as to which one to use to
compute energy-momentum distribution - given the suspicion these
would give different energy-momentum distributions for a given
space-time. The alternative concept of quasi-local mass is more
attractive because these are not restricted to the use of any special
coordinate system. There is a large number of definitions of
quasi-local masses. It has been shown\cite{Bergqv}  that for a
given space-time many quasi-local mass definitions do not give
agreed results. On the other hand previous results\cite{ACV96} and
our results \cite{Xulu3,Xulu4,Xulu5} show that for many
space-times several energy-momentum complexes give the same and
acceptable energy-momentum for a given space-time. The important
paper of Chang,\textit{et al}\cite{Changetal} dispels doubts
expressed about the physical meaning of these energy-momentum
complexes. Our results \cite{Xulu6,Xulu7} support Virbhadra's
conclusion that Einstein's energy-momentum complex is still the
best available method for computing energy-momentum in a given
space-time. These results also support the Cooperstock hypothesis
that the energy and momentum in a curved space-time  are confined
to the regions of non-vanishing energy-momentum tensor $T_i^{\ k}$
of the matter and all non-gravitational fields.

\newpage


\end{document}